\documentclass[12pt]{article}
\usepackage[utf8]{inputenc}
\usepackage{setspace}
\usepackage[letterpaper, margin=1in]{geometry}
\usepackage{amsthm, amsmath, mathtools, amsfonts, amssymb,bbm}
\usepackage{authblk}

\usepackage{color}
\usepackage{caption}
\usepackage{graphicx}
\graphicspath{ {figures/} }
\usepackage{subcaption}

\DeclareMathOperator*{\argmin}{arg\,min}
\DeclareMathOperator*{\argmax}{arg\,max}
\DeclareMathOperator*{\E}{\mathbb{E}}
\DeclareMathOperator*{\Prob}{\mathbb{P}}

\usepackage{algorithm}
\usepackage{algpseudocode}
\usepackage[rightbars, xcolor]{changebar}
\usepackage[normalem]{ulem}

\usepackage[authoryear]{natbib}

\RequirePackage[hidelinks]{hyperref}

\usepackage{cleveref}
\usepackage{enumitem}
\usepackage[title, titletoc]{appendix}
\usepackage{printlen}
\newtheorem{theorem}{Theorem}[section]
\newtheorem{lemma}[theorem]{Lemma}
\newtheorem{corollary}[theorem]{Corollary}
\newtheorem{proposition}[theorem]{Proposition}

\newtheorem{definition}[theorem]{Definition}

\newtheorem{remark}{Remark}[section]
\newtheorem{assumption}{Assumption}[section]

\usepackage{chngcntr} 
\counterwithin{figure}{section} 
\counterwithin{table}{section} 
\counterwithin{remark}{section}

\title{\Large From Isotonic to Lipschitz Regression:\\
A New Interpolative Perspective on Shape-restricted Estimation}
\author{Kenta Takatsu\thanks{Department of Statistics and Data Science, Carnegie Mellon University. ktakatsu@andrew.cmu.edu}, 
Tianyu Zhang\thanks{Department of Statistics and Applied Probability, University of California, Santa Barbara. tyz@ucsb.edu}, 
and Arun Kumar Kuchibhotla\thanks{Department of Statistics and Data Science, Carnegie Mellon University. arunku@cmu.edu}}
\date{}

\providecommand{\keywords}[1]
{
  \small{\textit{Keywords---} #1}
}
\begin{document}
\maketitle
\begin{abstract}
This manuscript bridges nonparametric smoothness-based and shape-restricted estimation, which may appear as two disjoint paradigms in the field. The proposed approach is motivated by a conceptually simple observation: every Lipschitz function is a sum of a monotonic and a linear function. This principle is further generalized to the higher-order monotonicity and multivariate settings. A family of estimators is proposed based on a sample-splitting procedure, inheriting desirable methodological, theoretical, and computational properties of shape-restricted estimators. The theoretical analysis provides convergence guarantees of the estimator under heteroscedastic and heavy-tailed errors, as well as adaptivity to the unknown ``complexity" of the true regression function. The generality of the proposed decomposition framework is demonstrated through new approximation results and numerical studies.
\end{abstract}

\newpage

\newpage
\section{Introduction}
The present article investigates the estimation of the conditional mean function based on observed covariates and response variables, without assuming a parametric form of the true function. Specifically, we consider $n$ independent and identically distributed (IID) observations $\{(X_i, Y_i)\}_{i=1}^n \sim P_0$, where the predictive covariate variable $X_i$ takes values in some measurable space $\Omega$, often a subset of $\mathbb{R}^d$, and the response $Y_i$ is a numerical random variable. We are interested in estimating the unknown conditional mean function $f_0(x):= \E[Y|X=x]$, which minimizes the Mean-Squared Error (MSE) among all the square-integrable functions of $X$. Defining the ``error variables'' as $\xi_i:= Y_i - f_0(X_i)$, we can always express the relationship between the covariates and the response as
\begin{align}\label{eq:regression-function}
    Y_i = f_0(X_i) + \xi_i \, \text{ for }\, i = 1, \dots, n.
\end{align}
We do not assume independence between $\xi_i$ and $X_i$, nor the normality of $\xi_i$. 

There are various approaches in nonparametric regression function estimation and we consider two of them in this work. One is based on smoothness assumptions and the other is on certain geometric properties. The former approach posits that the true signal $f_0$ resides within function spaces characterized by the degree of smoothness inherent in their functions. Examples of such spaces include H\"{o}lder, Sobolev, Besov, and bounded variation spaces \cite[Chapter 4]{gine2021mathematical}. Various estimation procedures have been proposed and proven to be minimax optimal. On the other hand, the latter approach imposes qualitative structures on $f_0$, such as monotonicity, convexity, and unimodality, often motivated by specific scientific insights. This has led to a significant body of literature on shape-restricted methodology, which seeks to approximate $f_0$ with a function that adheres to predefined shape constraints. 

One intriguing observation is that the minimax rates for estimating once-differentiable and monotone functions coincide. A similar observation can be made between twice-differentiable and convex functions. This leads to our belief that there is a unifying structure that relates these two function spaces based on smoothness and shape constraints. The central argument of this article is to identify such a structure that interpolates two function spaces. This new perspective gives rise to a family of estimators that take advantage of the properties of shape-restricted methodologies while operating within classical nonparametric function spaces based on smoothness. We demonstrate that, for some specific instances, these estimators are optimal for both smooth and shape-restricted functions \emph{simultaneously}.

The key concept can be summarized as follows: For any univariate $L$-Lipschitz function $f$, there exists a function $g$ such that 
\begin{equation}\label{eq:monotone-decompisition}
    f(x) = g(x) - Lx \, \text{ and $g$ is non-decreasing}.
\end{equation}
We will discuss this decomposition with greater generality in Section~\ref{sec:decomposition} beyond the context of univariate Lipschitz functions. The decomposition \eqref{eq:monotone-decompisition} suggests that nonparametric regression problems can be decomposed into two subproblems: (i) estimating the shape-restricted regression function $g$, and (ii) selecting the parameter $L$. The first estimation problem is relatively more challenging as it involves a function under shape constraints, which is an infinite-dimensional object. Therefore, the properties of estimation procedures leveraging this decomposition are expected to resemble those of shape-restricted estimators, as we demonstrate in this article.

Recent literature has revealed favorable properties of shape-restricted estimators that methods based on smoothness often lack. The proposed framework extends these benefits and the recent development of shape-restricted estimators to a broader context, as listed below:
\begin{itemize}
    \item \textbf{Broader nonparametric class.} 
The proposed estimator remains consistent over shape-constrained function spaces, but the addition of a monomial component substantially enlarges the target class. This extension allows the estimator to handle functions beyond standard shape-restricted models.
\item \textbf{Tuning-parameter-free nature.} 
Unlike many popular nonparametric estimators, such as kernel smoothing, shape-restricted estimators are often tuning-parameter free. The proposal in this work  solves two sets of optimization problems (one for $g$ and the other for $L$ in \eqref{eq:monotone-decompisition}). The estimation of the parameter $L$ can be treated as a model selection over a continuous domain. We establish oracle inequalities (\Cref{thm:oracle-cv,thm:oracle-cv-mom}) extending classical model selection theory \citep{massart2000some,gine2003ratio,vaart2006oracle,koltchinskii2011oracle} to heteroskedastic and heavy-tailed settings, while preserving a tuning-parameter-free implementation. 
\item \textbf{Heavy-tailed robustness.}  We establish convergence rates (measured in MSE as defined in \eqref{def:rate-of-convergence}) under heteroskedastic and heavy-tailed observations, including cases where the error has (i) finite $q$th moment or (ii) general sub-Weibull tails \citep{kuchibhotla2022moving}. Design and error assumptions are compared in Table~\ref{tab:shape-literature-design-noise}. For general function classes, least squares estimators necessarily exhibit tail-dependent rates \citep{HanWellner2019}. While such dependence disappears in the special case of isotonic regression \citep{han2018robustness}, explaining why finite variance suffices in classical results \citep{zhang2002risk}, it is a priori unclear whether comparable guarantees extend beyond isotonic spaces. We address this challenge by introducing a median-of-means tournament-based estimator that matches the isotonic regression rates under finite variance.
\item \textbf{Adaptive rates.} Shape-restricted estimators are known to exhibit adaptive risk bounds, converging faster than the global minimax rate when the true function is ``simple.''  For example, for a univariate monotone $f_0$, the shape-restricted least squares estimator attains the worst-case rate $n^{-2/3}$, while automatically adapting to the faster parametric rate $n^{-1}$ (up to polylogarithmic factors) when $f_0$ is monotone and piecewise constant. The proposed estimator retains this classical adaptivity, while strictly enlarging the class of functions over which such adaptive behavior holds. In particular, it is also adaptive for linear functions and more general low-complexity structures. The results are summarized in Table~\ref{tab:convergence-rate}.
        \item \textbf{Computation.} Despite their nonparametric nature, shape-restricted estimators can be computed efficiently. For instance, the least squares estimator over the monotone function class can be solved efficiently by the Pool Adjacent Violators Algorithm (PAVA), which ensures the overall efficiency of our proposal. 
        \item \textbf{Multivariate data.} Recent works have extended shape-restricted methods to multivariate covariates (\cite{anevski2011monotone, basu2016transformations, han2019isotonic, deng2020isotonic, kur2020suboptimality}). These methods can be used together with the proposed framework for multivariate estimation.
\end{itemize}
\begin{table}[!htbp]
\centering
\small
\renewcommand{\arraystretch}{1.2}
\setlength{\tabcolsep}{6pt}
\begin{tabular}{l l l} 
\hline
\textbf{Reference} &
\textbf{Design} &
\textbf{Noise assumptions} \\
\hline

\cite{zhang2002risk} &
Deterministic &
Homoscedastic with finite variance \\

\citeauthor{chatterjee2015risk} et al., \citeyear{chatterjee2015risk}&
Deterministic &
Homoscedastic with finite variance \\

\cite{bellec2018sharp} &
Deterministic &
Homoscedastic Gaussian \\

\cite{han2018robustness} &
Random, uniform &
Homoscedastic with finite $\|\cdot\|_{2,1}$-norm \\
\hline

Current work (MEAN) &
General random &
Heteroscedastic, finite $q$-th moment for  $q \ge 2$ \\

Current work (MEAN) &
General random &
Heteroscedastic, sub-Weibull of order $\beta$ \\

Current work (ROBUST) &
General random &
Heteroscedastic, finite variance \\
\hline
\end{tabular}
\caption{Designs and noise assumptions for representative shape-constrained regression results.}
\label{tab:shape-literature-design-noise}
\end{table}

\subsection*{Notation} 
Below, we present the glossary of notations we frequently use. For any $j \in \{1,\ldots,d\}$ with $d \ge 1$, we denote by $e_j$ the $d$-dimensional vector of zeroes with one at the $j$th position. For a univariate function $f$, $f^{(k)}$ with a positive integer $k$ denotes the $k$th derivative of $f$. We denote the marginal distribution $X$ as $P_X$. For each $P_X$-square integrable function $g$, we denote its $L_2(P_X)$-norm as $\|g\|_{L_2(P_X)} := \left(\int g ^2(x) \, dP_X(x)\right)^{1/2}$. For a real-valued function $f : \Omega \mapsto \mathbb{R}$, its supremum norm is denoted by $\|f\|_\infty := \sup_{x \in\Omega}|f(x)|$. For any deterministic sequences $\{x_n\}$ and $\{r_n\}$, we denote $x_n = O(r_n)$ if there exists a universal constant $C>0$ such that $|x_n| \le C|r_n|$ for large $n$. For a sequence of random variables $X_n$, we denote $X_n = O_P(r_n)$ if for any $\varepsilon > 0$, there exists $M > 0$ such that $\mathbb{P}(|X_n/r_n| > M ) < \varepsilon$ for large $n$. In particular, $X_n = O_P(1)$ means that the random variable $X_n$ is bounded in probability. For a vector $X := (X_{[1]},\ldots,X_{[d]})^\top\in\mathbb{R}^d$, we use $X_{[j]}\in\mathbb{R}$ to denote its $j$th entry. The set $\mathbb{N}$ denotes all non-negative integers, and $\mathbb{N}^+$ denotes strictly positive integers. We use $\mathrm{card}(I)$ to denote the cardinality of a finite set $I$. The ceiling of $x\in \mathbb{R}$, denoted by $\lceil x \rceil$, denotes the smallest integer greater or equal to $x$; similarly, the floor of $x$, denoted by $\lfloor x \rfloor$, is the greatest integer smaller or equal to $x$. Specifically, for an integer $x$, $\lceil x \rceil =\lfloor x \rfloor = x$. Throughout, we discuss the convergence rates $r_n$ of an estimator $\widehat f_n$ of $f_0$ in terms of the MSE under the marginal distribution of covariates $X \sim P_X$ such that 
\begin{align}\label{def:rate-of-convergence}
    \|\widehat f_n - f_0\|^2_{L_2(P_X)}  := O_P(r_n)\, \text { where } \, \|\widehat f_n - f_0\|^2_{L_2(P_X)} := \int |\widehat f_n(x) - f_0(x)|^2 \, dP_X(x).
\end{align}

\section{Literature Review}\label{sec:lit-review}
The proposed method in this article is closely related to nonparametric shape-restricted regression. The problem is particularly well-studied in the univariate covariate setting, where popular shape constraints include monotonicity and convexity. A common approach computes the least squares estimator (LSE) using the observed samples over the function class of interest \citep{guntuboyina2018nonparametric}. Under monotonicity, the LSE is commonly known as the isotonic regression, which can be efficiently computed using the \textit{Pool Adjacent Violators Algorithm} (PAVA) \citep{ayer1955empirical}. Some works considered estimating smooth, monotone conditional mean functions \citep{mammen1991estimating,dette2006simple}. The univariate regression estimator under the convexity has also been studied in the literature \citep{groeneboom2001estimation, guntuboyina2015global, ghosal2017univariate}. Recently, there has been a surge in theoretical analysis associated with the convergence rate of shape-restricted LSEs. The literature often emphasizes the remarkable adaptivity property of LSEs, where the estimator demonstrates a faster convergence rate depending on the local structure around the true regression function \citep{zhang2002risk, chatterjee2015risk, guntuboyina2018nonparametric, bellec2018sharp}. 

Multivariate applications of shape-restricted methods have been investigated in recent years although a comprehensive understanding of theoretical behaviors is still under development \citep{guntuboyina2018nonparametric}. \cite{han2019isotonic} studied the LSE for the multivariate isotonic regression while \cite{deng2020isotonic} proved that the LSE does not achieve minimax optimal adaptivity for all dimensions. Frequentist \citep{han2020limit} and Bayesian \citep{wang2023posterior} estimators have also been developed under coordinate-wise monotonicity constraints. The multivariate convex estimators have been studied in \cite{seijo2011nonparametric, han2016multivariate, kur2020convex, kur2022efficient, o2023spectrahedral}. Finally, additivity is a common structural assumption in nonparametric regression analysis to overcome the curse of dimensionality. In these works, studying shape constraints in conjunction with additivity helps maintain the theoretical properties of univariate shape-restricted methods. Some recent developments can be found in the works of \cite{mammen2007additive, chen2016generalized} and \cite{han2018robustness}. 

As outlined earlier, the proposed estimator leverages a decomposition of a nonparametric function into shape-restricted and parametric components. The decomposition of nonparametric functions into a parametric component has been explored in semiparametric regression. In particular, the literature has investigated two-stage estimation procedures that aim to improve the initial parametric estimator through nonparametric methods. This approach has been adopted to density and regression estimation~\citep{hjort1995nonparametric, hjort1996locally, eguchi2003local}, conditional distribution functions~\citep{veraverbeke2014preadjusted}, and additive models \citep{lin2009adaptive}. These methods often exhibit a faster rate of convergence when the initial parametric estimator is correctly specified. However, this differs from the adaptivity in the shape-restricted LSE literature, in the sense that adaptivity is \textit{not} automatic and often relies on the correct specification of the true data-generating distribution to attain a faster convergence rate. We do not make such an assumption in this article. 

We also note that the key decomposition we leverage has been known and studied in the optimization literature. In particular, \cite{zlobec2006characterization} analyzes functions with Lipschitz derivatives, which can be represented as the sum of convex and quadratic functions. This corresponds precisely to our second-order decomposition result, presented in Section~\ref{sec:decomposition} of this article. See Proposition~\ref{prop:k-monotone-decomposition} with $k=2$.

Two methods for selecting $\widehat L$ are proposed in this work. The first chooses $L$ by minimizing a standard average validation error, while the second is based on minimizing a criterion constructed using median-of-means (MOM) operators \citep{nemirovskiĭ1983problem}. It has been shown that using the MOM-based procedure can effectively improve the robustness of the estimator (compared with simply minimizing the empirical average loss). It mitigates the deterioration induced by heavy-tail noise and adversarial/atypical samples that are not from the population of interest. See, for instance, \cite{lecue2020robust, lugosi2019risk} and the references therein.

\section{Estimation Procedures}

\subsection{General two-stage estimation procedure}\label{section:general-estimation}

Let $\mathcal{C}$ be a user-specified function space, potentially motivated by some qualitative constraints\footnote{Although the proposed framework is largely motivated by the case when $\mathcal{C}$ satisfies some shape constraints, the result in this section allows $\mathcal{C}$ to be an arbitrary nonparametric function space.}. We use $\{h_L(x)\, :\, L\in\mathcal{L}\}$ to denote a collection of parametric functions parameterized by $\mathcal{L}\subset\mathbb{R}^\ell$ so that for each given $L$, the exact form of $h_L$ is specified. We will consider estimation with the following decomposition function space: for each $L\in\mathcal{L}$, define
\begin{equation}\label{eq:general-decomposition}
    \mathcal{M}(L; \Omega) = \{f:\Omega \mapsto\mathbb{R}  \mid  \exists g \in \mathcal{C}\text{ s.t. } f(x)=g(x)-h_L(x)\},
\end{equation}
where $\Omega\subset\mathbb{R}^d$. Here, we implicitly require $\mathcal{C}$ and $h_{L}$ to be defined over $\Omega$ as well. As will be shown in the following sections, these decomposition spaces contain several well-known classes for simple choices of $\mathcal{C}$ and $h_{L}$; we will go through a motivating example shortly in \Cref{sec:estimator}, with more general examples following in \Cref{sec:decomposition}.

We consider the following two-stage estimation procedure:
\begin{enumerate}
    \item Randomly split $\{1,2, \ldots, n\}$ into two disjoint subsets: $\mathcal{I}_1$ and $\mathcal{I}_2$. Denote resulting two sets of observations as $D_k := \{(X_i, Y_i)\, : \, i \in \mathcal{I}_k\}$ for $k=1, 2$.
    \item\label{ghat-step} For each $L \in \mathcal{L}$, define
    \begin{equation}\label{eq:fixed-L-gstar}
{g}^*_{L}:=\underset{g \in \mathcal{C}}{\arg \min } \, \mathbb{E}\left[\left\{Y+h_L(X)-g(X)\right\}^2\right].
    \end{equation}
    We assume $\mathcal{C}$ is regular to ensure the uniqueness of $g_L^*$. Define any estimator for $g^*_L$ as $\widehat{g}_L$ and its associated estimator of $f_0$ as
   \begin{equation}\label{eq:fixed-L-estimator-fL}
\widehat{f}_L(x):=\widehat{g}_L(x)-h_L(x).
    \end{equation}
    \item \label{cv-step}Optimize $L$ over $\mathcal{L}$ using $D_2$. We consider two approaches: 
   
        (\texttt{MEAN}) Let $\widehat L$ be (one of) the minimizers of the empirical squared risk, defined as 
        \begin{equation}\label{eq:estimator-of-L}
            \widehat L := \underset{L \in \mathcal{L}}{\arg \min }\,  \sum_{i \in \mathcal{I}_2}\{Y_i - \widehat{f}_L(X_i)\}^2.
        \end{equation}
        
        (\texttt{ROBUST}) Let $\texttt{MOM}_B(\cdot)$ denote the median of means operator related to an equipartition of size $B$ (formally defined in \Cref{def:mom}). Given $\mathcal{L}_J = \{L_1, L_2, \ldots, L_J\}$, a discretized approximation of $\mathcal{L}$ with cardinality $J$, let $\widehat L$ be the robust minimax parameter:
        \begin{equation}\label{eq:risk-estimator-robust}
            \widehat L  := \underset{L \in \mathcal{L}_J}{\arg \min }\,  \max_{K \in \mathcal{L}_J}\, \texttt{MOM}_B\bigg(\{(Y_i - \widehat{f}_L(X_i))^2 - (Y_i - \widehat{f}_{K}(X_i))^2\}_{i \in \mathcal{I}_2}\bigg).
        \end{equation}
    
    \item Return the estimator $\widehat{f}_n = \widehat{f}_{\widehat{L}}$.
\end{enumerate}

This procedure targets the projection of $f_0$ onto $\cup_{L \in \mathcal{L}} \mathcal{M}(L; \Omega)$.

Importantly, we estimate in Step \ref{ghat-step} the projection of $x\mapsto f_0(x) + h_L(x)$ onto the class $\mathcal{C}$. One example is the LSE; however, we allow for other estimators. This flexibility is important when the LSE is known to be sub-optimal, for instance, when the dimension of $X$ is large.

For many examples of $\mathcal{M}(L; \Omega)$ considered in this article, it is typical that $\mathcal{M}(L_1; \Omega) \subseteq \mathcal{M}(L_2; \Omega)$ when $L_1$ is smaller than $L_2$ under some natural ordering---one dimensional case usually yields $|L_1| < |L_2|$.
This implies that the estimator becomes more expressive as $\mathcal{L}$ is suitably enlarged. In practice, we only require $\mathcal{L}$ to be large enough so that there is an element in $\cup_{L\in\mathcal{L}}\mathcal{M}(L; \Omega)$ that approximates $f_0$ well or $f_0$ belongs to $\cup_{L\in\mathcal{L}}\mathcal{M}(L; \Omega)$. This can be usually satisfied by taking $\mathcal{L}$ as a large compact set.  The numerical details can almost always be handled by the standard optimizer, such as $\texttt{optim}$ in $\texttt{R}$. 

Among the two alternatives for optimizing $L$ in Step \ref{cv-step}, \texttt{MEAN} and \texttt{ROBUST}, each offer different advantages. Specifically, \texttt{MEAN} has computational advantages while \texttt{ROBUST} has theoretical ones.
\texttt{MEAN} can be implemented using a numerical program such as the Broyden-Fletcher-Goldfarb-Shanno (BFGS) algorithm \citep{byrd1989tool} over $\mathcal{L}\subset\mathbb{R}^\ell$. Consequently, we only need to compute the estimated risk measure in \eqref{eq:estimator-of-L} for finitely many values of $L$ that the optimizer evaluates. 

On the other hand, \texttt{ROBUST} is presented for its theoretical benefits. For instance, \Cref{thm:R2Q2-robust} in \Cref{sec:oracle-inequality} indicates that the resulting estimator exhibits a faster convergence rate than \texttt{MEAN} when the error distribution $\xi$ only has a finite (conditional) second moment. The practical implementation of \texttt{ROBUST} explicitly requires the choice of batch size $B$ and a discretized set $\mathcal{L}_J$, while this aspect was handled automatically in \texttt{MEAN}. For these reasons, the numerical study in \Cref{sec:simulation} only provides results based on \texttt{MEAN}.
\begin{remark}
    For technical reasons, we require $\mathcal{L}$ to be a bounded set whose diameter is allowed to depend on $n$. In practice, however, this requirement appears avoidable. 
    In the additional numerical study presented in \Cref{sec:tuning-parameter-free}, we observe that taking $\mathcal{L} = \mathbb{R}^\ell$ would not lead to overfitting thanks to the sample-splitting step.
\end{remark}

\subsection{Application: Univariate Isotonic Decomposition}\label{sec:estimator}

In this section, we consider one-dimensional covariates and provide more intuition for the estimation procedure. Assume $f_0$ is Lipschitz continuous: this is one of the less restrictive models when no other structural information on $f_0$ is available. Recall that a function $f: \Omega \mapsto \mathbb{R}$ is $L$-Lipschitz for a constant $L \ge 0$ if it satisfies
\begin{equation}\label{eq:def-lipschitz}
\left|f(x_2)-f(x_1)\right| \leq L\left|x_2-x_1\right| \text { for all } x_1, x_2 \in\Omega.
\end{equation}
We consider $X_i\in\Omega = [0,1]$ and omit $\Omega$ from the definition of function spaces, for instance, $\mathcal{M}(L) = \mathcal{M}(L; \Omega).$ Although the discussion in this section holds for more general $\Omega$, we establish the properties of the estimator for $\Omega = [0,1]$.

To estimate $f_0$, we choose 
\begin{equation}\label{eq:choice-C-hL}
    \mathcal{C} = \mathcal{C}(1) = \{f:[0,1]\mapsto \mathbb{R}\mid f\mbox{ is non-decreasing}\} \mbox{ and } h_L(x) = Lx,
\end{equation} with $\mathcal{L} = [-L_+,L_+] \subset\mathbb{R}$. Here, $L_+ \geq 0$ is usually a large number, potentially increasing with $n$ as well. Under these choices, the decomposition function space $\mathcal{F}$ is 
\begin{equation}\label{eq:1-decomp-function-space}
\mathcal{F}(1, L) := \left\{f:[0,1]\mapsto\mathbb{R}\mid \exists g\in\mathcal{C}(1)\text{ s.t. }f(x) = g(x) - Lx\right\}.
\end{equation}

As we stated in the Introduction, the simple summation of isotonic and linear functions creates a rich nonparametric class. Specifically, they contain Lipschitz continuous functions as stated in the following result.

\begin{proposition}\label{prop:simplied}
If $f$ is $L_0$-Lipschitz over $[0,1]$, then $f\in\mathcal{F}(1, L_0)$. Furthermore, $\{\mathcal{F}(1, L), L \geq 0\}$ is a collection of nested sets, i.e., $\mathcal{F}(1, L_0) \subseteq \mathcal{F}(1, L)$ for any $0 \le L_0 \le L$.
\end{proposition}

This result is a special case of Proposition~\ref{prop:k-monotone-decomposition} in Section~\ref{sec:decomposition}. It implies that $f_0\in \mathcal{F}(1,L)$ for any $L \ge L_0$ when $f_0$ is an $L_0$-Lipschitz function. 

\begin{remark}[Permissible values of $L$]
    Although the definition of the function space $\mathcal{F}(1, L)$ is motivated by the decomposition of $L$-Lipschitz function spaces, we allow $L$ to take negative values. By definition, Lipschitz constants cannot be negative. Allowing for negative values of $L$ yields function classes of higher expressiveness. 
\end{remark}
For any $L \ge L_0$, an $L_0$-Lipschitz function is also $L$-Lipschitz by definition. In this article, the ``true'' Lipschitz constant of a function $f$ refers to the smallest $L$ such that \eqref{eq:def-lipschitz} holds. Suppose the true constant for $f_0$ is $L_0$ and it is known, then we can construct the following estimator:

\begin{equation}\label{eq:fixed-L-estimator}
\widehat{f}_{L_0}(x) :=\widehat{g}_{L_0}(x)-L_0 x,
\end{equation}
where
\begin{equation}\label{eq: gL_fixed_L}
\widehat{g}_{L_0} :=\underset{g \in \mathcal{C}(1)}{\arg \min } \sum_{i=1}^n\left\{Y_i + L_0 X_i - g(X_i)\right\}^2.
\end{equation}

The isotonic LSE provides a left-continuous, piecewise-constant monotonic function based on the following closed-form formula (see Theorem 1.4.4 of \cite{robertson1988order}, for instance):
\begin{align}\label{eq:max-min-formula}
	\widehat g_{L_0}(x) := \min_{X_{(u)} \ge x}\, \max_{X_{(l)}\le x} \, \frac{1}{u-l+1} \sum_{i=l}^u Y_{(i)} + L_0 X_{(i)}.
\end{align}
Here, we denote by $X_{(1)} \le X_{(2)} \le \ldots \le X_{(n)}$ the order statistics and by $Y_{(1)}, Y_{(2)}, \ldots, Y_{(n)}$ their concomitants \footnote{The $k$th concomitant $Y_{(k)}$ is the corresponding response variable for $X_{(k)}$ from the original observation.}. This corresponds to the estimation procedure outlined in the beginning of Section~\ref{section:general-estimation} by fixing $\mathcal{L} = \{L_0\}$, and thus no model selection is performed during Step~\ref{cv-step}.

In practice, however, the Lipschitz constant $L_0$ is unknown and needs to be estimated. At first, it may be tempting to construct an LSE by minimizing $\sum_{i=1}^n \{Y_i - f(X_i)\}^2$ over all functions $f\in\cup_{L\geq 0}\mathcal{F}(1, L)$. This is not an effective strategy because any finite set of sample points can be perfectly interpolated using a Lipschitz function with a sufficiently large constant $L$. As a result, this class of functions always achieves zero training error for any observed data set. In short, we need some protective measure from selecting an arbitrarily large $L$. We further provide empirical evidence in \Cref{sec:est-wo-splitting}, confirming that the estimator without sample-splitting will result in over-fitting. This explains the sample-splitting procedure within the proposal.

Instead of assuming a known $L_0$, we consider a data-driven estimate of $\widehat L$, defined as \eqref{eq:estimator-of-L}. The final estimator is $\widehat f_n = \widehat f_{\widehat L}$. While the proposed procedure can be interpreted as performing cross-validation over the parameter $L$, it also possesses some desirable properties. Firstly, the core decomposition based on Proposition~\ref{prop:simplied} holds for any $L \ge L_0$ without requiring $L = L_0$. This means that a ``reasonable'' estimate of the regression function $f_0$, in terms of mean squared errors, can be obtained without precisely identifying the true value of $L_0$. Our experiments indicate that the performance of the estimators is robust with over-specified $L_0$ (\Cref{fig:L-robust}). This stands in contrast to many other nonparametric estimators, such as Gaussian kernel ridge regression or orthogonal basis estimators, where critical tuning parameters depend on sample size and often need to be precisely determined. This property also generalizes to other settings beyond Lipschitz regression.

\section{Rates of Convergence and Oracle Inequalities}\label{sec:oracle-inequality}
This section presents the oracle inequalities associated with the estimator defined in \Cref{section:general-estimation}. The oracle inequality is a prevalent theoretical statement in the LSE literature, providing a risk bound without requiring the working statistical model to be correctly specified. All results in this section adopts the notation $\|\cdot\| =\|\cdot\|_{L_2(P_X)}$. We assume that the estimator $\widehat f_n$, defined in Section~\ref{section:general-estimation}, is obtained with $\mathrm{card}(\mathcal{I}_1) = \lceil n/2 \rceil$. The results continue to hold as they are if there exist $c$ and $C$ such that $0 < c \le \mathrm{card}(\mathcal{I}_1)/\mathrm{card}(\mathcal{I}_2)\le C < \infty$. 

\subsection{General two-stage oracle inequalities}

First, we present a general result concerning the estimator defined in \Cref{section:general-estimation} constructed from a generic combination of $\mathcal{C}$ and $h_L(x)$. We then apply this result to the specific instance where $\mathcal{C}$ and $h_L(x)$ are given by \eqref{eq:choice-C-hL}. In the following assumptions, we will use $M(n,\epsilon)$ to denote a non-stochastic function of $n \geq 1$ and $\epsilon \in (0,1)$. We define the event
\begin{equation}
\mathcal{E}_M^{\mathtt{bdd}}:=\left\{\sup _{L \in \mathcal{L}}\|\widehat{f}_L\|_{\infty} \leq M(n, \epsilon)\right\},
\end{equation}
for estimators being bounded by $M(n,\epsilon)$. 

\begin{assumption}\label{as:oracle-est}
Let $\mathfrak{C} > 0$ be a universal constant. For each $\epsilon \in (0, 1)$ and $L\in \mathcal{L}$, there exists $\mathrm{R}_{L}^{\mathrm{Est}} = \mathrm{R}_{L,n,\epsilon}^{\mathrm{Est}}$ depending on $L,n,\epsilon$ such that the estimators defined in \eqref{eq:fixed-L-estimator-fL} satisfy 
\begin{equation}\label{eq:oracle-est}
\mathbb{P}\left(\|\widehat{f}_L-f_0\|^2 \geq \inf_{f \in \mathcal{M}(L)}\mathfrak{C}\left\|f-f_0\right\|^2+\mathrm{R}_{L}^{\mathrm{Est}}\mid \mathcal{E}^{\mathtt{bdd}}_{M}\right) \leq \epsilon.
\end{equation}
\end{assumption}

\begin{assumption}\label{as:oracle-est-adaptive}
Let $\mathfrak{C} > 0$ be a universal constant. For each $L\in\mathcal{L}$, let $\{\mathcal{M}_k(L): k \in \mathcal{K}\subset\mathbb{N}\}$ be a collection of subsets of $\mathcal{M}(L)$. For each $\epsilon \in (0,1)$ and $k \in \mathcal{K}$, there exists $\mathrm{R}_{L,k}^{\mathrm{Est}} = \mathrm{R}_{L,k,n,\epsilon}^{\mathrm{Est}}$ depending on $L, k, n, \epsilon$ such that the estimators defined in \eqref{eq:fixed-L-estimator-fL} satisfy
    \begin{equation}
\mathbb{P}\left(\|\widehat{f}_L-f_0\|^2 \geq \inf_{k \in\mathcal{K} }\left\{\inf_{f\in \mathcal{M}_k(L)}\mathfrak{C}\| f-f_0\|^2 + \mathrm{R}_{L,k}^{\rm Est}\right\}\mid  \mathcal{E}_M^{\mathtt{bdd}}\right) \leq \epsilon.
\end{equation}
\end{assumption}

\begin{remark}
    If \Cref{as:oracle-est-adaptive} holds with $\mathcal{M}_k(L) = \mathcal{M}(L)$, then \Cref{as:oracle-est} follows as a special case. While \Cref{as:oracle-est} establishes a global oracle inequality, \Cref{as:oracle-est-adaptive} provides a local analog. For certain choices of $\mathcal{M}_k(L)$ with lower complexity, the estimator can be shown to achieve a faster, that is, adaptive rate of convergence. We explore such an example in more detail in \Cref{sec:oracle-inequality-isotonic}.
\end{remark}
\begin{remark}
Once $L$ is fixed, the explicit form of the error $\mathrm{R}_{L}^{\mathrm{Est}}$ can be determined on a case-by-case basis. At first glance, existing oracle inequalities for shape-restricted regression may seem directly applicable for establishing \Cref{as:oracle-est} or \Cref{as:oracle-est-adaptive}, since $L$ is fixed in these assumptions. However, a subtle issue arises because we are regressing $Y + h_L(X)$ on $X$, which induces dependence between the ``error variable" and $X$. This means existing results established under independence, such as  \cite{zhang2002risk, chatterjee2015risk, bellec2018sharp, han2019convergence}, do not apply in this context. In particular, we are not aware of the corresponding results for monotone regression, and we developed new oracle inequalities without independence in \Cref{sec:model-selection}.
\end{remark}
 We also require the estimators $\widehat{f}_L$ to be continuous in $L$.
\begin{assumption}For the estimators defined in \eqref{eq:fixed-L-estimator-fL}, there exists a norm $\|\cdot\|_{\mathcal{L}}$ such that
\label{as:Lipschitz-in-parameter}
\begin{equation}
        \|\widehat{f}_{L_1}-\widehat{f}_{L_2}\|_{\infty} \le \left\|L_1-L_2\right\|_{\mathcal{L}} \text { for all } L_1, L_2 \in \mathcal{L}.
\end{equation}
\end{assumption}

\begin{assumption}\label{as:stochastic-bounded}
The class of estimators $\{\widehat f_L\, :\, L\in\mathcal{L}\}$ satisfies $\mathbb{P}\left(\mathcal{E}_M^{\mathtt{bdd}}\right) \geq 1-\epsilon$ for any $\epsilon \in (0,1)$.
\end{assumption}

The main results are established under weaker assumptions on the error variable, which guarantees the performance of the proposed procedure in some heavy-tailed and heteroscedastic noise settings. Typically, only one of the following two assumptions is used, and the choice results in different theoretical guarantees of the estimator.
\begin{assumption}\label{as:q-moment}
    There exists constants $q \geq 2$ and $C_q \in (0,\infty)$ such that 
    \begin{equation*}
        \left(\E[|\xi_i|^q\mid X_i]\right)^{1/q}\le C_q \text{ almost surely}.
    \end{equation*}
\end{assumption}
\begin{assumption}\label{as:beta-weibull}
    There exist constants $\beta>0$ and $C_\beta \in(0, \infty)$ such that for all $r \geq 1$
    \begin{equation*}
\left(\mathbb{E}\left[|\xi_i|^r\mid X_i\right]\right)^{1 / r} \leq C_\beta r^{1 / \beta}\text{ almost surely}.
\end{equation*}
\end{assumption}
Neither \ref{as:q-moment} nor \ref{as:beta-weibull} requires $\xi_i$ to be independent from $X_i$, and \ref{as:beta-weibull} is strictly stronger than \ref{as:q-moment}. The assumption~\ref{as:q-moment} only requires the finite $q$-th conditional moments, and the assumption~\ref{as:beta-weibull} assumes that the error variable possesses a sub-Weibull tail \citep{kuchibhotla2022moving}. Sub-Weibull random variables generalize light-tail conditions such as sub-Gaussian and sub-exponential tails. When $\beta=1$, this corresponds to assuming $\xi_i$ is sub-exponential, and $\beta=2$ corresponds to a sub-Gaussian tail. When $\beta<1$, the distribution has a heavier yet still exponential tail. While these assumptions play a role in model selection literature, shape-restricted functions can often be estimated under much weaker assumptions \citep{zhang2002risk}.

The $\eta$-covering number $\mathcal{N}(\eta, \mathcal{L}, \|\cdot\|_{\mathcal{L}})$ is the smallest number of $\|\cdot\|_{\mathcal{L}}$-balls of radius $\eta$ whose union contains $\mathcal{L}$. We are now ready to provide our first main result:

\begin{theorem}\label{thm:R2Q2}
    Suppose that \texttt{MEAN} is employed during Step \ref{cv-step} and Assumptions \ref{as:Lipschitz-in-parameter} and \ref{as:stochastic-bounded} hold. Then there exists a constant $C_\epsilon$ only depending on $\epsilon \in (0,1)$, such that 
            \begin{equation}
            \begin{aligned}
        \|\widehat{f}_n-f_0\|^2 &\le  \inf_{L\in \mathcal{L}}\overline{\mathrm{R}}_L^{\mathtt{Est}} + C_\epsilon\inf _{\eta>0}\left\{\log \left(\mathcal{N}\left(\eta, \mathcal{L}, \|\cdot\|_{\mathcal{L}}\right)\right)\left(\mathrm{R}_n M(n,\epsilon)+\frac{M(n,\epsilon)^2}{n }\right)+\eta^2\right\},
            \end{aligned}
    \end{equation}
    with probability greater than $1-\epsilon$, where
    \begin{equation}\label{eq:estimationg_rate}
        \overline{\mathrm{R}}_{L}^{\mathtt{Est}} = 
        \begin{cases}
        \inf_{f\in \mathcal{M}(L)}\mathfrak{C}\|f - f_0\|^2 + \mathrm{R}_{L}^{\mathrm{Est}} & \textrm{under \Cref{as:oracle-est} } \\
       \inf_{k \in\mathcal{K} }\left\{\inf_{f\in \mathcal{M}_k(L)}\mathfrak{C}\| f-f_0\|^2 + \mathrm{R}_{L,k}^{\rm Est}\right\} & \textrm{under \Cref{as:oracle-est-adaptive} }
        \end{cases}
    \end{equation}
    and
    \begin{align}\label{eq:rate_under_LSE}
    	\mathrm{R}_n := \begin{cases}
    		n^{-1+1/q} &\text{ under }\Cref{as:q-moment},\\
    		n^{-1}(\log n)^{1/\beta} &\text{ under }\Cref{as:beta-weibull}.
    	\end{cases}
    \end{align}
\end{theorem}
The proof of \Cref{thm:R2Q2} is presented in \Cref{app: two_level_oracle}. \Cref{thm:R2Q2} is an improvement of the existing oracle inequalities for model selection to the heavy-tailed and heteroscedastic noises \citep{massart2000some, gine2003ratio, vaart2006oracle, koltchinskii2011oracle}. Under the alternative model selection procedure, we obtain the following result:

\begin{theorem}\label{thm:R2Q2-robust}
    Suppose that \texttt{ROBUST} is employed during Step \ref{cv-step} with $J = \mathcal{\mathcal{N}\left(\eta, \mathcal{L}, \|\cdot\|_{\mathcal{L}}\right)}$ for $\eta > 0$, $\mathcal{L}_J$ corresponding to an $\eta$-net of $\mathcal{L}$, and $B = 4 \lceil \ln J \rceil$. Suppose Assumptions \ref{as:Lipschitz-in-parameter}, \ref{as:stochastic-bounded} and \Cref{as:q-moment} with $q=2$ hold. Then for $n \ge B$, there exists a constant $C_\epsilon$ only depending on $\epsilon \in (0,1)$, such that 
            \begin{equation}
            \begin{aligned}
        \|\widehat{f}_n-f_0\|^2 &\le \inf_{L\in \mathcal{L}}\overline{\mathrm{R}}_L^{\mathtt{Est}} + 
 C_\epsilon  \left(\log \left(\mathcal{N}\left(\eta, \mathcal{L}, \|\cdot\|_{\mathcal{L}}\right)\right)\frac{M(n,\epsilon)^2}{n} + \eta^2\right),
            \end{aligned}
    \end{equation}
    with probability greater than $1-\epsilon$ where $\overline{\mathrm{R}}^{\mathtt{Est}}_L$ is defined in \eqref{eq:estimationg_rate}.
\end{theorem}

To extend these general results to the specific choices of $\mathcal{C}$ and $h_L$, it remains to verify \Cref{as:oracle-est}--\ref{as:stochastic-bounded}, which has to be done on a case-by-case basis. In the following section, we provide analysis when $\mathcal{C}$ and $h_L$ are given by \eqref{eq:choice-C-hL}.

\subsection{Oracle inequalities for isotonic decomposition}\label{sec:oracle-inequality-isotonic}
We now present the convergence rates of the proposed estimator under the specific choice of $\mathcal{C}$ and $x \mapsto h_L(x)$ given by \eqref{eq:choice-C-hL}. We state results with $\Omega = [0,1]$. For a set $\mathcal{L} \subset \mathbb{R}$, we use $|\mathcal{L}|$ to denote its diameter, and let $L_{+}:=\sup _{L \in \mathcal{L}}|L|$. We will use $\sigma^2 < \infty$ to denote a bound on the second conditional moment of the error variable: $\mathbb{E}\left[\xi^2\mid X\right] \leq \sigma^2$ with probability one. Throughout we focus on the result associated with the \texttt{MEAN} method.

Our main result is the oracle inequality stated in \Cref{th: general_isotonic}. Consider the submodels of $\mathcal{C}$ as

\begin{equation}\label{eq:adaptive-class}
	\mathcal{F}_{m}(1,L) := \left\{f\mid f(x) = \sum_{j=1}^m a_j1(x\in I_j)-Lx,\, a_j < a_k \text { for } j<k\right\},
\end{equation}
and $m \in \{1,\ldots,n\}$. For each $m$, $\{I_j\,: j = 1,\ldots,m\}$ denotes an $m$-piece disjoint partition of $[0,1]$. The intervals $I_j$'s are ordered, so that for any $a\in I_i$ and $b\in I_k$, $a<b$ whenever $i<k$.

\begin{theorem}\label{th: general_isotonic}
Consider the IID setting \eqref{eq:regression-function}.
Suppose one of the Assumptions \ref{as:q-moment} or \ref{as:beta-weibull} hold. Then, there exists a constant $N_\epsilon$ only depending on $\epsilon\in(0,1)$ such that for any $n \ge N_\epsilon$, with probability greater than $1-\epsilon$,
\begin{equation*}
\begin{aligned}
\|\widehat{f}_n-f_0\|^2  \leq \inf _{L \in \mathcal{L}}\inf _{1 \leq m \leq n}\left\{ \, \inf _{f \in \mathcal{F}_m(1, L)}\left\|f-f_0\right\|^2+ 
\mathrm{R}_{L,m}^{\mathrm{Est}}\right\} +\mathrm{R}^{\rm CV}_{\rm iso},
\end{aligned}
\end{equation*}
where
\begin{equation*}
    \begin{aligned}
\mathrm{R}_{L,m}^{\mathrm{Est}}& =C_\epsilon m \sigma^2 B_n^2 \log ^2\left(n \right)n^{-1},\\
        \mathrm{R}^{\rm CV}_{\rm iso}
 & =  C_\epsilon \log (1+ L_{+} n^{1 / 2})\left( B_n\mathrm{R}_n + B_n^2n^{-1} \right), 
 \end{aligned}
 \end{equation*}
 and
\begin{equation*}
\mathrm{R}_n =
\begin{cases}
   n^{-1+1/q}, & \text{under } \Cref{as:q-moment}, \\[6pt]
   n^{-1} (\log n)^{1/\beta}, & \text{under } \Cref{as:beta-weibull}.
\end{cases}
\end{equation*}
We used $B_n=\left\|f_0\right\|_{\infty}+\sigma^2+L_{+}$, which may depend on $n$ though $L_+$.
\end{theorem}

The proof of \Cref{th: general_isotonic} is given in \Cref{app: general_isotonic_proof}. This result states that the difference between $\widehat f_n$ and $f_0$ is not much worse than the best approximation in the union of all submodels---infimum over $L,m$ and $f$. The estimation error $\mathrm{R}_{L,m}^{\rm Est}$ changes with $m$. When the true regression function belongs to one of the submodels $\mathcal{F}_{m}$, the approximation error $\left\|f-f_0\right\|^2$ vanishes, and we can obtain the following (almost) parametric-rate result, which is also known as local adaptivity in the literature \citep{guntuboyina2018nonparametric}.

 \begin{corollary}[Low complexity adaptation with \texttt{MEAN}]\label{cor:low-complexity} Consider the IID setting \eqref{eq:regression-function} and assume $f_0 \in \mathcal{F}_m(1, L_0)$ for $L_0 \ge 0$. We assume $L_+\rightarrow \infty$ as $n \rightarrow\infty$ and one of the noise conditions \ref{as:q-moment} or \ref{as:beta-weibull} holds. Then, for any $\epsilon \in(0,1)$ and large enough $n$, we have
\begin{equation*}
\|\widehat{f}_n-f_0\|^2 \leq C_\epsilon\left(\frac{m L_{+}^2 \log ^2 n}{n}+\log \left(1+L_{+} n\right)\left(L_{+} \mathrm{R}_n+L_{+}^2 n^{-1}\right)\right),
\end{equation*}
with probability greater than $1-\epsilon$. Here, $C_\epsilon > 0$ is a constant depending on $\epsilon, L_0, \|f_0\|_\infty, \sigma^2$ as well as constants in \ref{as:q-moment} or \ref{as:beta-weibull}, and $\mathrm{R}_n$ is defined in \Cref{eq:rate_under_LSE}.
\end{corollary}

\Cref{cor:low-complexity} becomes relevant when the true signal $f_0$ is a linear function ($m=1$), a monotone piecewise constant function ($L=0$), or the sum of monotone piecewise constant and linear functions. Similar to shape-constrained estimation procedures, oracle inequalities like \Cref{th: general_isotonic}
also imply convergence results for the truth with minimal structures, as stated below. One needs to find a proper choice of $m$ to balance the approximation and estimation error.

\begin{corollary}[Worst case guarantee with \texttt{MEAN}]\label{cor:worst-case} 
Assume the same setting as in \Cref{cor:low-complexity}, but with $f_0\in\mathcal{F}(1,L_0)$ for $L_0 \ge 0$.
Then, for any $\epsilon\in(0,1)$ and large enough $n$, we have
\begin{equation*}
\|\widehat{f}_n-f_0\|^2 \leq C_\epsilon\left(\frac{L_{+}^{4 / 3}(\log n)^{4 / 3}}{n^{2 / 3}}+\log \left(1+L_{+} n\right)\left(L_{+} \mathrm{R}_n+L_{+}^2 n^{-1}\right)\right),
\end{equation*}
with probability greater than $1-\epsilon$. Here, $C_\epsilon >0$ is a constant depending on $\epsilon, L_0, \|f_0\|_\infty, \sigma^2$ as well as constants in \ref{as:q-moment} or \ref{as:beta-weibull}, and $\mathrm{R}_n$ is defined in \Cref{eq:rate_under_LSE}.
\end{corollary}

\Cref{cor:worst-case} provides sufficient conditions to imply the near minimax optimality of $\widehat f_n$ for the Lipschitz class (up to a polylogarithmic factor); With the choice of $L_+=O(\log n)$, the estimator is near optimal under \Cref{as:q-moment} with $q \ge 3$ or under \Cref{as:beta-weibull}. Although the motivating example is estimating an $L_0$-Lipschitz $f_0$ with bounded (but unknown) $L_0$, we also note that \Cref{cor:worst-case} holds for any $f_0\in\mathcal{F}(1,L_0)$, including some discontinuous functions that are not Lipschitz.

The proof of \Cref{cor:low-complexity} and \ref{cor:worst-case} is presented in \Cref{app: proof of isotonic cases}. These results imply that the estimator exhibits different convergence rates according to the complexity of the true signal and the distribution of the error variable. \Cref{tab:convergence-rate} summarizes the rates in four different regimes with the choice $L_+ = O(\log n)$. The displayed rates suppress polylogarithmic factors for ease of exposition.

\begin{table}
\renewcommand*{\arraystretch}{1.2}
\centering
\begin{tabular}{c c c c }
\hline 
Step \ref{cv-step} & Error distribution & Worst case & Low complexity     \\
\hline 
\texttt{MEAN} & \ref{as:q-moment}&  $n^{-2/3} \vee n^{-1+1/q}$        &  $\max\{m\wedge n^{1/3}, n^{1/q}\}n^{-1}$ \\
\texttt{MEAN} &  \ref{as:beta-weibull}  &$n^{-2/3} $  & $(m\wedge n^{1/3})n^{-1}$ \\
\texttt{ROBUST} &  \ref{as:q-moment} or \ref{as:beta-weibull}  &$n^{-2/3} $  & $(m\wedge n^{1/3})n^{-1}$ \\
 \hline 
\end{tabular}
\caption{The convergence rates of the estimator under different regimes, logarithm terms omitted. The estimator exhibits different rates of convergence depending on the complexity of $f_0$ and the distribution of the error variables.}\label{tab:convergence-rate}
\end{table}

Adaptation to the unknown signal $f_0$ with low complexity is one of the intriguing characteristics of shape-restricted estimators that the proposed estimator also possesses \citep{zhang2002risk, chatterjee2015risk, guntuboyina2018nonparametric, bellec2018sharp}. The proposed estimator achieves adaptivity simultaneously with respect to a wider class of ``simple structures'', including both linear and piecewise constant $f_0$.

\begin{remark}[Corresponding results with \texttt{ROBUST}]
    In the supplement, we obtain the corresponding results when \texttt{ROBUST} is employed during Step \ref{cv-step} with $J = \mathcal{\mathcal{N}\left(\varepsilon, \mathcal{L}, \|\cdot\|_{\mathcal{L}}\right)}$ and $B = 4 \lceil \ln J \rceil$. Although \Cref{thm:R2Q2-robust} already foreshadows such results, we can obtain the following guarantees. With \texttt{ROBUST} method, \Cref{cor:low-complexity} becomes
    \begin{align}
        \|\widehat f_n - f_0\|^2 = O_P\left( \frac{m (\log n)^2}{n} + \log (1+ L_+n) \left(\frac{L_++L_+^2}{n}\right)\right), \quad \text{and}
    \end{align}\Cref{cor:worst-case} becomes
    \begin{align}
		\|\widehat f_n - f_0\|^2= O_P\left( 
		 \frac{(\log n)^{4/3}}{n^{2/3}} + \log (1+ L_+ n) \left(\frac{L_++L_+^2}{n}\right) \right), 
	\end{align}
    both of which hold only assuming \ref{as:q-moment} with $q=2$. See \Cref{tab:convergence-rate} for the summary of the convergence rates with the choice $L_+ = O(\log n)$.
\end{remark}

\begin{remark}[Adaptation to non-increasing piecewise functions]\label{remark:alt-adaptivity}
The definition of the function space $\mathcal{F}(1, L)$ is motivated by the fact that any univariate $L$-Lipschitz function can be decomposed into non-decreasing and linear functions. Following the proof of \Cref{prop:simplied}, one can also decompose an $L$-Lipschitz function into non-increasing and linear functions. All the methods and results discussed can be extended to the complementary function class $-\mathcal{F}(1, L)$. The resulting estimators adapt to non-increasing functions, piecewise constant non-increasing functions, and linear functions with positive slopes.  
\end{remark}

\begin{remark}[Rates under discrete $\mathcal{L}$ or known $L_0$]\label{remark:known-L}
Although $\mathcal{L}\subset\mathbb{R}$ is defined as a continuous interval, \Cref{cor:low-complexity} or  \Cref{cor:worst-case} still holds when $\mathcal{L}$ is a discrete set of finite cardinality, replacing the $\log \left(1+L_{+} n\right)$ term by its log-cardinality $\log({\rm card}(\mathcal{L}))$. This is achieved by directly applying \Cref{thm:R2Q2} or \Cref{thm:R2Q2-robust} to the covering number. When $\mathcal{L}$ is a singleton set $\{L_0\}$ containing the a true $L_0$, the results in this section imply

\begin{align}\label{eq:singleton}
\|\widehat f_n - f_0\|^2 = O_P\bigg(\inf _{1 \leq m \leq n}\left\{ \, \inf _{f \in \mathcal{F}_m(1, L_0)}\left\|f-f_0\right\|^2+ 
\mathrm{R}_{L_0,m}^{\mathrm{Est}}\right\}\bigg).
\end{align}
The model selection term disappears as expected. Since the remainder term $\mathrm{R}_n$, defined in \Cref{eq:rate_under_LSE}, is no longer involved, this result does not require \ref{as:q-moment} or \ref{as:beta-weibull}; The only requirement is $\E[\xi^2_i\mid X_i]\le \sigma^2$ with probability one for $1 \le i \le n$. Equation~\eqref{eq:singleton} extends Theorem 3 of \cite{han2019convergence} to the heteroscedastic error. 
\end{remark}

\section{Generalization to k-Monotone and Multivariate Decompositions}\label{sec:decomposition}

So far in this article, we have discussed the application of the univariate, Lipschitz-monotonic decomposition (Proposition~\ref{prop:simplied}). This section generalizes this basic relationship to higher-order monotonicity and multivariate covariates and establishes corresponding estimators.

The proposed estimation procedure from \Cref{section:general-estimation} as well as the oracle inequalities given by \Cref{thm:R2Q2} and \Cref{thm:R2Q2-robust} remain relevant to these generalizations. Depending on the specific setting and modeling assumptions, one needs to select an appropriate nonparametric-parametric pair in \eqref{eq:fixed-L-estimator-fL} and follow the general estimation procedure. In addition, we also compare the decomposition spaces with better-studied function spaces. Their known minimax convergence rate implies that of our model space as well.

\subsection{Decomposition spaces based on shape-restricted components}\label{subsec:decomposition}
In this section, we will formally introduce our model space of higher order and relate it to function spaces widely discussed in the literature. 

In summary, we will present a chain of inclusion properties: for each $L\ge0$ and $k\in\mathbb{N}^+$,
\begin{align*}
    \mathrm{BL}_1(k, L) ~\subsetneq~\mathcal{F}^\dag(k, L)~\subsetneq~ \mathrm{BV}(k, 5L),
\end{align*}
when the domain is $[0,1]$.

Here $\mathcal{F}^\dag(k, L)$ is a slightly more restricted version of our model space defined in order to make the comparison interesting (defined in \eqref{eq:F-dag-definition} below), and the rest of the function spaces are 
general bounded Lipschitz space (BL) and bounded variation space (BV). All of them will be defined formally.

The shape-restricted component we considered in \eqref{eq:1-decomp-function-space} is monotonic. In general, for $L\ge0$, $k \in \mathbb{N}^+$ and $\Omega \subset \mathbb{R}$, we define the $k$-th order decomposition space:
\begin{equation}\label{eq:k-decomp-function-space}
\mathcal{F}(k, L;\Omega):=\left\{f:\Omega \mapsto \mathbb{R}\mid  \exists g \in \mathcal{C}(k) \text{ s.t. } f(x)=g(x)-(L / k!) x^k\right\},
\end{equation}
where $\mathcal{C}(k)$ is the collection of univariate $k$-monotone functions~\citep{chatterjee2015risk}. Formally,
\begin{equation*}
\mathcal{C}(k) :=\left\{g:\Omega\mapsto \mathbb{R}\mid  \forall x \in \Omega, h>0, \Delta_h^k(g, x) \geq 0\right\},
\end{equation*}
where
\begin{equation*}
\Delta_h^k(g, x):= 
\begin{cases}
\sum_{m=0}^k\binom{k}{m}(-1)^{k-m} g(x+m h), & \text{if } x + m h \in \Omega \text{ for all } m  \\
0, & \text{otherwise.}
\end{cases}
\end{equation*}
The resulting decomposition space $\mathcal{F}(k, L;\Omega)$ is a special instance of the class described as \eqref{eq:general-decomposition}.

In particular, all univariate functions whose $(k-1)$th derivative is non-decreasing belong to $\mathcal{C}(k)$. The concept of $k$-monotonicity generalizes the common shape constraints such as monotonicity $(k=1)$ and convexity $(k=2)$. In the literature $\Delta_h^k$ is referred to as the ($k$-th) forward difference operator e.g., \cite[section 1.1]{ditzian2012moduli}. When $k=1$, it reduces to the forward difference: $\Delta_h(g,x) = g(x+h)-g(x)$. It is direct to verify that the decomposition spaces are nested in $L$: $\mathcal{F}(k,L)\subset \mathcal{F}(k,L')$ for $L \leq L'$ (See \Cref{lemma: F_nested}). See \cite{meyer2008inference} for estimation and inference procedures under $k$-monotone $f_0$.

Bounded Lipschitz space is commonly considered in literature, and we will relate some generalized versions of it to our decomposition space. For $L>0$, $k \in \mathbb{N}^+$ and $\Omega \subset \mathbb{R}^d$, we define bounded Lipschitz space as
    \begin{equation}\label{eq:Bounded-Lipschitz-space}
\begin{aligned}
& \mathrm{BL}_d(k, L ; \Omega) \\
& \quad:=\left\{f: \Omega \mapsto \mathbb{R} \left\lvert\, \sum_{0 \leq|m| \leq k-1}\left\|D^m f\right\|_{\infty}
+\sum_{|m|=k-1} \sup _{x \neq y, x, y \in \Omega} \frac{\left|D^m f(y)-D^m f(x)\right|}{\|x-y\|} \leq L\right.\right\},
\end{aligned}
\end{equation}
where the differential operator $D^k$ is defined as 
\begin{equation*}
D^k f:=\frac{\partial^{|k|} f(x)}{\partial^{k_1} x_1 \ldots \partial^{k_d} x_d} \quad \text{and}\quad D^0 f:=f
\end{equation*}
for $d$-dimensional index $k = (k_1,\ldots,k_d)$ where each $k_i \in\mathbb{N}$. We used the notation
$|k|= \sum_{i=1}^d k_i$. $\mathrm{BL}_1(1, L;\mathbb{R})$ is a bounded subspace of the collection of all Lipschitz functions. They are also closely related to the H\"{o}lder space, which are also widely discussed in the statistics literature, e.g., \citep[Equation (4.113)]{gine2021mathematical}.

We define the bounded decomposition space 
\begin{equation}\label{eq:F-dag-definition}
\mathcal{F}^{\dagger}(k, L;\Omega):=\mathcal{F}(k, L;\Omega) \cap\left\{f:\Omega\mapsto\mathbb{R}\mid \sum_{0 \leq m \leq k-1}\left\|D^m f\right\|_{\infty}\leq L\right\}.
\end{equation}
We have the following inclusion result. 

\begin{proposition}
\label{prop:k-monotone-decomposition}
For each $k\in\mathbb{N}^+$, $L \geq 0$ and $\Omega \subset \mathbb{R}$,
\begin{equation*}
    \mathrm{BL}_1(k, L;\Omega) ~\subsetneq~ \mathcal{F}^\dagger(k, L;\Omega) ~\subsetneq~ \mathcal{F}(k, L;\Omega).
\end{equation*}
\end{proposition}
The proof of this proposition is provided in \Cref{appsec:auxiliary-results}. \Cref{prop:simplied} is a special case of this result when $k=1$. Since $\mathrm{BL}_1(1, L; \Omega)$ is a subset of $\mathcal{F}^\dag(1,L; \Omega)$ as stated in \Cref{prop:k-monotone-decomposition}, the estimator from Section~\ref{sec:estimator} is a consistent estimator for any $f_0\in \mathrm{BL}_1(1, L; \Omega)$ while adapting to unknown parameter $L$. In fact, $\mathrm{BL}_1(k, L; \Omega)$ is not only strictly smaller than $\mathcal{F}^{\dagger}(k, L; \Omega)$, but it is also possible to show that it is not dense in the latter (under the $L_2$-norm). We give a constructive proof for $k=1$ and $\Omega = [0,1]$ in \Cref{appsec:auxiliary-results}. This indicates the decomposition model space is much bigger than the typical smooth classes.

\begin{remark}
\Cref{prop:k-monotone-decomposition} is stated only for integer-order spaces since $k$-monotone functions $\mathcal{C}(k)$ are only intuitively defined for integer values of $k$. Fortunately, we can still approximate $\mathrm{BL}_1( s, L;[0,1])$ for non-integer $ s$ with $\mathrm{BL}_{1}(\lceil s\rceil, L';[0,1])$ for some diverging $L' > L$. See \Cref{prop:k-monotone-approximation} for a formal discussion. Since $\mathrm{BL}_{1}(\lceil s\rceil, L')$ is a subset of $\mathcal{F}^\dagger(\lceil s\rceil, L')$, this in turn implies an estimator with $\mathcal{C} = \mathcal{C}(\lceil s\rceil; [0,1])$ and $h_{L_n}(X) = L_n x$ with a diverging parameter $L_n$.
\end{remark}

We now demonstrate that $\mathcal{F}^\dagger(k,L; [a,b])$ can be regarded as a strict subset of the $k$th bounded variation class \citep{mammen1997locally, tibshirani2014adaptive}. The total variation of a function $f:[a,b]\subset 
\mathbb{R} \mapsto \mathbb{R}$ is defined as
$$
\operatorname{TV}(f):=\sup\, \left\{\sum_{i=1}^m\left|f(z_{i+1})-f(z_i)\right|\,: z_1<\ldots<z_m \text { is a partition of }[a,b]\right\}
$$
where the supremum is taken over all partitions. The $k$th bounded variation space for $k \in \mathbb{N}^+$ over $[a,b]$ is defined as
\begin{equation*}
 \operatorname{BV}(k, L;[a,b]):=\left\{f:[a,b]\mapsto\mathbb{R}\mid  \operatorname{TV}(D^{k-1}_wf) \leq L\right\}.
\end{equation*}

Here $D_w^{k-1}f$ denotes the $(k-1)$-th weak derivative of $f$---see, for example, \cite{adams2003sobolev}[Item 1.62] for its formal definition.

\begin{proposition}\label{prop:k-bv}
For given $k\in\mathbb{N}^+$ and $L > 0$, we have
\begin{align}
\mathcal{F}^\dag(k, L;[a,b])~\subsetneq~ \mathrm{BV}(k, C(a,b)\cdot L;[a,b]).
    \end{align}
    where $C(a,b) := (2+b-a+2|a| + 2|b|)$.
\end{proposition}
The proof of this proposition is provided in \Cref{app: bv proof}. In particular, this result implies that minimax optimal estimators for $\mathrm{BV}$ classes, such as those in \cite{mammen1997locally} and \cite{tibshirani2014adaptive}, provide the minimax risk upper bound for the class $\mathcal{F}^\dagger(k,L; [a, b])$.

Although the decomposition spaces, as well as the smaller Lipschitz classes, are strict subsets of BV spaces for each \textit{fixed} $L$, it is still possible to approach any element in $\operatorname{BV}(1,L;[a,b])$ using a sequence of functions with diverging Lipschitz parameters. This relationship between Lipschitz and BV functions is a known result in constructive approximation literature, e.g., \citep[Section 6.6.2, Theorem 2]{evans2018measure}. We state a version that is relevant to statistical interests:
\begin{proposition}\label{prop: BV approximation}
For any function $f_0 \in \operatorname{BV}(1, M;[a,b])$, there exists a function $f_{L}\in\mathrm{BL}_1(1, L ; \Omega)\subset\mathcal{F}^\dagger(1,L;[a,b])$ such that 
\begin{equation*}
\int_a^b\left(f_L(x)-f_0(x)\right)^2 d x \leq CM^3 L^{-1},
\end{equation*}
where $C>0$ is a universal constant.
\end{proposition}
A proof of this result is provided in \Cref{app: bv proof}. This establishes that the isotonic two-step estimation procedure introduced in \Cref{sec:estimator} remains consistent even when $f_0$ is only assumed to have bounded total variation. Specifically, we can show that 
\begin{equation*}
\|\widehat{f}_n-f_0\|^2=O_P\left((\log n)^{4 / 9} n^{-2 / 9}\right)
\end{equation*}
when $L_+ = n^{2/9} \log^{-4/9} n$ plus other regularity conditions. A complete statement of this result is in \Cref{cor: BV case}. This theoretical guarantee is not minimax rate-optimal; it remains an open question whether it is due to a proof artifact or the proposal needs to be systematically modified to achieve more favorable results.

\subsection{Generalization to multivariate settings}\label{subsec:multi-variate}

While \Cref{prop:k-monotone-decomposition} is presented for univariate function classes, analogous results also hold for certain multivariate functions. This section provides three possible extensions: Coordinate-Wise (CW) $k$-monotonicity, $k$-monotone gradients, and additive $k$-monotone functions. 

We first consider CW $k$-monotone functions. Given the index set $k:= (k_1, \ldots, k_d)\in \mathbb{N}^d$ and $\Omega \subseteq \mathbb{R}^d$, a function $g : \Omega \mapsto \mathbb{R}$ is CW $k$-monotone if for each $j \in \{1, 2, \ldots, d\}$, the univariate mapping $t \mapsto g(x + te_j)$ is $k$-monotone. We similarly define the multivariate decomposition space as follows.
Given $k\in\mathbb{N}^d$ and  $L :=(L_1, \ldots, L_d)\in \mathbb{R}^d$:
\begin{equation*}
\mathcal{F}_d(k, L;\Omega):=\left\{f : \Omega \mapsto\mathbb{R}\mid  \exists g \text { is CW } k \text {-monotone s.t. } f(x):=g(x)-\sum_{i=1}^d \frac{L_i x_{[i]}^{k_i}}{k_{i}!}\right\}
\end{equation*}
where $x=(x_{[1]},\ldots, x_{[d]})^\top\in \Omega \subseteq \mathbb{R}^d$. Furthermore, we define $L \preceq L'$ when $L_{j} \le L'_j$ for all $1 \le j \le d$. It is direct to verify that $\{\mathcal{F}_d(k, L; \Omega)\}_{L \succeq 0}$ are nested sets, i.e., $\mathcal{F}_d(k, L; \Omega) \subseteq \mathcal{F}_d(k, L'; \Omega)$ for any $L \preceq L'$. 

\begin{proposition}[CW Lipschitz derivatives]\label{prop:coordinate-deriv}
Given $k = (k_1,\ldots,k_d)\in\mathbb{N}^d$. For a function $f: \Omega \mapsto \mathbb R$, and each $j \in \{1,\ldots,d\}$, we define a univariate mapping $f_{j,x}(t):\mathbb{R}\rightarrow\mathbb{R}$
    \begin{equation*}
        f_{j,x}(t) =  \frac{\partial^{k_j-1}}{\partial t^{k_j-1}}f(x+t e_j).
    \end{equation*}
    Suppose for each $j \in \{1,\ldots,d\}$, $f_{j, x}(\cdot)$ is $L_j$-Lipschitz continuous for all $x\in\Omega$, then $f\in\mathcal{F}_d(k, L; \Omega)$ with $L = (L_1,\ldots,L_d)$.
\end{proposition}

The proof is presented in \Cref{app: proof of coordinate-deriv}. In particular, we take $k_1 = \ldots = k_d = 1$ in Proposition~\ref{prop:coordinate-deriv} and obtain the following result:
\begin{proposition}[CW Lipschitz functions]\label{example1}
Suppose $f : \Omega\subset\mathbb{R}^d\rightarrow \mathbb{R}$ is a CW Lipschitz function with Lipschitz constants $L := (L_1, \ldots, L_d)$, meaning for all $x, y\in\Omega$:
\begin{equation*}
\left|f(y)-f(x)\right| \leq \sum_{j=1}^d L_je_j^\top |y-x|.
\end{equation*}
Then $f \in \mathcal{F}_d(\mathbf{1},L; \Omega)$ (where $\mathbf{1} = \{1\}^d$) i.e. there is a CW monotone $g$ such that $f(x) = g(x) - L^\top x$.
\end{proposition}

A similar result to Proposition~\ref{prop:coordinate-deriv} holds when the gradient of $f: \Omega \mapsto \mathbb{R}$ is Lipschitz continuous in Euclidean norm. The following result is reminiscent of multivariate convex functions.

\begin{proposition}[Functions with Lipschitz Gradient]\label{prop:multivariate-convex} Consider a function $f : \Omega \to \mathbb{R}$. Suppose there exists $L > 0$ such that the gradient of $f$ is $L$-Lipschitz in $\|\cdot\|_2$: \begin{equation}\label{eq:lipschitz-in-L2} \|\nabla f(y) - \nabla f(x)\|_2 \leq L \|y - x\|_2, \quad \text{for all } x, y \in \Omega. \end{equation} Define the function class 
\begin{equation*}
\mathcal{F}'_d(2, L; \Omega) := \left\{f: \Omega \mapsto \mathbb{R}\mid \exists \text{convex } g: \Omega \mapsto \mathbb{R} \text{ s.t. } f(x) = g(x) - L \|x\|_2^2 \right\}. 
\end{equation*} Then any function $f$ satisfying \eqref{eq:lipschitz-in-L2} belongs to $\mathcal{F}'_d(2, L; \Omega)$. Moreover, $\mathcal{F}'_d(2, L; \Omega) \subseteq \mathcal{F}'_d(2, L'; \Omega)$ for $L \leq L'$.
\end{proposition}

The proof is also presented in \Cref{app: proof of coordinate-deriv}. We note that the parameter $L$ is a $d$-dimensional vector in $\mathcal{F}_d(k, L; \Omega)$ while it is a scalar in $\mathcal{F}'_d(2, L; \Omega)$. 

Finally, \cite{mammen2007additive} and \cite{chen2016generalized} studied the generalized additive index model under shape constraints. In this model, the multivariate function admits the additive decomposition. For any $x\in\mathbb{R}^d$, 
\begin{equation*}
f(x)=\sum_{i=1}^m f_i(s_i^{\top} x)
\end{equation*}
with some $ s_i \in \mathbb{R}^d$, $i = 1, \ldots ,m$. Since each $f_i$ is a univariate function, we can extend \Cref{prop:k-monotone-decomposition} to the context of the additive model as well. For $k:=(k_1, \ldots, k_m)\in\mathbb{N}^m$ and $L:=(L_1,\ldots,L_m)\in\mathbb{R}^m$, the additive decomposition space is  
\begin{equation*}
\begin{aligned}
\mathcal{F}_d^{\mathrm{add }}(k, L; \Omega):=\bigg\{f: \Omega \mapsto \mathbb{R}  \mid   \exists g_i &\in \mathcal{C}\left(k_i\right), s_i \in \mathbb{S}^{d-1} \\
& \text{ s.t. } f(x)=\sum_{i=1}^m g_i(s_i^{\top} x)-\frac{L_i(s_i^{\top} x)^{k_i}}{k_{i}!}\bigg\},
\end{aligned}
\end{equation*}
where $\mathbb{S}^{d-1}\subset \mathbb{R}^d$ contains all vectors of unit $2$-norm.

\begin{proposition}[Shape-restricted multi-index model]\label{example:additive-index}
Suppose there exists $s_i \in \mathbb{S}^{d-1}$ and $f_i$ such that $f(x) = \sum_{i=1}^m f_i( s_i^\top x)$ for all $x\in\Omega$. If we further assume each $f_i$ has an $L_i$-Lipschitz $(k_i-1)$th derivative, then $f \in \mathcal{F}^{\textrm{add}}_d(k, L; \Omega)$ with $L = (L_1,\ldots,L_m) \in \mathbb{R}^m$.
In particular, if $k_i = 1$ for all $1\le i\le m$, then there exists non-decreasing functions $g_{i}(\cdot)$ such that $f(x) = \sum_{i=1}^m g_{i}( s_i^{\top}x) - \ell^{\top}x$ where $\ell=\sum_{i=1}^m L_i s_i\in\mathbb{R}^d$.
\end{proposition}
The ``standard" additive model with $s_i = e_i$ and $m=d$ is a special case of Proposition~\ref{example:additive-index}. \Cref{supp:additive} outlines a backfitting procedure for additive models with Lipschitz coordinate functions.

The general estimation procedure described in \Cref{section:general-estimation} naturally extends to the higher-order and multivariate function classes introduced in \Cref{subsec:decomposition} and \Cref{subsec:multi-variate}. For example, to estimate functions in $\mathcal{F}(2, L; [0,1])$, one may use the convex least squares estimator as $\widehat g_L$ with $h_L(x) = Lx^2/2$. Similarly, for functions in $\mathcal{F}_d(1, L; [0,1])$, one may use the block max-min estimator from \cite{han2020limit} in the same step, with $h_L(x) = L^\top x$. The corresponding convergence rates can be derived using \Cref{thm:R2Q2} or \Cref{thm:R2Q2-robust}, provided that \Cref{as:oracle-est}--\Cref{as:stochastic-bounded} are verified. We leave the verification of these assumptions for general values of $k$ and $d$, as well as for other classes of estimators, to future work.

\subsection{Minimax rates in the literature}
This section is purposed to highlight the minimax optimality of the procedure from Section~\ref{sec:estimator} and to illustrate some open problems. 

We begin with a minimax result for the univariate decomposition space, which follows directly from the inclusion properties established earlier.
\begin{theorem}[Minimax rates for $\mathcal{F}^\dag(k, L)$]\label{theorem:minimax-rate} Consider the IID setting \eqref{eq:regression-function} with $\xi_i$ following IID sub-Gaussian errors. For given $k\in\mathbb{N}$ and $L > 0$, consider $\mathcal{F}^\dag(k, L)$ defined in \eqref{eq:F-dag-definition}, we have 
\begin{align*}
    cn^{-2k/(2k+1)}\le \inf_{\widehat f}\, \sup_{f \in \mathcal{F}^\dagger(k,L)}\, \E 
    [\|\widehat f - f\|_{L_2(P_X)}^2]\le  Cn^{-2k/(2k+1)},
\end{align*}
where $C$ and $c > 0$ are constants not depending on $n$. The lower bound also holds for $\mathcal{F}^\dagger(k,L)$ replaced with $\mathcal{F}(k, L)$.
\end{theorem}

\begin{proof}
The lower bound is due to the inclusion property \Cref{prop:k-monotone-decomposition}. The smaller space $\mathrm{BL}_1(k, L)$ is further a super set of the $k$-th order H\"{o}lder class $\Sigma_k$ \citep[page 351 and the references therein]{gine2021mathematical}. So we know an estimation problem in $\mathcal{F}^{\dagger}(k, L)$ should be no easier than one in $\Sigma_k$. Given the well-known minimax bounds for $\Sigma_k$---which scale as $n^{-2k/(2k+1)}$ (see, i.e., \citep{stone1982optimal} and \citep{ibragimov1984asymptotic}), we know the minimax rate should be no faster than $c n^{-2 k /(2 k+1)}$. The upper bound is due to another inclusion property (Proposition~\ref{prop:k-bv}) and the minimax rate associated with $\mathrm{BV}$ spaces. See, for instance, \citet[Theorem 10]{mammen2007additive} and \citet[Theorem 1]{tibshirani2014adaptive}. In particular, these works analyze their estimators under the assumption that the error is sub-Gaussian whereas the results in this article can be extended to heavy-tailed errors.
\end{proof}

\begin{remark}

 Theorem~\ref{theorem:minimax-rate} implies that the proposed estimator from Section~\ref{sec:estimator} is (almost) minimax optimal under \ref{as:q-moment} for $q \ge 3$ or \ref{as:beta-weibull}, attaining the convergence rate of $n^{-2/3}$ up to a polylogarithmic factor. To the best of our knowledge, we are unaware of the general minimax lower bound under \ref{as:q-moment} with $q < 3$, particularly for the decomposition space $\mathcal{F}(k, L)$.   

 The risk upper bound of the generic estimator for $k > 2$ is underdeveloped. When $k=2$, Theorem 3.1 of \cite{kuchibhotla2022least} and Theorem 3 of \cite{han2018robustness} together suggest that analogous results to \Cref{th: general_isotonic} may be obtained.
\end{remark}

For the multivariate function class corresponding to Proposition~\ref{prop:coordinate-deriv}, we no longer have the luxury of deducing minimax rates from inclusion properties alone since the optimality of function estimation in the $k$th bounded variation class on general $d$-dimensional covariates is currently less understood; See \cite{sadhanala2017higher} and \cite{hu2022voronoigram} on current developments in this direction. When $k=1$, it is known that the LSEs do not adapt to the low complexity truth at the minimax optimal rate in $L_2$ for $d \ge 3$ \citep{han2019isotonic}. Hence, an alternative approach such as the block max-min estimator must be considered to recover minimax optimal adaptivity \citep{han2020limit}. We anticipate that the convergence rate of the block max-min estimator provided in that work can be extended to our setting. A similar phenomenon is studied for convex regression in \cite{kur2020suboptimality}, which showed that LSEs are not globally minimax optimal for $d \ge 6$. See their Corollary 2 for further details. The general statements for arbitrary $k$ and any $d$ are unknown in the literature. 

\section{Numerical Studies}\label{sec:simulation}
We conduct numerical studies to assess the finite-sample properties of the proposed procedures. The aim of this section is twofold: First, we confirm the theoretical results, namely \ref{cor:low-complexity} and \Cref{cor:worst-case}, such that the estimator attains the minimax rate and the faster rate for the low complexity case. Second, we implement the generalized procedures for convex functions and additive models. The aim is to highlight the flexibility of the proposed methodological framework and present preliminary empirical evidence for the general theory that one may expect. Throughout, we perform numerical studies with \texttt{MEAN} for its practical advantages. Additional numerical results are provided in the supplementary material.

\subsection{Univariate nonparametric regression}\label{sec:univariate-setting}
We first consider univariate covariates where $X$ follows a uniform distribution $\mbox{Unif}[0,1]$. The response variables are generated according to $Y_i = f(X_i) + \xi_i$ for $i = 1, 2,\ldots, n$ with several different regression functions $f$. The error terms $\xi_i$'s are IID, normally distributed $N(0, 0.1^2)$ variables across all scenarios. The sample sizes vary from $10^2$ to $10^4$. For each scenario, we replicate the experiment 300 times. This article provides results associated with the settings where $f_0 \in \mathcal{F}(k, L)$ for $k = 1, 2$ and $f_0 \in \mathcal{F}^{\textrm{add}}_d(k, L)$ for $k=1$ and $d=2, 5$ (See Proposition~\ref{example:additive-index} with $s_i=e_i$ and $m=d$).

\begin{figure}[t]
     \centering
    \includegraphics[width=5.5in]{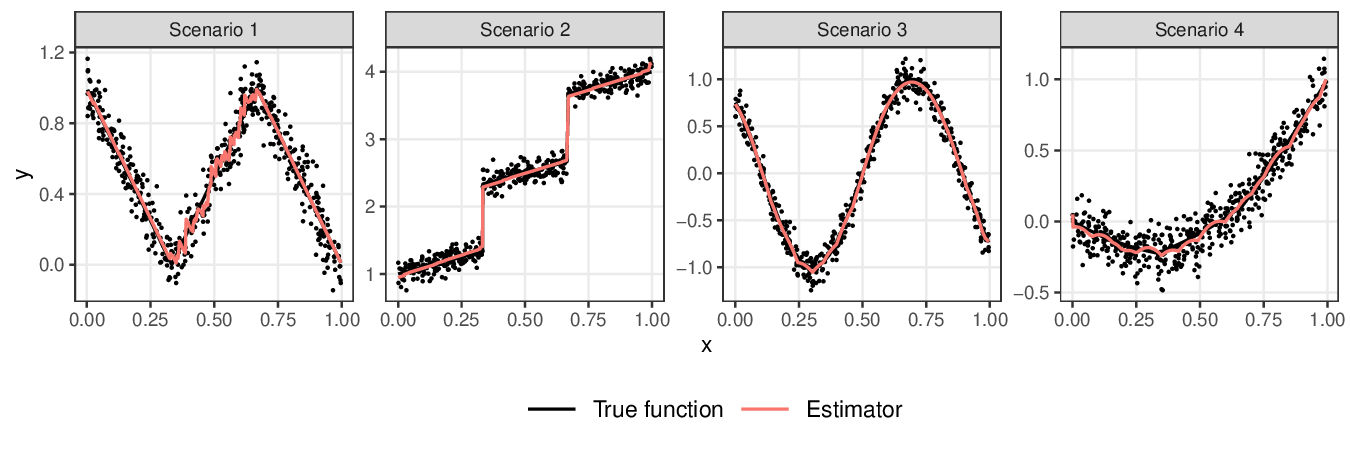}
    \caption{Estimated regression functions based on $500$ samples. Scenario 1 \& 2, isotonic regression plus a linear component; Scenario 3 \& 4, convex regression plus quadratic.}\label{fig:estimated-curve}
\end{figure}

For the proposed method, we split the index set $\{1, 2, \dots, n\}$ into two disjoint subsets with $\mathrm{card}(\mathcal{I}_1) = \lceil n/2 \rceil$. Throughout, the proposed method is referred to as \texttt{LSE+Parameteric}. We also compare the proposed method with the following nonparametric estimators in the literature: Kernel ridge regression (\texttt{KRR}), Gradient boosting machines (\texttt{GBM}), Random forest regression (\texttt{RF}), and Penalized sieve estimator with cosine basis (\texttt{Sieve}) \citep{randomforrestPackage, zhang2022regression}. We provide the implementation details in \Cref{supp: numerical study detail}. The replication of the results is available at \url{https://github.com/Kenta426/sim-npparam}. For sample sizes exceeding 2000, we omit the results obtained from \texttt{KRR} and \texttt{RF} due to computational limitations. We consider the four ``true'' regression functions, which are formally described in \Cref{supp: regression description}. Their plots are presented in \Cref{fig:estimated-curve};
\begin{itemize}
    \item Scenario 1 corresponds to the ``worst case'' for $\mathcal{F}(1,L)$ class. The proposed estimator, along with other nonparametric regression estimators, is expected to converge at a rate of $n^{-2/3}$. 
    \item Scenario 2 corresponds to the ``low complexity case'' for $\mathcal{F}(1,L)$ class. The proposed estimator is expected to be adaptive, converging at a parametric rate of $n^{-1}$ (up to a polylogarithmic factor). 
    \item Scenario 3 corresponds to the ``worst case'' for $\mathcal{F}(2,L)$ class. The proposed estimator is anticipated to converge at a rate of $n^{-4/5}$ in view of Theorem 3.1 of \cite{kuchibhotla2022least} and Theorem 3 of \cite{han2018robustness}.
    \item Scenario 4 corresponds to the ``low complexity case'' for $\mathcal{F}(2,L)$ class where the proposed estimator is expected to be adaptive. See \cite{guntuboyina2015global, han2018robustness} on the low complexity adaption of the convex LSE. 
\end{itemize}

Figure~\ref{fig:estimated-curve} displays a single realization of observations from each scenario along with the estimated regression function using a sample size of $n=500$. The estimator appears to be consistent with the true curves, including the regression functions that contain non-differentiable points (Scenario 1) as well as discontinuities (Scenario 2). 

\begin{figure}[t]
    \centering
   \includegraphics[width=5.5in]{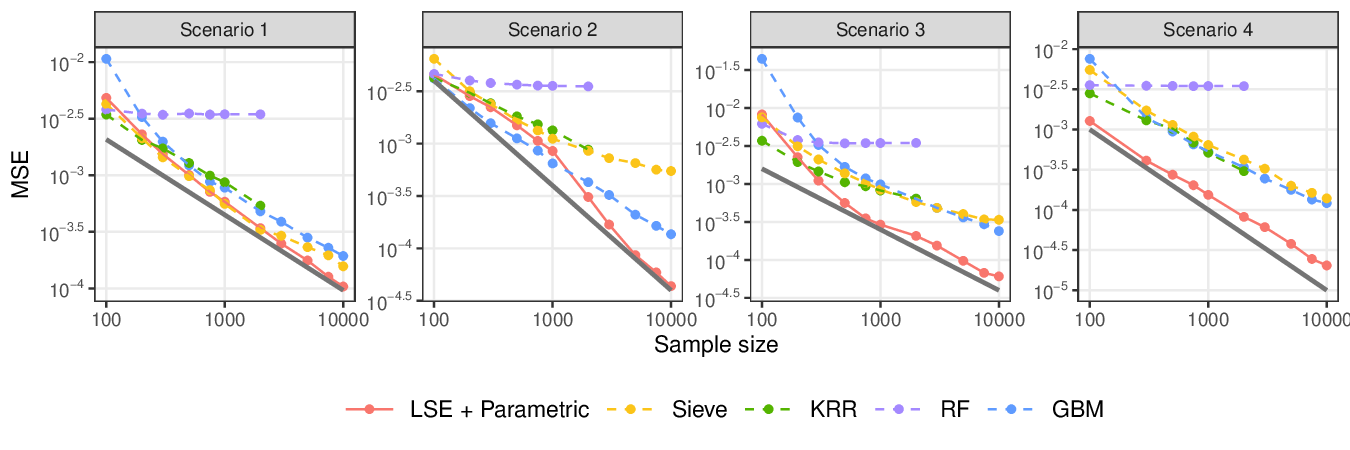}
 \caption{Average MSEs across 300 replications for different sample sizes. The solid black line represents the expected convergence rate of the proposed method.}\label{fig:mse-rate}

\includegraphics[width=5.5in]{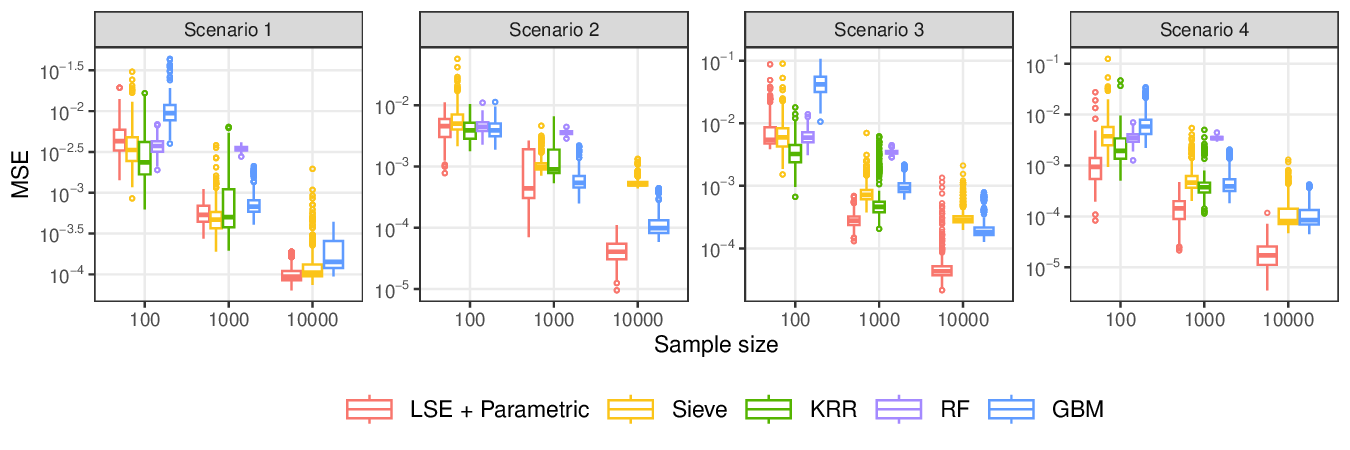}
  \caption{Boxplot of MSEs across 300 replications for sample sizes $n=10^2, 10^3$, and $10^4$.}\label{fig:mse-box}
\end{figure}

\begin{table}
\centering
\begin{tabular}{l|cccc}
                 & Scenario 1 & Scenario 2 & Scenario 3 & Scenario 4 \\ \hline
Theoretical rate & $-$0.667     & $-$1         & $-$0.8       & $-$1         \\
LSE + parametric & $-$0.738     & $-$1.201     & $-$0.799     & $-$0.895     \\
Sieve            & $-$0.449     & $-$0.276     & $-$0.349     & $-$0.710     \\
GBM              & $-$0.568     & $-$0.720     & $-$0.590     & $-$0.645    
\end{tabular}
\caption{Estimated slopes of Figure~\ref{fig:mse-rate} for each method. The displayed values are the estimated exponent $r$ of the convergence rate $n^r$.  The values in this table are computed by linear regression using $(\log_{10} n, \log_{10} \textrm{MSE})$ from the points in Figure~\ref{fig:mse-rate}. Only the points for sample sizes $n \ge 2\times 10^3$ are considered. 
}\label{tab:est-slope}
\end{table}

Next, we study the convergence rate of mean squared errors (MSE) for various methods as the sample sizes vary. To estimate the MSEs, we generate additional $10^5$ sample paths to approximate $\|\widehat f_n - f_0\|^2_{L_2(P_X)}$. Figure~\ref{fig:mse-rate} displays the average MSEs for various sample sizes, while Figure~\ref{fig:mse-box} displays box plots representing the MSE distributions specifically for sample sizes of $n=10^2, 10^3$, and $10^4$. The solid black lines in Figure~\ref{fig:mse-rate} represent the expected rates of convergence for the proposed method, namely $n^{-2/3}$ for Scenario 1, $n^{-4/5}$ for Scenario 3, and $n^{-1}$ (ignoring a polylogarithmic rate) for Scenarios 2 and 4. Additionally, Table~\ref{tab:est-slope} presents the estimated slope for each method based on linear regression using $(\log_{10} n, \log_{10} \textrm{MSE})$ from the Figure~\ref{fig:mse-rate} with sample sizes $n \ge 2\times 10^3$.

In the small-sample regimes, the proposed method slightly deviates from the theoretical convergence rate. As the sample size increases, the proposed method aligns more closely with the expected convergence rate. In Scenario 1, most nonparametric regression methods demonstrate comparable performance, with the proposed method performing particularly well for larger sample sizes. Similar conclusions can be drawn for Scenario 3, with the performance of the proposed method being especially pronounced. In two scenarios where the proposed method is expected to be adaptive, and hence to converge at a near parametric rate, it outperforms all other methods for sample sizes larger than 1000. These findings highlight the practical benefits of the method with the adaptivity property.

\subsection{Multivariate regression under additive structure}
This section considers the multivariate covariates under an additive structure, as discussed in Proposition~\ref{example:additive-index} from Section~\ref{subsec:multi-variate}. The implementation is based on the algorithm presented in \Cref{supp: additive algorithm}. This set of numerical studies also includes the results based on the Generalized Additive Model (\texttt{GAM}). We use the implementation based on R package \texttt{mgcv} \citep{mgcv} with the number of the basis of each smoothing spline is set to $30$. We first consider two examples with 2-dimensional covariates $x=(x_{[1]}, x_{[2]})$ and two examples with 5-dimensional covariates $x=(x_{[1]}, \ldots,x_{[5]})$. All components in each setting follow independent uniform distributions over $[0,1]$. Similar to the univariate case, we only provide a brief description. Each function is formally defined in \Cref{supp: regression description}.
\begin{itemize}
    \item Scenario 1 (2d): Each additive component is identical to Scenario 1 from the univariate case, which corresponds to the ``worst case'' for $\mathcal{F}(1,L)$ class.
    \item Scenario 2 (2d): Each additive component is identical to Scenario 2 from the univariate case, which corresponds to the ``low complexity case'' for $\mathcal{F}(1,L)$ class.
    \item Scenario 3 (5d): Each component is identical to Scenario 1 from the univariate case.
    \item Scenario 4 (5d): Each component is identical to Scenario 2 from the univariate case.
\end{itemize}
Similar to the univariate cases, we anticipate the proposed method to converge essentially at a rate of $n^{-2/3}$ for Scenario 1,3 and $n^{-1}$ for Scenario 2,4. We discuss the justification behind this argument in \Cref{supp: regression description}.

\begin{figure}[!t]
    \centering
   \includegraphics[width=5.5in]{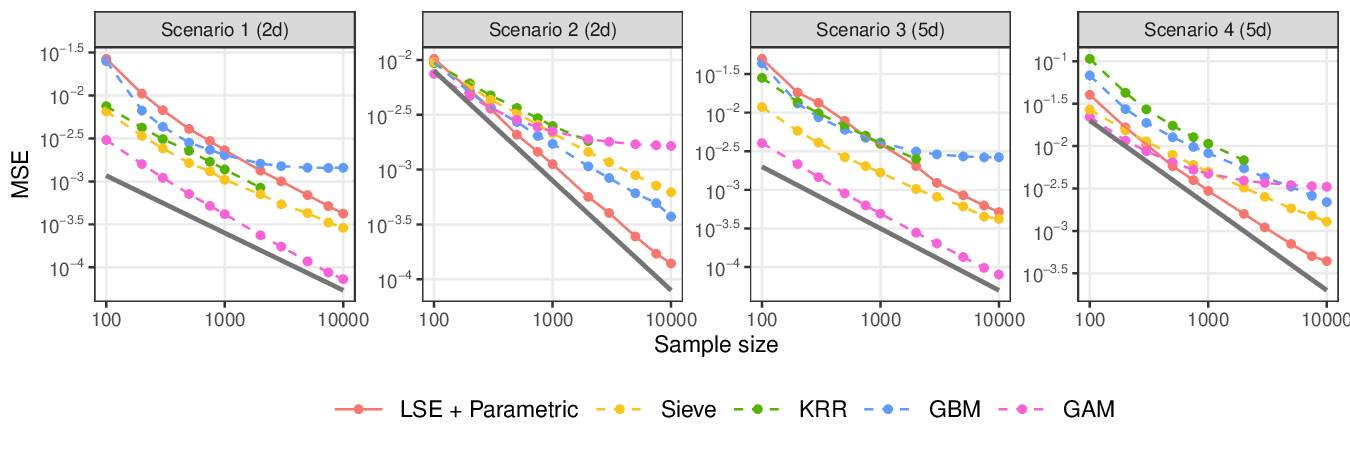}
 \caption{Average MSEs across 300 replications. The solid black line represents the expected convergence rate of the proposed method.}\label{fig:mse-additive-rate}
   \includegraphics[width=5.5in]{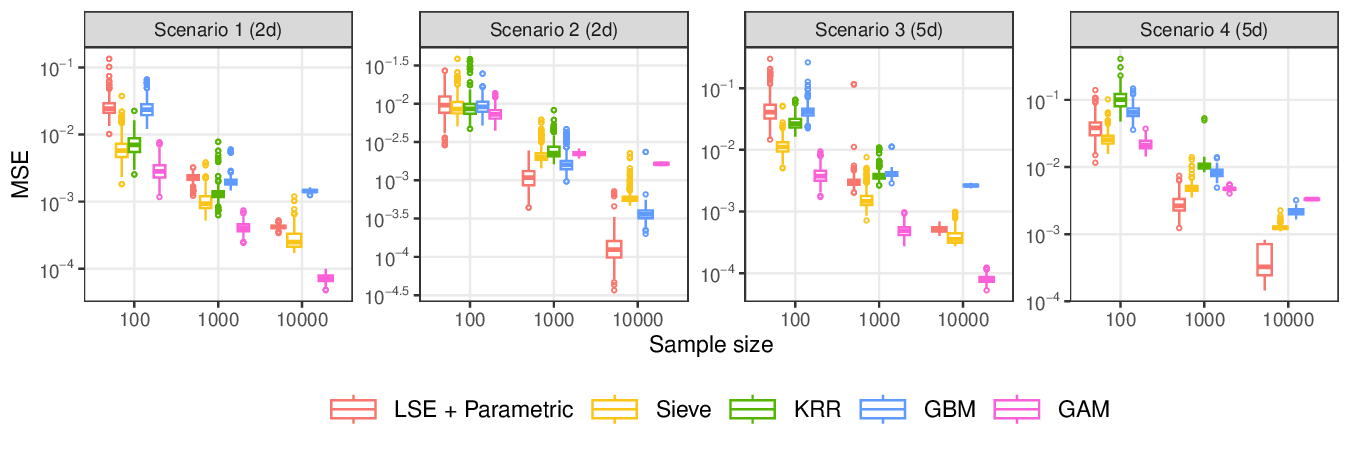}
  \caption{Boxplot of MSEs across 300 replications for sample sizes $n=10^2, 10^3$, and $10^4$.}\label{fig:mse-additive-box}
\end{figure}

Figure~\ref{fig:mse-additive-rate} displays the average MSEs while Figure~\ref{fig:mse-additive-box} displays box plots representing the distribution of observed MSEs. Additionally, Table~\ref{tab:est-additive-slope} presents the estimated slope based on linear regression using linear regression using $(\log_{10} n,$ $ \log_{10} \textrm{MSE})$ from the Figure~\ref{fig:mse-additive-rate} with sample sizes $n \ge 2000$. Similar to the univariate cases, the theoretical convergence rate aligns with the empirical behavior for larger sample sizes. We observe that \texttt{gam} performs well for regression functions even with non-differentiable points (Scenarios 1 and 3). Most methods share the same convergence rates but the constant is better for \texttt{gam}. However, \texttt{gam} struggles to accurately estimate functions with discontinuities (Scenarios 2 and 4). In contrast, the proposed method performs well when each additive component is a non-decreasing piecewise constant function plus a linear term. We emphasize that the proposed method is practically tuning parameter free while the number of the basis plays an important role for \texttt{gam}.

\begin{table}[t]
\centering
\begin{tabular}{l|cccc}
                 & Scenario 1 (2d) & Scenario 2 (2d) & Scenario 3 (5d)& Scenario 4 (5d)\\ \hline
Theoretical rate & $-$0.667     & $-$1         & $-$0.667       & $-1$         \\
LSE + parametric & $-$0.717     & $-$0.885     & $-$0.828     & $-$0.814     \\
Sieve            & $-$0.553     & $-$0.528     &  $-$0.577    & $-$0.571     \\
GBM              & $-$0.068     & $-$0.633     & $-0$.107     & $-$0.563        \\
GAM              & $-$0.738     & $-$0.089     & $-$0.780     &  $-$0.104 
\end{tabular}
\caption{Estimated slopes of Figure~\ref{fig:mse-additive-rate} for each method. The displayed values are the estimated exponent $r$ of the convergence rate $n^r$. The values in this table are computed by linear regression using $(\log_{10} n, \log_{10} \textrm{MSE})$ from the points in Figure~\ref{fig:mse-additive-rate}. Only the points for sample sizes $n \ge 2\times 10^3$ are considered. 
}
\label{tab:est-additive-slope}
\end{table}

\section{Conclusions}\label{sec:conclusion}
This article introduces a new approach to nonparametric regression estimation by leveraging shape-restricted regression methods. The proposed approach takes advantage of the general decomposition of nonparametric functions into shape-restricted and parametric components. We propose an estimation procedure based on sample-splitting, which practically eliminates the burden of tuning hyperparameters. The proposed method inherits favorable properties from shape-restricted estimators, including minimax-optimal convergence rate (up to a polylogarithmic factor), adaptivity, and efficient computation.

\begin{figure}[t]
 \centering
    \includegraphics[width=3.25in]{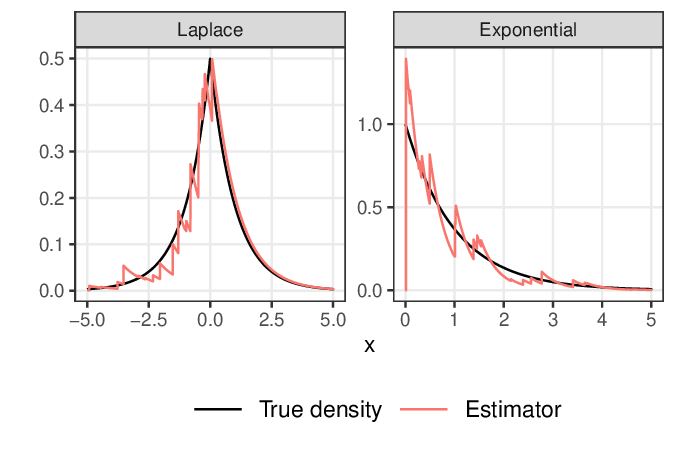}
    \caption{Estimated density functions based on $500$ observations (1:1 split ratio) generated from two scenarios. The left panel corresponds to the data drawn from a Laplace distribution where the estimator is expected to converge at the minimax rate. The right panel corresponds to the data drawn from an exponential distribution where the estimator is expected to converge at the adaptive rate. We refer to \Cref{supp:density} for further details.}\label{fig:estimated-density}
\end{figure}

Although this article focuses on nonparametric regression, the proposed shape-restricted decomposition is more general and allows for further methodological advances beyond regression settings. They include but are not limited to density estimation, quantile regression, instrumental variables regression, and classification. As a proof of concept, we extend the proposed framework to the ``log-Lipschitz'' density estimation, that is, the logarithm of true density admits the decomposition \eqref{eq:monotone-decompisition}. Based on this decomposition, we implement the nonparametric maximum likelihood estimator for the density function. The estimated functions under two different scenarios, Laplace and exponential distributions, are displayed in Figure~\ref{fig:estimated-density}. We refer to \Cref{supp:density} for further details.

We discuss several open problems. One interesting direction is uncertainty quantification. When the parametric component $L$ is known, we anticipate that existing confidence sets for shape-restricted regression \citep{dumbgen2003optimal, bellec2016adaptive,yang2019contraction} are directly applicable. These methods provide uniform confidence bands whose width shrinks at the adaptive rates, depending on the complexity of the regression function. Extending this idea to the general $\mathcal{F}(k,L)$ class is of great interest. Constructing confidence sets including the uncertainty from selecting $L$ will be more challenging. We anticipate that the recently developed HulC procedure \citep{kuchibhotla2021hulc} can be applicable in this setting.

Several properties of $\mathcal{F}(k,L)$ remain open. For instance, Theorem~\ref{theorem:minimax-rate} provides the minimax rate for $\mathcal{F}^\dag(k,L)$ but not for $\mathcal{F}(k,L)$. In addition, minimax-optimal estimators in general dimensions are not currently available for multivariate $k$-monotone estimation beyond the cases $k=1,2$. Extending the algorithms in \Cref{section:general-estimation} to general orders appears nontrivial.

\section*{Acknowledgements}
We thank the Reviewers and Associate Editors for their constructive comments, which helped to improve this work. We also thank Sivaraman Balakrishnan for suggesting the analysis of the robust model selection based on the median of means estimator, which led to the development of \Cref{thm:R2Q2-robust}. Kenta Takatsu and Arun Kumar Kuchibohotla gratefully acknowledge support from National Science Foundation DMS-2210662. Part of this work was completed while Tianyu Zhang was at Carnegie Mellon University. 

    \appendix
    \section{Appendix}\label{sec:model-selection}

General two-level oracle inequalities, \Cref{thm:R2Q2} and \Cref{thm:R2Q2-robust}, rely on new results in model selection under heteroscedastic and heavy-tailed errors. As the corresponding results are new to our knowledge, we present them below for our readers' ease of reference. 

In applications, the model collection $\mathcal{F}$ below is often a random function class depending on the training data, with each element corresponding to a hyperparameter (in our case, an element of $\mathcal{L}$). The randomness, including the sample size $n$ below, is related to the validation samples $D_2$ in \Cref{section:general-estimation}. Notably, our treatment did not require $\mathcal{F}$ to be convex. We need this generality because $\mathcal{F}$ is often a random set depending on the training data $D_1$ and may not satisfy such constraints.

\begin{theorem}[Model selection --- \texttt{MEAN}]\label{thm:oracle-cv}
	Consider the IID setting in \eqref{eq:regression-function}. Let $\mathcal{F}$ be a set of functions and $F := \sup_{f \in \mathcal{F}}\|f\|_\infty$. Denote the least squares estimator:
	\begin{equation*}
\widehat{f}^{\mathtt{MEAN}}_n:=\underset{f \in \mathcal{F}}{\arg \min }  \, \sum_{i=1}^n\left(Y_i-f(X_i)\right)^2.
\end{equation*}
Let $\mathfrak{C} > 0$ be a universal constant. Then for any $\epsilon \in (0,1)$, 

\begin{equation}
\begin{aligned}
\|\widehat{f}^{\mathtt{MEAN}}_n-f_0\|^2 \le  \mathfrak{C}\inf_{f \in \mathcal{F}} \left\|f-f_0\right\|^2 +C_\epsilon\inf _{\eta>0}\left\{\log \left(\mathcal{N}\left(\eta, \mathcal{F}, L_2\left(\mathbb{P}_n\right)\right)\right)\left(\mathrm{R}_n F+\frac{F^2}{n}\right)+\eta^2\right\}
\end{aligned}
\end{equation}
with probability greater than $1-\epsilon$ where
\begin{align}\label{eq:oracle-remainder}
    \mathrm{R}_n := \begin{cases}
        n^{-1+1/q} &\text{ under }\Cref{as:q-moment}\\
        n^{-1}(\log n)^{1/\beta} &\text{ under }\Cref{as:beta-weibull}
    \end{cases}
\end{align}
and $C_\epsilon$ is a constant depending on $\epsilon, C_q, C_{\beta}, \|f_0\|_\infty, q,\beta$ and $\sigma^2$.
\end{theorem}

We also provide the corresponding result under weaker moment assumptions when applying \texttt{ROBUST}.

\begin{theorem}[Model selection --- \texttt{ROBUST}]\label{thm:oracle-cv-mom}
	Consider the IID setting in \eqref{eq:regression-function}. Let $\mathcal{F}_J$ be a finite set of functions with cardinality $J$ and $F := \sup_{f \in \mathcal{F}_J}\|f\|_\infty$. Consider the following estimator based on median of means (See \Cref{def:mom} in \Cref{supp:oracle-robust}):
	\begin{equation*}
\widehat{f}_n^{\mathtt{MOM}}:=\underset{f \in \mathcal{F}_J}{\arg \min } \, \max_{h \in \mathcal{F}_J} \, \mathtt{MOM}_B \left(\{\left(Y_i-f(X_i)\right)^2 - \left(Y_i-h(X_i)\right)^2\}_{i=1}^n\right)
\end{equation*}
with $B = 4\lceil \ln J \rceil$. Let $\mathfrak{C} > 0$ be a universal constant. It then holds for $n \ge B$ and any $\epsilon \in (0,1)$, 
	\begin{equation}
\begin{aligned}
\|\widehat{f}^{\mathtt{MOM}}_n-f_0\|^2 \leq  \mathfrak{C}\inf_{f\in\mathcal{F}_J}\|f-f_0\|^2 + \frac{C_\epsilon F^2\lceil \ln J \rceil}{n}
\end{aligned}
\end{equation}
with probability greater than $1-\epsilon$ under \Cref{as:q-moment} with $q=2$ and $C_\epsilon$ is a constant depending on $\epsilon, \|f_0\|_\infty$ and $\sigma^2$.
\end{theorem}

\bibliographystyle{abbrvnat}
\bibliography{ref.bib}
\newpage
\begin{appendices}
\newpage
\section{Proofs of Two-level Oracle Inequalities}\label{app: two_level_oracle}

The general decomposition class in this section is defined as 
\begin{equation*}
    \mathcal{M}(L; \Omega) = \{m:\Omega \mapsto\mathbb{R} \mid \exists g \in \mathcal{C}\text{ s.t. } m(x)=g(x)-h_L(x)\}.
\end{equation*}
First, we present results without assuming specific forms of $\mathcal{C}$ or $x \mapsto h_L(x)$. \Cref{app: proof of isotonic regression using new structure} presents the application of \Cref{thm:R2Q2} and \Cref{thm:R2Q2-robust} to the case when $\mathcal{C} = \mathcal{C}(1)$ and $h_L(x) = Lx$. Define an event $\mathcal{E}^{\mathtt{bdd}}_{M} := \{\sup_{L \in \mathcal{L}}\|\widehat{f}_L\|_\infty \le M(n, \epsilon)\}$ for a given $n \ge 1$ and $\epsilon \in (0,1)$. We provide the following result.
\begin{lemma}\label{lemma:general-tower-prop}
    Assume \Cref{as:stochastic-bounded} and either one of \Cref{as:oracle-est} or \Cref{as:oracle-est-adaptive}. Furthermore, we assume that for any $\epsilon \in (0,1)$, there exits a $\mathrm{R}_{\mathcal{L}, n, \epsilon}^{\mathtt{CV}}$ such that
\begin{equation}\label{eq:oracle-CV}
    \mathbb{P}\left(\|\widehat{f}_n - f_0\|^2\ge \mathfrak{C}\|\widehat{f}_{L} - f_0\|^2 + \mathrm{R}_{\mathcal{L}, n, \epsilon}^{\mathtt{CV}} \mid\mathcal{I}_1, \mathcal{E}^{\mathtt{bdd}}_{M}\right) \le \epsilon
\end{equation}
holds for all $L \in \mathcal{L}$. Here $\mathfrak{C}$ denotes a universal constant.

Then we know that
            \begin{equation*}
            \begin{aligned}
        \|\widehat{f}_n-f_0\|^2 &\le   \inf_{L \in \mathcal{L}}\overline{\mathrm{R}}_{L,n,\epsilon}^{\mathtt{Est}}  + \mathrm{R}_{\mathcal{L}, n,\epsilon}^{\mathtt{CV}}
            \end{aligned}
    \end{equation*}
    with probability greater than $1-\epsilon$ where 
    \begin{equation*}
        \overline{\mathrm{R}}_{L,n,\epsilon}^{\mathtt{Est}} = 
        \begin{cases}
        \inf_{f\in \mathcal{M}(L)}\mathfrak{C}\|f - f_0\|^2 + \mathrm{R}_{L,n,\epsilon}^{\mathrm{Est}} & \textrm{under \Cref{as:oracle-est} } \\
       \inf_{k \in\mathcal{K} }\left\{\inf_{f\in \mathcal{M}_k(L)}\mathfrak{C}\| f-f_0\|^2 + \mathrm{R}_{L,k,n,\epsilon}^{\rm Est}\right\} & \textrm{under \Cref{as:oracle-est-adaptive}.}
        \end{cases}
    \end{equation*}
\end{lemma}
\begin{proof}[Proof of \Cref{lemma:general-tower-prop}]
Let $L\in\mathcal{L}$ be any fixed element. Let $\mathcal{E}_{L,\epsilon}^{\mathrm{Est}}$ be an event defined for each $\epsilon$ as:
    \begin{align*}
        \mathcal{E}_{L,\epsilon}^{\mathrm{Est}} := \left\{\|\widehat{f}_L-f_0\|^2 \leq \overline{\mathrm{R}}_{L,n,\epsilon}^{\mathtt{Est}}\right\}.
    \end{align*}
For any $t \ge 0$, it follows that 
\begin{equation*}
\begin{aligned}
    &\mathbb{P}(\|\widehat f_n - f_0\|^2 \ge t) \\
    &\qquad\le \mathbb{P}(\|\widehat f_n - f_0\|^2 \ge t \mid \mathcal{E}^{\mathtt{bdd}}_{M}) + (1-\mathbb{P}(\mathcal{E}^{\mathtt{bdd}}_{M})) \\
    &\qquad\le \mathbb{P}(\|\widehat f_n - f_0\|^2 \ge t \cap \mathcal{E}^{\mathrm{Est}}_{L,\epsilon/4} \mid  \mathcal{E}^{\mathtt{bdd}}_{M}) + (1-\mathbb{P}(\mathcal{E}^{\mathrm{Est}}_{L,\epsilon/4} \mid \mathcal{E}^{\mathtt{bdd}}_{M})) + (1-\mathbb{P}(\mathcal{E}^{\mathtt{bdd}}_{M})) \\
    &\qquad\le \mathbb{P}(\|\widehat f_n - f_0\|^2 \ge t \cap \mathcal{E}^{\mathrm{Est}}_{L,\epsilon/4} \mid  \mathcal{E}^{\mathtt{bdd}}_{M} ) + \epsilon/2
\end{aligned}
\end{equation*}
where the last step follows from \Cref{as:stochastic-bounded} and either \Cref{as:oracle-est} or \Cref{as:oracle-est-adaptive}. It remains to analyze the first term. Let $ t = \mathfrak{C}\overline{\mathrm{R}}_{L,n,\epsilon}^{\mathtt{Est}} + \mathrm{R}_{\mathcal{L}, n,\epsilon/2}^{\mathtt{CV}}$. Then it follows
\begin{align*}
    &\mathbb{P}(\|\widehat f_n - f_0\|^2 \ge t \cap \mathcal{E}^{\mathrm{Est}}_{L,\epsilon/4} \mid  \mathcal{E}^{\mathtt{bdd}}_M) \\
    &\qquad \le \mathbb{P}(\|\widehat f_n - f_0\|^2 \ge   \mathfrak{C}\|\widehat{f}_L-f_0\|^2  + \mathrm{R}_{\mathcal{L}, n, \epsilon/2}^{\mathtt{CV}} \mid  \mathcal{E}^{\mathtt{bdd}}_M) \\
    &\qquad = \mathbb{E}[\mathbb{P}(\|\widehat f_n - f_0\|^2 \ge  \mathfrak{C}\|\widehat{f}_L-f_0\|^2  + \mathrm{R}_{\mathcal{L}, n, \epsilon/2}^{\mathtt{CV}} \mid  \mathcal{E}^{\mathtt{bdd}}_M, \mathcal{I}_1)] \le \epsilon/2
\end{align*}
where the last step follows from \eqref{eq:oracle-CV}. We thus conclude that for any $L \in \mathcal{L}$ and $\epsilon \in (0,1)$ that 
\begin{align*}
    \mathbb{P}(\|\widehat f_n - f_0\|^2 \ge \mathfrak{C}\overline{\mathrm{R}}_{L,n,\epsilon}^{\mathtt{Est}}+ \mathrm{R}_{\mathcal{L}, n,\epsilon}^{\mathtt{CV}}) \le \epsilon.
\end{align*}
Since the above inequality holds for any $L \in \mathcal{L}$, let $L_*$ be the element in $\mathcal{L}$ that attains the infimum $L_* = \argmin_{L \in \mathcal{L}}\overline{\mathrm{R}}_{L,n,\epsilon}^{\mathtt{Est}}$. This concludes the claim such that
\begin{align*}
    \mathbb{P}(\|\widehat f_n - f_0\|^2 \ge \mathfrak{C}\overline{\mathrm{R}}_{L_*,n,\epsilon}^{\mathtt{Est}}+ \mathrm{R}_{\mathcal{L}, n,\epsilon}^{\mathtt{CV}}) \le \epsilon.
\end{align*}

\end{proof}
\begin{remark}
    Similar bounds in expectation can be provided under additional assumptions, for instance, $\|\widehat f_n - f_0\|_\infty = O_P(1)$.
\end{remark}

\subsection{Proofs of \Cref{thm:R2Q2} and \Cref{thm:R2Q2-robust}}
The proofs of \Cref{thm:R2Q2} and \Cref{thm:R2Q2-robust} follow directly from \Cref{lemma:general-tower-prop} once the oracle inequality \eqref{eq:oracle-CV} is verified for \texttt{MEAN} and \texttt{ROBUST}. We denote the set of random functions as 
 	\begin{align*}
		\mathcal{M}(\mathcal{L}) := \left\{\widehat f_{L} := \widehat g_{L}(x) - h_L(x)\,; \, L \in \mathcal{L}\right\}.
	\end{align*}
Conditioning on $\mathcal{I}_1$, each element in $\mathcal{M}(\mathcal{L})$ becomes no longer random. The problem is then reduced to standard model selection over functions indexed by $\mathcal{L}$.
\begin{proof}[Proof of \Cref{thm:R2Q2}]
    First, we establish \eqref{eq:oracle-CV}. Under \Cref{as:Lipschitz-in-parameter}, it follows that 
    \[\|L_1 - L_2\|_\mathcal{L} \le \eta \Rightarrow \|\widehat f_{L_1}-\widehat f_{L_2}\|_\infty \le \eta.\]
    Hence, the $\eta$-covering number of $\mathcal{M}(\mathcal{L})$ with respect to $\|\cdot\|_\infty$-norm is bounded by the $\eta$-covering number of $\mathcal{L}$ with respect to $\|\cdot\|_{\mathcal{L}}$-norm. This readily implies 
    \begin{align*}
        \mathcal{N}(\eta, \mathcal{M}(\mathcal{L}), L_2(\mathbb{P}_n)) \le \mathcal{N}(\eta, \mathcal{L}, \|\cdot\|_{\mathcal{L}}).
    \end{align*}

For notational convenience, we introduce $\mathbb{P}_C(\cdot) = \mathbb{P}(\cdot | \mathcal{I}_1, \mathcal{E}_{M}^{\tt bdd})$ and $M = M(n, \epsilon)$. By \Cref{thm:oracle-cv}, we have 
\begin{align*}
&\mathbb{P}_C\left(\|\widehat{f}_n - f_0\|^2\ge \mathfrak{C}\|\widehat{f}_{L} - f_0\|^2 + C_\epsilon\inf _{\eta>0}\left\{\log \left(\mathcal{N}\left(\eta, \mathcal{L}, \|\cdot\|_{\mathcal{L}}\right)\right)\left(\mathrm{R}_n M+\frac{ M^2}{n}\right)+\eta^2\right\}\right) \\
    &\le \mathbb{P}_C\left(\|\widehat{f}_n - f_0\|^2\ge \mathfrak{C}\|\widehat{f}_{L} - f_0\|^2 + C_\epsilon\inf _{\eta>0}\left\{\log \left(\mathcal{N}\left(\eta, \mathcal{M}(\mathcal{L}), L_2\left(\mathbb{P}_n\right)\right)\right)\left(\mathrm{R}_n M+\frac{ M^2}{n}\right)+\eta^2\right\}\right) \\
    &\le \epsilon.
\end{align*}
where $\mathrm{R}_n$ is defined in \Cref{thm:oracle-cv}. Hence this establishes \eqref{eq:oracle-CV} with
    \begin{align*}
    \mathrm{R}_{\mathcal{L}, n, \epsilon}^{\mathtt{CV}} = C_\epsilon \log \left(\mathcal{N}\left(\eta, \mathcal{L}, \|\cdot\|_{\mathcal{L}}\right)\right)\left(\mathrm{R}_n M+\frac{M^2}{n }\right)+\eta^2.
\end{align*}
	
The claim is concluded in view of \Cref{lemma:general-tower-prop}.
\end{proof}
\begin{proof}[Proof of \Cref{thm:R2Q2-robust}]
    The proof is almost analogous to that of \Cref{thm:R2Q2}. Instead of \Cref{thm:oracle-cv}, we invoke \Cref{thm:oracle-cv-mom} with the choice of $J = \mathcal{N}\left(\eta, \mathcal{L}, \|\cdot\|_{\mathcal{L}}\right)$ and $B = 4 \lceil \ln J \rceil$. The set $\mathcal{L}_J$ corresponds to the $\eta$-net of $\mathcal{L}$. In other words, for any $L \in \mathcal{L}$, there is a closest element in $\{L_1, \ldots, L_J\}$ to $L$, which we denote as $L^\dag$. Then by \Cref{as:Lipschitz-in-parameter}, it follows that
    \begin{align*}
        \|f_{L^\dag} - f_0\|_\infty^2 \le 2\|f_{L} - f_0\|_\infty^2 + 2\|f_{L^\dag} - f_{L}\|_\infty^2 \le 2\|f_{L} - f_0\|_\infty^2 + 2\eta^2.
    \end{align*}
    \Cref{thm:oracle-cv-mom} then implies \eqref{eq:oracle-CV} with 
    \begin{align*}
    \mathrm{R}_{\mathcal{L}, n, \epsilon}^{\mathtt{CV}} = C_\epsilon \left(\log \left(\mathcal{N}\left(\eta, \mathcal{L}, \|\cdot\|_{\mathcal{L}}\right)\right)\frac{M(n,\epsilon)^2}{n} + \eta^2\right).
\end{align*}
The claim is concluded in view of \Cref{lemma:general-tower-prop}.
\end{proof}

\newpage
\section{Proofs Related to Isotonic Decomposition}\label{app: proof of isotonic regression using new structure}

In this section, we present proof of the oracle inequalities for univariate regression in $\mathcal{F}(1,L)$ spaces.

\subsection{Proof of \Cref{th: general_isotonic}}\label{app: general_isotonic_proof}

\begin{proof}[Proof of \Cref{th: general_isotonic}]
This result is a special case of \Cref{thm:R2Q2}. In the following sections, we verify that the estimator satisfies the required Assumptions \ref{as:oracle-est-adaptive}--\ref{as:stochastic-bounded}. In \Cref{supp:boundedness}, we show that Assumption \ref{as:stochastic-bounded} is satisfied when $M(n,\epsilon) = C\epsilon^{-1}B_n$, where $B_n = \left\|f_0\right\|_{\infty}+\sigma^2+L_+$ and $L_+$ may grow slowly with $n$. \Cref{as:Lipschitz-in-parameter} is verified in \Cref{suppsec:application-model-selection} (defining $2|L_1 - L_2| = \|L_1 - L_2\|_{\mathcal{L}}$). 

 For each $L \in \mathcal{L}$, the projection of the regression function $f_0$ onto $\mathcal{F}(1,L)$ is denoted by
\begin{equation}\label{eq:fixed-L-oracle}
    f_L^* := \argmin_{f \in \mathcal{F}(1,L)}\, \|f_0 - f\|^2.
\end{equation}

In \Cref{th: orcale theorem}, we showed that for any $\epsilon >0$, there exists $C_\epsilon$ such that:
\begin{equation*}
\mathbb{P}\left(\|\widehat{f}_L-f_L^*\|_{L_2(P)} \geq \delta_n \mid\|\widehat{f}_L\|_{\infty} \leq M(n,\epsilon)\right) \leq \epsilon
\end{equation*}
where
\begin{equation*}
\begin{aligned}
  \delta_n &:=\inf _{m \in \mathbb{N}}\left(\inf _{f \in \mathcal{F}_m(1, L)}\left\|f-f_L^*\right\|_{L_2(P)}+C_\epsilon \sigma M(n,\epsilon) (\log n) \sqrt{\frac{m}{n}}\right)\\
  & \leq  \inf _{m \in \mathbb{N}}\left(\inf _{f \in \mathcal{F}_m(1, L)}\left\|f-f_0\right\|_{L_2(P)}+C_\epsilon \sigma M(n, \epsilon) (\log n) \sqrt{\frac{m}{n}}\right)+ \|f_L^* - f_0\|.
\end{aligned}
\end{equation*}

When $\|\widehat{f}_L - f_L^*\| \leq \delta_n$, we have
\begin{equation*}
\begin{aligned}
\|\widehat{f}_L - f_0\|^2 
& \leq 2\left( \|\widehat{f}_L - f_L^*\|^2 + \|f_L^* - f_0\|^2 \right) \\
& \leq 2 \left( \delta_n^2 + \|f_L^* - f_0\|^2 \right) \\
& \lesssim 
 \inf _{m \in \mathbb{N}}\left(\inf _{f \in \mathcal{F}_m(1, L)}\left\|f-f_0\right\|^2+C_{\epsilon}^2\frac{m \sigma^2 M^2(n, \epsilon)\log ^2\left(n \right)}{n}\right)+ \left\|f_L^*-f_0\right\|^2\\
& \lesssim \inf _{m \in \mathbb{N}}\left(\inf _{f \in \mathcal { F } _m ( 1 , L )}\left\|f-f_0\right\|^2+ C_{\epsilon}^2\frac{m \sigma^2 M^2(n, \epsilon)\log ^2\left(n\right)}{n}\right).
\end{aligned}
\end{equation*}
In the last step, we used the fact that $f_L^*$ minimizes $\|f - f_0\|$ over $\mathcal{F}(1,L)\supset \mathcal{F}_{m}(1,L)$. Then we know  \Cref{as:oracle-est-adaptive} is satisfied with 
\begin{equation*}
\mathrm{R}_m^{\mathrm{Est}}  = C_\epsilon^2 \frac{m \sigma^2 B_n^2 \log ^2(n)}{ n}.
\end{equation*}
Then we can apply \Cref{thm:R2Q2} and get
\begin{equation*}
\begin{aligned}
   \left\|\widehat{f}_n-f_0\right\| 
    & \leq \inf_{L \in \mathcal{L}} \inf _{1 \leq m \leq n}\left\{\inf _{f \in \mathcal{F}_m(1, L)}\left\|f-f_0\right\|^2+ C_\epsilon \frac{m \sigma^2 B_n^2 \log ^2\left(n\right)}{n} \right\}\\
   & + C_\epsilon \inf _{\eta>0}\left\{\log \left(\mathcal{N}\left(\eta, \mathcal{L},\|\cdot\|_{\mathcal{L}}\right)\right)\left(C\mathrm{R}_n \epsilon^{-1}B_n+\frac{C\epsilon^{-2}B_n^2}{n}\right)+\eta^2\right\} 
\end{aligned}
\end{equation*}
Take $\eta = n^{-1/2}$,
\begin{equation*}
\begin{aligned}
   &\log \left(\mathcal{N}\left(\eta, \mathcal{L},\|\cdot\|_{\mathcal{L}}\right)\right)\left(C \mathrm{R}_n \epsilon^{-1} B_n+\frac{C \epsilon^{-2} B_n^2}{n}\right)+\eta^2\\
    &\leq C\log(1+2L_+ n^{1/2})(\epsilon^{-1}R_nB_n + \epsilon^{-2}n^{-1}B_n^2)
\end{aligned}
\end{equation*}
\end{proof}

\subsection{Technical Results for Verifying \Cref{as:oracle-est-adaptive}}\label{app: oracle theorem fixed L}
Establishing point-wise oracle inequality for the isotonic decomposition estimator is the most technical part of the proof. For $P$-integrable function $g$, we denote by $Pg = \int g \, dP$ and by $\mathbb{G}_n g := n^{1/2}(\mathbb{P}_n -P)g$ its empirical process. 
\begin{lemma}
	\label{th: orcale theorem}
    Let $\widehat f_{L}$ be the fixed-$L$ estimator given by \eqref{eq:fixed-L-estimator}, and let $f^*_{L}$ be the fixed-$L$ oracle function defined in \eqref{eq:fixed-L-oracle}. Consider the regression model \eqref{eq:regression-function}, and assume that $\mathbb{E}[\xi_i^2 | X_i] \le \sigma^2$. Define
    \[
    B_L := \|f_0\|_{\infty} + \sigma^2 + 2L_+.
    \]
    And let $M(\varepsilon) \geq 1$ be a function of $\varepsilon\in (0,1)$. 
    Then for large enough $n$:
    \begin{equation*}
        \mathbb{P}\left(\|\widehat{f}_L-f_L^*\|_{L_2(P)} \geq \delta_n \mid\|\widehat{f}_L\|_{\infty} \leq M(\varepsilon) B_L\right) \leq CD^{-1}
    \end{equation*}
    where $D\geq 1$ and $C>0$ are two constants. And the sequence $\delta_n$ is
    \begin{equation*}
        \delta_n = \inf _{m \in \mathbb{N}}\left(\inf _{f \in \mathcal{F}_m(1, L)}\left\|f-f_L^*\right\|_{L_2(P)}+3D\sigma M(\epsilon) B_L (\log n)\sqrt{\frac{m}{n}}\right).
    \end{equation*}
\end{lemma}
\begin{proof}[Proof of \Cref{th: orcale theorem}]
We will apply the peeling device provided in \Cref{peeling lemma} to bound the probability of 
\begin{equation*}
    \mathbb{P}\left(\|\widehat{f}_L-f_L^*\|_{L_2(P)} \geq \delta_n \mid\|\widehat{f}_L\|_{\infty} \leq M(\epsilon) B_L\right)
\end{equation*}
The most important step of applying \Cref{peeling lemma} is finding a $\phi_n$ that establishes a bound on the local multiplier process for the function class of interest. In \Cref{multiplier process lemma} we show that 
\begin{equation*}
\mathbb{E}\left[\sup _{\substack{f \in \mathcal{F}(1, L):\|f-f_L^*\|_{L_2(P)} \leq \delta \\\|f-f_L^*\|_{\infty} \leq M(\epsilon) B_L}}\left|\mathbb{G}_n \xi\left(f-f_L^*\right)\right|\right] \leq \phi_n\left(\delta\right)
\end{equation*}
with
\begin{equation*}
\phi_n(\delta) = \sqrt{m} \sigma\left(\delta+\left\|f_{m, L}-f_L^*\right\|_{L_2(P)}\right) \log \left(3 \delta^{-1} M(\varepsilon) B_L\right),
\end{equation*}
$f_{m, L}$ is any $m$-piecewise function-they are elements of the piecewise function space $\mathcal{F}_m(1, L)$ defined in \eqref{eq:adaptive-class}.

With this choice of $\phi_n$, we know
\begin{equation*}
\mathbb{P}\left(\|\widehat{f}_L-f_L^*\|_{L_2(P)} \geq D \delta_n' \mid\|\widehat{f}_L\|_{\infty} \leq M(\epsilon) B_L\right) \leq 
\frac{C M(\epsilon)B_L}{D^2 \delta_n'^2 \sqrt{n}} \sum_{j=0}^{\infty} \frac{\phi_n\left(2^{j+1} D \delta_n'\right)}{2^{2 j}}.
\end{equation*}
When we choose
\begin{equation*}
    \delta_n' = 
\left\|f_{m, L}-f_L^*\right\|_{L_2(P)}/D + 3\sigma M(\epsilon)B_L  (\log n)\sqrt{\frac{m}{n}},
\end{equation*}
the probability of the event of interest can be bounded by a constant times $D^{-1}$. The detail of this last step is presented in \Cref{lemma: tail probability}.
\end{proof}

\begin{lemma}\label{lemma: tail probability}
Let $\phi(\delta) = C(\delta + A)\log(B/\delta)$ for numbers $A, B> 0$, $C >1$. Let $D \geq 1$. If 
\[
\delta_n \geq \frac{A}{D} + \frac{BC (\log n)}{\sqrt{n}},
\]
then for $n\geq 3$
\[
\mathcal{A} = \frac{B}{D^2 \delta_n^2 \sqrt{n}} \sum_{j=0}^{\infty} \frac{\phi(2^{j+1} D \delta_n)}{2^{2j}} \leq {\rm constant }\times D^{-1}.
\]
\end{lemma}

\begin{proof}[Proof of \Cref{lemma: tail probability}]
Use the definition of $\phi$:
\begin{equation*}
\begin{aligned}
    \phi(2^{j+1} D\delta_n) 
    & = C(2^{j+1} D\delta_n + A) \log(B2^{-j-1}D^{-1}\delta_n^{-1})\\
    & \leq C(2^{j+1} D\delta_n + A)(\log n),
\end{aligned}
\end{equation*}
note that $\delta_n \geq B/n^{-1/2}$. Then the quantity of interest, $\mathcal{A}$, can be bounded by
\begin{equation*}
    \mathcal{A} \leq \frac{BC(\log n)}{D\delta_n \sqrt{n}} \sum_{j=0}^\infty 2^{-j+1} + \frac{BCA(\log n)}{D^2\delta_n^2\sqrt{n}} \sum_{j=0}^\infty 2^{-2j}.
\end{equation*}
We want to further bound the quantity above by a constant times $D^{-1}$. A pair of sufficient conditions are:
\begin{equation*}
    \begin{aligned}
        \sqrt{n} \delta_n &\geq BC(\log n)\\
        \sqrt{n}D\delta_n^2 & \geq BCA (\log n).
    \end{aligned}
\end{equation*}
A further sufficient condition is
\begin{equation*}
    \delta_n \geq \frac{A}{D} + \frac{BC(\log n)}{\sqrt{n}}.
\end{equation*}
For the last step, we used
\begin{equation*}
    \sqrt{\frac{A}{D}}\cdot \frac{\sqrt{BC(\log n)}}{n^{1/4}} \leq \frac{A}{D} +\frac{BC(\log n)}{n^{1/2}}.
\end{equation*}
\end{proof}

\subsection{Verifying \Cref{as:Lipschitz-in-parameter}}\label{suppsec:application-model-selection}
We will verify that the estimators defined in \Cref{sec:estimator} satisfy the Lipschitz condition.
\begin{lemma}\label{lm:lipshitz_in_param}
Let 
\begin{equation}
\widehat{f}_{L}(x):=\widehat{g}_{L}(x)-L x
\end{equation}
be a regression estimator with $\widehat g_L$ calculated as in \eqref{eq:max-min-formula}. Assume $x\in[0,1]$, we have
\begin{align*}
    \|\widehat{f}_{ L_1} - \widehat{f}_{L_2}\|_\infty \le 2|L_1 - L_2|.
\end{align*}
\end{lemma}
\begin{proof}[Proof of \Cref{lm:lipshitz_in_param}] We denote by $|D_1| =n_1$. For each $L \in \mathcal{L}$, the corresponding estimator $\widehat{f}_{L}$ is defined explicitly by the max-min formula \citep{robertson1988order} as
\begin{align*}
    \widehat{f}_{L}(x) :=  \widehat{g}_{L}(x; D_1) - L x = \min_{X_{(u)} \ge x} \, \max_{X_{(l)} \le x} \, \frac{1}{u-l+1}\sum_{i = l}^u (Y_{(i)}+LX_{(i)}) - Lx
\end{align*}
where $X_{(1)}\, \ldots X_{(n_1)}$ is the order statistics of $X$ according to $D_1$ and $Y_{(1)}\, \ldots Y_{(n_1)}$ denotes their corresponding observations without breaking the initial pairs of $(X_i, Y_i)$ for each $i \in \mathcal{I}_1$. Given two estimators $\widehat{f}_{L_1}(x)$ and $\widehat{f}_{L_2}(x)$ constructed from shared data $D_1$, we define the indices that attain the minimum and maximum at in the above display as $(u_1, l_1)$ and $(u_2, l_2)$ respectively. Note that $(u_1, l_1)$ and $(u_2, l_2)$ depend on $x$. To illustrate concretely, we have 
\begin{align*}
    \widehat{f}_{L_1}(x) =  \frac{1}{u_1-l_1+1}\sum_{i = l_1}^{u_1} (Y_{(i)}+L_1X_{(i)}) - L_1x.
\end{align*}
Since two estimators are obtained from the same training set with the only difference being the values of $L$, the order statistics $\{X_{(i)}, Y_{(i)}\}_{i \in \mathcal{I}_1}$ are also shared. For each $x$, it now follows that
    \begin{align*}
       &\widehat{f}_{L_1}(x) - \widehat{f}_{L_2}(x) \\
       &\qquad = (\widehat{g}_{L_1}(x) - L_1 x) - (\widehat{g}_{L_2}(x) - L_2 x) \\
        &\qquad = \min_{X_{(u)} \ge x} \, \max_{X_{(l)} \le x} \, \frac{1}{u-l+1}\sum_{i = l}^u (Y_{(i)}+L_1X_{(i)}) \\
        &\qquad\qquad- \min_{X_{(u)} \ge x} \, \max_{X_{(l)} \le x} \, \frac{1}{u-l+1}\sum_{i = l}^{u} (Y_{(i)}+L_2X_{(i)}) \\
        &\qquad\qquad-(L_1-L_2)x \\
        &\qquad \le \max_{X_{(l)} \le x} \, \frac{1}{u_2-l+1}\sum_{i = l}^{u_2} (Y_{(i)}+L_1X_{(i)}) - \max_{X_{(l)} \le x} \, \frac{1}{u_2-l+1}\sum_{i = l}^{u_2} (Y_{(i)}+L_2X_{(i)}) \\
        &\qquad\qquad- (L_1-L_2)x\\
        &\qquad \le \frac{1}{u_2-l_1+1}\sum_{i = l_1}^{u_2} (Y_{(i)}+L_1X_{(i)}) - \frac{1}{u_2-l_1+1}\sum_{i = l_1}^{u_2} (Y_{(i)}+L_2X_{(i)}) - (L_1-L_2)x \\
        &\qquad = \frac{|L_2-L_1|}{u_2-l_1+1}\sum_{i = l_1}^{u_2} X_{(i)}+ |L_1-L_2|x.
    \end{align*}
    The lower bound can be derived analogously after swapping the roles of $(u_1, l_1)$ and $(u_2, l_2)$. Finally, we conclude the claim by taking the supremum over $x \in [0,1]$.
\end{proof}
\subsection{Verifying \Cref{as:stochastic-bounded}}\label{supp:boundedness}
We show that our proposed estimators are uniformly bounded in probability. The following lemma is a more general result than what is known in current literature, as it does not require the covariate $X_i$ to be independent of the error $\xi_i$.

\begin{lemma}\label{lemma:boundedness_estimator}
Assume the regression model \eqref{eq:regression-function} where $\|f_0\|_\infty < \infty$ and $\E[\xi^2|X] \leq \sigma^2$ almost surely. Let $\widehat{f}_{L}$ be a fixed-$L$ estimator given by \eqref{eq:fixed-L-estimator}. Assume $x\in[0,1]$, it then follows that $\|\widehat f_{L}\|_\infty = O_P(\|f_0\|_\infty + \sigma^2 + |L|)$. 

Furthermore, for any set $\mathcal{L} \subset \mathbb{R}$ whose largest element is denoted by $L_+$, it follows that 
\begin{equation}
\mathbb{P}\left(\sup _{L \in \mathcal{L}}\|\widehat{f}_L\|_{\infty} \geq \frac{C\left(\|f\|_{\infty}+\sigma^2+2 L_{+}\right)}{\epsilon}\right) \leq \epsilon,
\end{equation}
for some large enough constant $C$.
\end{lemma}

\begin{proof}[Proof of Lemma~\ref{lemma:boundedness_estimator}]
For the result under fixed $L$, we only need to establish the uniform boundedness of the monotonic component $\widehat g_{L}$, which directly implies the boundedness of $\widehat f_{L}(x) = \widehat g_{L}(x) + Lx$. For the ease of notation, we will drop the $L$'s in the subscripts. 

We denote by $X_{(1)} \leq X_{(2)} \leq \dots \le X_{(i)} \leq \ldots \leq X_{(n)}$ the order statistics of $X_1,\ldots,X_n$. Similarly, we denote $Y_{(i)}$ and $\xi_{(i)}$ as the corresponding outcome and noise variables of $X_{(i)}$. It is direct to check that $\xi_{(i)}$ and $\xi_{(j)}$ are not independent from each other when $i \neq j$, but we have $\xi_{(i)} \perp \!\!\! \perp \xi_{(j)} | X_1,\ldots,X_n$ for any $i \neq j$. This fact is also known as the conditional independence of \emph{concomitant statistics} or induced order statistics. We demonstrate this fact in the case $n = 2$ and when $X$ follows a discrete distribution. We need to show for any $x,y\in \mathbb{R}$, $x_1,x_2 \in [0,1]$,
\begin{equation}\nonumber
\begin{aligned}
    &\Prob(\xi_{(1)} \leq x, \xi_{(2)} \leq y | X_1 = x_1, X_2 = x_2)\\
    &\qquad = \Prob(\xi_{(1)} \leq x | X_1 = x_1, X_2 = x_2)\Prob(\xi_{(2)} \leq y | X_1 = x_1, X_2 = x_2).
\end{aligned}
\end{equation}
When $x_1 \leq x_2:$
\begin{equation}\nonumber
    \begin{aligned}
     & \Prob(\xi_{(1)} \leq x, \xi_{(2)} \leq y | X_1 = x_1, X_2 = x_2)\\
     & \qquad = \Prob(\xi_{1} \leq x, \xi_{2} \leq y | X_1 = x_1, X_2 = x_2)\\
     & \qquad= \Prob(\xi_{1} \leq x, \xi_{2} \leq y, X_1 = x_1, X_2 = x_2)/\Prob(X_1 = x_1, X_2 = x_2)\\
     & \qquad= \Prob(\xi_{1} \leq x,  X_1 = x_1)\Prob(\xi_{2} \leq y,X_2 = x_2)/\{\Prob(X_1 = x_1)\Prob( X_2 = x_2)\}\\
     & \qquad= \Prob(\xi_{1} \leq x | X_1 = x_1)\Prob(\xi_{2} \leq y | X_2 = x_2)\\
     & \qquad= \Prob(\xi_{1} \leq x | X_1 = x_1 , X_2 = x_2)\Prob(\xi_{2} \leq y | X_1 = x_1, X_2 = x_2)\\
     & \qquad= \Prob(\xi_{(1)} \leq x | X_1 = x_1 , X_2 = x_2)\Prob(\xi_{(2)} \leq y | X_1 = x_1, X_2 = x_2).
    \end{aligned}
\end{equation}
We have a almost identical argument when $x_1 > x_2$, in this case $\xi_{(1)} = \xi_2$ and $\xi_{(2)} = \xi_1$. We will use this conditional independence later.

Since $\widehat g$ is a piecewise constant non-decreasing function, we only need to control its evaluation on the smallest and largest covariates. The exact formula for these values is given by the max-min formula (See Theorem 1.4.4 of \cite{robertson1988order}, for instance) as follows:
\begin{equation}\nonumber
\begin{aligned}
   \widehat g_n(X_{(1)}) = \min_{v \geq 1}\, \left\{v^{-1}\sum_{i=1}^v Y_{(i)} - LX_{(i)}\right\}\textrm{ and }\\
    \widehat g_n(X_{(n)}) = \max_{u \leq n}\, \left\{u^{-1}\sum_{i=n-u+1}^n Y_{(i)} - LX_{(i)}\right\}.
\end{aligned}
 \end{equation}
We need to show that these objects are bounded in probability. We only examine $\widehat g(X_{(1)})$ as the argument for $\widehat g(X_{(n)})$ is analogous. Let $v^*$ be the value of $v$ that attains the minimum of the first equation. It then follows that
\begin{equation}
\label{bound on minimal value}
\begin{aligned}
    \left|\widehat g_n(X_{(1)})\right| & = \left|\min_{v \geq 1}v^{-1}\sum_{i=1}^v Y_{(i)} - LX_{(i)}\right| \\
    & = \left|v^{-1}_*\sum_{i=1}^{v_*} Y_{(i)} - LX_{(i)}\right|\\
    & = \left|v^{-1}_*\sum_{i=1}^{v_*} f_0(X_{(i)}) + \xi_{(i)} - LX_{(i)}\right|\\
    & \leq \|f_0\|_\infty +  \left|\max_{1\leq k \leq n}k^{-1}\sum_{i=1}^{k}\xi_{(i)} \right|  
    + |L| \left|v^{-1}_*\sum_{i=1}^{v_*}X_{(i)}\right| \\ 
    \Longrightarrow \E\left[ \left|\widehat g_n(X_{(1)})\right|\right]& \leq \|f_0\|_\infty +  \E\left|\max_{1\leq k \leq n}k^{-1}\sum_{i=1}^{k}\xi_{(i)} \right|\nonumber 
    + |L| .
\end{aligned}
\end{equation}
It remains to examine the middle term. Denoting by $W(1,n) := \max_{1 \leq k \leq n} k^{-1}\left|\sum_{i=1}^{k} \xi_{(i)}\right |$, we observe 
\begin{equation}
\label{relate expectation to probabilit w1n}
\begin{aligned}
    \E[W(1,n)] & = \E[\E[W(1,n) \mid X_1,\ldots,X_n]] \\
    & = \E\left[\int_0^\infty \Prob(W(1,n) \geq t \mid X_1,\ldots,X_n)\, dt \right]\\
    & \leq 1 + \E\left[\int_1^\infty \Prob(W(1,n) \geq t \mid X_1,\ldots,X_n)\, dt \right].
\end{aligned}
\end{equation}
Now we provide the upper bound of the tail probability $\Prob(W(1,n) \geq t \mid X_1,\ldots,X_n)$ for $t\geq 1$. Without loss of generality, we assume $\log_2 n$ is an integer. In the presentation below, we use $ \Prob_c(\cdot)$ to denote the conditional probability $\Prob(\cdot | X_1,\ldots,X_n)$. It then follows that
\begin{equation}\nonumber
    \begin{aligned}
   &\mathbb{P}_c(W(1,n) \geq t) \nonumber\\
   &\qquad\leq \mathbb{P}_c\left(\exists j \in \{1,2,\ldots,\log_2 n\}, \max_{2^{j-1} \leq k < 2^j} k^{-1}\left|\sum_{i=1}^{k} \xi_{(i)}\right | \geq t\right) +
     \mathbb{P}_c\left(\left|\sum_{i=1}^n \xi_{(i)}\right | \geq nt\right)\\
    & \qquad\leq \sum_{j = 1}^{\log_2 n}  \mathbb{P}_c\left(\max_{2^{j-1} \leq k < 2^j} k^{-1}\left|\sum_{i=1}^{k} \xi_{(i)}\right | \geq t\right) + (nt)^{-2}\E\left[\left(\sum_{i=1}^n \xi_{(i)}\right)^2\mid  X_1,\ldots,X_n\right]\\
    &\qquad \leq \sum_{j = 1}^{\log_2 n}  \mathbb{P}_c\left(\max_{2^{j-1} \leq k < 2^j} \left|\sum_{i=1}^{k} \xi_{(i)}\right | \geq 2^{j-1}t\right) + (nt)^{-2}\E\left[\left(\sum_{i=1}^n \xi_{(i)}\right)^2 \mid X_1,\ldots,X_n\right]\\
    &\qquad \leq \sum_{j = 1}^{\log_2 n}(2^{2j-2}t^2)^{-1} \E\left[\max_{2^{j-1} \leq k < 2^j} \left(\sum_{i=1}^{k} \xi_{(i)}\right )^2 \mid X_1,\ldots,X_n\right] + (nt^2)^{-1}\sigma^2\\
    &\qquad \stackrel{(1)}{\lesssim} \sum_{j = 1}^{\log_2 n} (2^{2j}t^2)^{-1} \sum_{i=1}^{2^j} \E[\xi_{(i)}^2 \mid X_1,\ldots,X_n] + (nt^2)^{-1}\sigma^2 \\ 
    &\qquad \leq \sigma^2 t^{-2}\sum_{j = 1}^{\log_2 n} 2^{-j} +  (nt^2)^{-1}\sigma^2 \lesssim \sigma^2t^{-2}.
    \end{aligned}
\end{equation}
In the above derivation, we use the conditional independence of $\xi_{(i)}$ and $\xi_{(j)}$ for $i \neq j$ as shown earlier. In step $(1)$ we applied a Rosenthal-type inequality \citep{merlevede2013rosenthal}. We have demonstrated $\E[W(1,n)]$ in \eqref{relate expectation to probabilit w1n} can be bounded by $1+\sigma^2$ up to a constant. We hence conclude that 
\begin{equation*}
    \E|\widehat g(X_{(1)})| \leq C (\|f_0\|_\infty + \sigma^2 + 2|L|)
\end{equation*}
which implies the boundedness in probability in view of Markov's inequality.

The uniform statement follows by observing that 
\begin{equation*}
\sup _{L \in \mathcal{L}}\|\widehat{f}_L\|_{\infty} \leq \sup _{L \in \mathcal{L}}\left\|\widehat{g}_L\right\|_{\infty}+\sup _{L \in \mathcal{L}} L x \leq \sup _{L \in \mathcal{L}}\left\|\widehat{g}_L\right\|_{\infty}+L_{+} .
\end{equation*}
Again, the max-min formula implies that
\begin{align}
	\sup_{L\in\mathcal{L}}\, |\widehat g(X_{(1)})| &\le \|f_0\|_\infty +  \left|\max_{1\leq k \leq n}k^{-1}\sum_{i=1}^{k}\xi_{(i)} \right|  
    + \sup_{L\in\mathcal{L}}\,L \left|v^{-1}_*\sum_{i=1}^{v_*}X_{(i)}\right|\nonumber\\
    &\le \|f_0\|_\infty +  \left|\max_{1\leq k \leq n}k^{-1}\sum_{i=1}^{k}\xi_{(i)} \right| + L_+\nonumber.
\end{align}
Using the analogous argument for controlling the middle term of the above display, which does not depend on $L$, we conclude that 
\[\E\left[\sup_{L\in\mathcal{L}}\, \|\widehat f_{L}\|_\infty\right] \leq C( \|f\|_\infty + \sigma^2 + 2L_+).\]
We conclude the claim in view of Markov's inequality:
\begin{equation*}
\mathbb{P}\left(\sup _{L \in \mathcal{L}}\|\widehat{f}_L\|_{\infty} \geq \frac{C\left(\|f\|_{\infty}+\sigma^2+2 L_{+}\right)}{\epsilon}\right) \leq \epsilon.
\end{equation*}
\end{proof}

\subsection{Proofs of \Cref{cor:low-complexity} and \Cref{cor:worst-case}}\label{app: proof of isotonic cases}

Both results are special cases of \Cref{th: general_isotonic}. \Cref{cor:low-complexity} is straightforward because there exists an element $f \in \mathcal{F}_m(1, L_0)$ such that $\left\|f-f_0\right\|^2 = 0$. We do not have an additional approximation error. 

To prove \Cref{cor:worst-case}, we need to consider function spaces $\mathcal{F}_m(1, L_0)$ with increasing $m$ as $n\rightarrow\infty$. Define $g_0 = f_0 + L_0x$. By definition we know $\|g_0\|_\infty \leq \|f_0\|_{\infty} + L_0$. Denote $\Delta = g_{0}(1)-g_{0}(0)$. We consider the following partition of $[0,1]$ into $m$ disjoint sets:
\begin{equation*}
    I_j = \{ x\in[0,1] \mid g_0(x) \in [(j-1) \epsilon, j \epsilon], \epsilon = \Delta/m\}.
\end{equation*}
The corresponding approximating function $f$ is 
\begin{equation*}
\begin{aligned}
    f(x) & := g_m (x) - L_0 x\\
 & := \sum_{j=1}^m(j-1) (\Delta/m) 1\left(x \in I_j\right) - L_0 x .
\end{aligned}
\end{equation*}
Under this choice, 
\begin{equation*}
    f(x) - f_0(x) = (g_m(x) - L_0 x) - (g_0(x) - L_0 x) = g_m(x) - g_0(x),
\end{equation*}
so the approximation error is
\begin{equation*}
    \begin{aligned}
        \left\|f-f_0\right\|^2 &= \int_{[0,1]} (g_m(x) - g_0(x))^2 dP_X(x)\\
        & = \sum_{j=1}^m \int_{I_j} (g_m(x) - g_0(x))^2 dP_X(x)\\
        & = \sum_{j=1}^m \int_{I_j} (g_m(x) - g_0(x)\cdot 1(x \in I_j))^2 dP_X(x).
    \end{aligned}
\end{equation*}

By the definition of $g_m$ and $I_j$:
\begin{equation*}
    \begin{aligned}
        |g_m(x) - g_0(x) \cdot 1(x \in I_j)|
        & = |(j-1)\Delta/m - g_0(x)| \cdot 1(x \in I_j)\\
        & \leq \Delta m^{-1}\cdot 1(x \in I_j).
    \end{aligned}
\end{equation*}

So we have 

\begin{equation*}
    \begin{aligned}
        \left\|f-f_0\right\|^2 \leq \Delta^2 m^{-2} \sum_{j=1}^m \int_{I_j} 1(x\in I_j) dP_X(x) = \Delta^2 m^{-2}.
    \end{aligned}
\end{equation*}

To balance $\|f - f_0\|^2$ and $\mathrm{R}_{L,m}^{\text {Est }}$ in \Cref{th: general_isotonic}, we need to find $m$ minimizing:
\begin{equation*}
(\left\|f_0\right\|_{\infty}+L_0)^2 m^{-2} + C_\epsilon m \sigma^2 B_n^2 \log ^2(n) n^{-1}.
\end{equation*}
We choose $m = n^{1/3} B_n^{-2/3}\log^{-2/3} n$. Then the summation above is less than 
\begin{equation*}
    C_\epsilon B_n^{4/3} \log^{4/3}(n) n^{-2/3},
\end{equation*}
where $C_\epsilon$ depends on the other constants like $L_0$.

\newpage
\section{Supporting lemmas for oracle inequalities}\label{supp:oracle-lemma}
The proof of \Cref{th: orcale theorem} depends on the oracle inequality for the estimator $\widehat f_L$ under some \emph{fixed} $L$. In this section, we provide details of the technical lemmas invoked in its proof.   

\subsection{The Peeling Device and Local Multiplier Process}\label{suppsec:main-lemma}
Two Lemmas~\ref{peeling lemma} and \ref{multiplier process lemma} play crucial roles in proving \Cref{th: orcale theorem}. Lemma~\ref{peeling lemma} provides a high-probability upper bound of the estimation error of an empirical risk minimizer via the supremum of localized empirical processes. This result is commonly known as the \textit{peeling device} in the literature. Although this result is not novel, we present the details for the ease of the readers' reference. The following result is presented under general settings.

We define $f^*$ and $\widehat f_n$ as the $L_2(P)$ and $L_2(\mathbb{P}_n)$-projections of the true regression function $f_0$ onto $\mathcal{F}$ such that 
\begin{align}
	f^* &:= \argmin_{f\in \mathcal{F}}\, \|Y-f\|^2_{L_2(P)} = \argmin_{f\in \mathcal{F}} \, \mathbb{E}\left(f_0(X)+\xi-f(X)\right)^2,\quad \textrm{and}\nonumber\\
    \widehat f_n &:= \argmin_{f\in \mathcal{F}} \, \|Y-f\|^2_{L_2(\mathbb{P}_n)} = \argmin_{f\in \mathcal{F}} \, n^{-1}\sum_{i = 1}^n\left(f_0(X_i)+\xi_i-f(X_i)\right)^2.\nonumber
\end{align}

We also use 
\begin{equation*}
\mathcal{F}\left(\delta_n\right):=\left\{f \in \mathcal{F} ;\left\|f-f^*\right\|_{L_2(P)} \leq \delta_n\right\}
\end{equation*}
to denote a collection of functions that are close to $f^*$.

\begin{lemma}
\label{peeling lemma}
Let $\mathcal{F}$ be a convex set of functions and $\widehat f_n$ minimizes $\mathbb{P}_n(Y - f(X))^2$ over $f\in\mathcal{F}$. Consider the regression model \eqref{eq:regression-function}.
We assume that 

\begin{itemize}
    \item for any $f\in\mathcal{F}$, there exists $B^* < \infty$ such that $\|f - f^*\|_\infty \leq B^*$;
    \item for some $\phi_n : \mathbb{R}\rightarrow\mathbb{R}$ the following inequalities hold:
\begin{equation}
\label{three processes equation}
\begin{aligned}
&\mathbb{E} \left[\sup _{f \in \mathcal{F}(\delta_n)}\,\left| \mathbb{G}_n\xi\left(f-f^*\right)\right| \right] \leq \phi_n(\delta_n),\\
&\mathbb{E} \left[ \sup _{f \in \mathcal{F}(\delta_n)}\,\left|\mathbb{G}_n\epsilon\left(f-f^*\right)\right|\right] \leq \phi_n(\delta_n),\quad\textrm{and}\\
&\mathbb{E}  \left[\sup _{f \in\mathcal{F}(\delta_n)}\,\left|\mathbb{G}_n\epsilon\left(f-f^*\right)\times \left(f_0-f^*\right)\right|\right] \leq \phi_n(\delta_n),
\end{aligned}
\end{equation}
where $\{\epsilon_i\}_{i=1}^n$ is an IID Rademacher random variable.
\end{itemize} 

Then, there exists a constant $C>0$ such that for any given $D>0$ and $\delta_n > 0$:
\begin{equation}
\mathbb{P}\left(\left\|\widehat{f}_n-f^*\right\|_{L_2(P)} \geq D \delta_n\right) \leq \frac{C B^*}{D^2 \delta_n^2 \sqrt{n}} \sum_{j=0}^{\infty} \frac{\phi_n\left(2^{j+1} D \delta_n\right)}{2^{2 j}}.
\end{equation}
\end{lemma}
\begin{proof}[Proof of Lemma~\ref{peeling lemma}]
The event we aim to control is
\begin{equation}\nonumber
    \left\{\|\widehat f_n - f^*\|_{L_2(P)} \geq D\delta_n\right\}.
\end{equation}

We apply the following peeling device:
\begin{equation}\nonumber
    \begin{aligned}
    & \left\{\|\widehat f_n - f^*\|_{L_2(P)} \geq D\delta_n\right\}\nonumber\\
    &\qquad \stackrel{(I)}{\subseteq}\left\{ \inf_{\|f - f^*\|_{L_2(P)} \geq D\delta_n} \,\mathbb{P}_n(Y- f(X))^2 \leq \mathbb{P}_n(Y- f^*(X))^2\right\}\\
    & \qquad = \bigcup_{j=0}^{\infty} \left\{\inf_{f\in {\rm slice}_j} \mathbb{P}_n(Y- f(X))^2 - \mathbb{P}_n(Y- f^*(X))^2 \leq 0\right\}.
    \end{aligned}
\end{equation}
where ${\rm slice}_j = \{f\in\mathcal{F} \mid 2^j D \delta_n \leq\left\|f-f^*\right\|_{L_2(P)} \leq 2^{j+1} D \delta_n\}$. In step $(I)$, we used that $\widehat f_n$ is the risk minimizer. Therefore, we have
\begin{equation}\label{eq:afterpeeling}
\begin{aligned}
    & \mathbb{P}(\|\widehat f_n - f^*\|_{L_2(P)}\geq D\delta_n) 
    \\
    &\qquad\leq \sum_{j=0}^{\infty} \mathbb{P}\left(\inf_{f\in {\rm slice}_j} \mathbb{P}_n(Y- f(X))^2 - \mathbb{P}_n(Y- f^*(X))^2 \leq 0\right)\\
    &\qquad = \sum_{j=0}^{\infty} \mathbb{P}\left(\inf _{f \in \operatorname{slice}_j} \mathbb{K}_n\left(f, f^*, f_0\right) \leq-P\left(f^*-f\right)^2-2 P\left(f_0-f^*\right)\left(f^*-f\right)\right)
\end{aligned}
\end{equation}
where
\begin{equation*}
  \begin{aligned}
    \mathbb{K}_n(f,f^*,f_0) & := \mathbb{P}_n(Y- f(X))^2 - 
    \mathbb{P}_n(Y- f^*(X))^2 \\
    & \quad - P(f^* - f)^2 - 2P(f_0 - f^*)(f^* - f).
\end{aligned}  
\end{equation*}

The last term of the inequality \eqref{eq:afterpeeling} can be further bounded as follows:
\begin{equation}\label{eq:useMineq}
    \begin{aligned}
     &\mathbb{P}(\|\widehat f_n - f^*\|\geq D\delta_n)\\
    &\qquad \stackrel{(1)}{\leq} \sum_{j=0}^{\infty} \mathbb{P}\left(\inf_{f\in {\rm slice}_j}\,\mathbb{K}_n(f,f^*,f_0) \leq - P(f^* - f)^2\right)\\
    &\qquad  \leq  \sum_{j=0}^{\infty} \mathbb{P}\left(\inf_{f\in {\rm slice}_j}\,\mathbb{K}_n(f,f^*,f_0) \leq -2^{2j}(D\delta_n)^2 \right)\\
    &\qquad  \leq \sum_{j=0}^{\infty} \mathbb{P}\left(\sup_{f\in {\rm slice}_j}\,|\sqrt{n}\mathbb{K}_n(f,f^*,f_0)| \geq \sqrt{n}2^{2j}(D\delta_n)^2 \right)\\
    & \qquad \leq \sum_{j=0}^{\infty} \mathbb{E}\left[\sup_{f\in {\rm slice}_j}\,|\sqrt{n}\mathbb{K}_n(f,f^*,f_0)|\right]/(\sqrt{n}2^{2j}(D\delta_n)^2)\\
    &\qquad  \leq \sum_{j=0}^{\infty} \mathbb{E}\left[\sup_{ f\in {\rm ball}_j}\, |\sqrt{n}\mathbb{K}_n(f,f^*,f_0)|\right]/(\sqrt{n}2^{2j}(D\delta_n)^2).
    \end{aligned}
\end{equation}
where ${\rm ball}_j = \left\{f \in \mathcal{F}\, ;\, \|f-f^*\|_{L_2(P)} \leq 2^{j+1}D\delta_n\right\}$

In step (1), we use that $\mathcal{F}$ is convex, which implies $P(f_0 - f^*)(f^* - f)>0$. To elaborate this claim, for any $0 < \delta < 1$, we observe that 
\begin{align*}
    P(f_0-f^*)^2 
    &\stackrel{(2)}{\leq} P(f_0-(1-\delta)f^*-\delta\widehat{f}_{n})^2\\
    &=P(f_0-f^*+\delta(f^*-\widehat{f}_{n}))^2 \\
    &=P(f_0-f^*)^2+2\delta P(f_0-f^*)(f^*-\widehat{f}_{n}) + \delta^2 P(f^*-\widehat{f}_{n})^2\\
    \Longrightarrow 2 &P(f_0-f^*)(f^*-\widehat{f}_{n}) \geq -\delta P(f^*-\widehat{f}_{n})^2
\end{align*}
and thus we conclude $P(f_0-f^*)(f^*-\widehat{f}_{n})\geq 0$ by taking $\delta \rightarrow 0$. In step (2) above, we use the definition of $f^*$ as a $L^2(P)$-projection of $f_0$ onto $\mathcal{F}$. Also, as $\mathcal{F}$ is convex, $(1-\delta)f^* + \delta\widehat{f}_{n}$ is an element of $\mathcal{F}$. 

We now rearrange the last expressions in \eqref{eq:useMineq}, and we relate them with the empirical processes in our assumption. To show this, we first observe that 
\begin{equation}
    \begin{aligned}\nonumber
    &\mathbb{P}_n(Y- f(X))^2 - \mathbb{P}_n(Y- f^*(X))^2\\
    &\qquad= \mathbb{P}_n(Y-f_0(X))^2 +  \mathbb{P}_n(f_0 - f)^2 + 2\mathbb{P}_n(Y - f_0(X))(f_0 - f)\\
    &\qquad \qquad - \mathbb{P}_n(Y-f_0(X))^2 - \mathbb{P}_n(f_0 - f^*)^2 - 2\mathbb{P}_n(Y-f_0(X))(f_0 - f^*)\\
    & \qquad= \mathbb{P}_n(f_0 - f)^2- \mathbb{P}_n(f_0 - f^*)^2 +2\mathbb{P}_n(Y-f_0(X))(f^*_0 - f)\\
    &\qquad = \mathbb{P}_n(f_0 - f^*)^2 + \mathbb{P}_n(f^* - f)^2  +2\mathbb{P}_n(f_0-f^*)(f^*_0 - f)\\
    &\qquad\qquad - \mathbb{P}_n(f_0 - f^*)^2 + 2\mathbb{P}_n\xi(f^*- f)\\
    &\qquad = \mathbb{P}_n(f^* - f)^2 +2\mathbb{P}_n(f_0-f^*)(f^*_0 - f)+ 2\mathbb{P}_n\xi(f^* - f).
    \end{aligned}
\end{equation}
Subtracting $P(f^* - f)^2 + 2P(f_0 - f^*)(f^* - f)$ from both sides, we obtain
\begin{equation}
\begin{aligned}\nonumber
    \mathbb{K}_n(f, f^*, f_0) 
    &= \mathbb{P}_n(f^* - f)^2 +2\mathbb{P}_n(f_0-f^*)(f^*_0 - f)+ 2\mathbb{P}_n\xi(f^* - f)\\
   &\qquad - P(f^* - f)^2 - 2P(f_0 - f^*)(f^* - f)\\
   \Longrightarrow  \sqrt{n}\mathbb{K}_n(f, f^*, f_0) 
   &= \mathbb{G}_n(f^* - f)^2 + 2\mathbb{G}_n(f_0 - f^*)(f^* - f) + 2\mathbb{G}_n\xi(f^* - f).
\end{aligned}
\end{equation}
Continuing from \eqref{eq:useMineq}, it follows that
\begin{equation}
\label{eq:notcleanempiricalp}
\begin{aligned}
    &\mathbb{P}(\|\widehat f_n - f^*\|\geq D\delta_n) \leq 2\sum_{j=0}^{\infty}(\sqrt{n}2^{2j}D^2\delta_n^2)^{-1}\left( \mathbb{E}\left[\sup_{ f\in {\rm ball}_j}\, |\mathbb{G}_n(f^* - f)^2|\right]\right.\\ & \qquad +\left.\mathbb{E}\left[\sup_{ f\in {\rm ball}_j} \,|\mathbb{G}_n(f_0 - f^*)(f^* - f)|\right]  + \mathbb{E}\left[\sup_{ f\in {\rm ball}_j}\, |\mathbb{G}_n\xi(f^* - f)|\right] \right).
\end{aligned}
\end{equation}
Using a standard symmetrization argument (See, for instance, \cite[Section 3.3]{sen2018gentle}), we have
\begin{equation}\nonumber
\begin{aligned}
        & \mathbb{E}\left[\sup_{ f\in {\rm ball}_j}\, |\mathbb{G}_n(f^* - f)^2|\right] \leq 2\mathbb{E}\left[\sup_{ f\in {\rm ball}_j}\, \left|\mathbb{G}_n \epsilon (f^*\ - f)^2\right|\right]\quad \textrm{and} \nonumber
\end{aligned}
\end{equation}
\begin{align}
        & \mathbb{E}\left[\sup_{ f\in {\rm ball}_j}\, |\mathbb{G}_n(f_0 - f^*)(f^* - f)|\right] \leq 2\mathbb{E}\left[\sup_{ f\in {\rm ball}_j}\left|\mathbb{G}_n \epsilon (f_0 - f^*)(f^* - f)\right|\right].\nonumber
\end{align}
By our assumption that $\|f-f^*_0\|_\infty \le B^*$, it also implies $|f^*(X_i) - f(X_i)| \leq B^*$. Therefore, $(f^*(X_i) - f(X_i))^2$ is a $2B^*$-Lipschitz function of $f^*(X_i) - f(X_i)$. Applying Talagrand’s contraction inequality (See, for instance, Theorem 4.12 of \cite{ledoux1991probability}), we have
\begin{equation}
\begin{aligned}\nonumber
        &\mathbb{E}\left[\sup_{ f\in {\rm ball}_j} \,\left|\mathbb{G}_n \epsilon (f^* - f)^2\right|\right] \le 2B^*\mathbb{E}\left[\sup_{ f\in {\rm ball}_j} \,\left|\mathbb{G}_n \epsilon (f^* - f)\right|\right].
\end{aligned}
\end{equation}
Hence, the terms in the right-hand side of \eqref{eq:notcleanempiricalp} can be bounded under our assumptions that:
\begin{equation}
    \begin{aligned}
      &\mathbb{P}(\|\widehat f_n - f^*\|\geq D\delta_n) \\
     & \qquad \lesssim \sum_{j=0}^{\infty}(\sqrt{n}2^{2j}(D\delta_n)^2)^{-1}\left( B^*\mathbb{E}\left[\sup_{ f\in {\rm ball}_j}\, \left|\mathbb{G}_n\epsilon(f - f^*)\right|\right] \right.\\ 
     & \qquad +\left.\mathbb{E}\left[\sup_{ f\in {\rm ball}_j} \,\left|\mathbb{G}_n\epsilon(f_0 - f^*)(f^* - f)\right|\right] +\mathbb{E}\left[\sup_{ f\in {\rm ball}_j}\,\left|\mathbb{G}_n\xi(f^* - f)\right|\right] \right)\\
     & \lesssim B^*D^{-2}\delta_n^{-2}n^{-1/2}\sum_{j=0}^{\infty}\frac{\phi_n(2^{j+1}D\delta_n)}{2^{2j}}.
    \end{aligned}
\end{equation}
\end{proof}

In the proof of \Cref{th: orcale theorem}, we combine \Cref{peeling lemma} with the following bound on the supremum of the empirical process of interest. We note that the proof is only provided for the third term of \Cref{peeling lemma}, which is most dispersed and the proof for other terms follow by analogous derivation. 
\begin{lemma}
\label{multiplier process lemma}
Assume the same conditions as in \Cref{th: orcale theorem}, 

then we know for any $\delta_n \geq n^{-1} > 0, B > 0$,
\begin{equation}
\begin{aligned}
    &\mathbb{E}\left[ \sup _{\substack{f \in \mathcal{F}(1, L):\|f-f^*_{L}\|_{L_2(P)} \leq \delta_n \\\|f-f^*_{L}\|_{\infty} \leq B}}\left|\mathbb{G}_n \xi\left(f-f^*_{L}\right)\right|\right]\nonumber\\
    &\qquad \lesssim \sqrt{m}\sigma (\delta_n + \|f_{m,L} - f^*_{L}\|_{L_2(P)}) \log\{\delta_n^{-1}(B+\|f_{0}\|_\infty + 2L)\}.
\end{aligned}
\end{equation}
\end{lemma}

\begin{proof}[Proof of Lemma~\ref{multiplier process lemma}]
For ease of notation, we drop the $L$ in the subscript for $f_{m,L}$ and $f^*_{L}$. For any given $f_m$, we have
\begin{equation}
\label{transform the supremum}
\begin{aligned}
&\mathbb{E} \left[\sup _{\substack{f \in \mathcal{F}(1, L):\|f-f^*\|_{L_2(P)} \leq \delta_n \\\|f-f^*\|_{\infty} \leq B}}\, \left|\mathbb{G}_n \xi(f-f^*)\right|\right]\\
&\qquad \leq \mathbb{E} \left[\sup _{\substack{f \in \mathcal{F}(1, L):\|f-f^*\|_{L_2(P)} \leq \delta_n \\\|f-f^*\|_{\infty} \leq B}}\left|\mathbb{G}_n \xi(f-f_m)\right|\right]+\mathbb{E}\left|\mathbb{G}_n\xi(f_m-f^*)\right|\\
& \qquad\leq \mathbb{E} \left[
\sup _{\substack{f \in \mathcal{F}(1, L):\|f-f_m\|_{L_2(P)} \leq \delta_n+\|f_m-f^*\|_{L_2(P)} \\\|f-f_m\|_{\infty} \leq B+\|f^* - f_m\|_{\infty}}}
\left|\mathbb{G}_n \xi(f-f_m)\right|\right]+\mathbb{E}\left|\mathbb{G}_n \xi(f_m-f^*)\right|\\
& \qquad\leq \mathbb{E} \left[
\sup _{\substack{f \in \mathcal{F}(1, L):\|f-f_m\|_{L_2(P)} \leq \delta_n+\|f_m-f^*\|_{L_2(P)} \\\|f-f_m\|_{\infty} \leq B+\left\|f^* - f_m\right\|_{\infty}}}
\left|\mathbb{G}_n \xi(f-f_m)\right|\right]+\sigma\|f_m - f^*\|_{L_2(P)}.
\end{aligned}
\end{equation}
Denoting by $B_0 := B + 2\|f^*\|_{\infty}$ and $\tilde\delta_n := \delta_n + \|f_m - f^*\|_{L_2(P)}$, we need to bound the following term:
\begin{equation}\nonumber
 \mathbb{E}\left[ 
\sup _{\substack{f \in \mathcal{F}(1, L):\|f-f_m\|_{L_2(P)} \leq \tilde\delta_n
\\\|f-f_m\|_{\infty} \leq B_0}}
\left|\mathbb{G}_n \xi(f-f_m)\right|\right].
\end{equation}
By definition of $f \in \mathcal{F}(1, L)$ and $f_m$, we have
\begin{align}
	f(x) = g(x)-Lx\quad\textrm{and}\quad f_m(x) = \sum_{j=1}^m a_j1(x\in I_j)-Lx\nonumber,
\end{align}
where $g$ is a non-decreasing function and $\{a_j\}_{j=1}^m$ is a sequence such that $a_{i_1} < a_{i_2}$ for any $i_1 < i_2$. Hence, this quantity is identical to the following expression:
\begin{equation}\nonumber
     \mathbb{E}\left[ 
\sup _{\substack{g \in \mathcal{C}:\|g-g_m\|_{L_2(P)} \leq \tilde\delta_n
\\\|g-g_m\|_{\infty} \leq B_0}}
\left|\mathbb{G}_n \xi(g-g_m)\right|\right],
\end{equation}
where $ \mathcal{C}$ is a convex cone consisting of all non-decreasing functions from $[0,1]$ to $\mathbb{R}$ and $g_m(x) :=\sum_{j=1}^m a_j1(x\in I_j)$. We now define a new function space $\mathcal{F}_{\textrm{target}} \equiv \mathcal{F}_{\textrm{target}}(\tilde \delta_n, B_0)$ whose element is defined as $g-g_m$ for any non-decreasing function $g$ and fixed $g_m$, such that, 
\begin{align}
	\mathcal{F}_{\textrm{target}}:= \left\{g-g_m\, \bigg| \, g \in \mathcal{C}, \|g-g_m\|_{L_2(P)}\le \tilde \delta_n, \|g-g_m\|_\infty \le B_0\right\}.\nonumber
\end{align}
Each function in this space is an $m$-piece piecewise isotonic function. Hence, it is equivalent to analyzing the following:
\begin{equation*}
\mathbb{E}\left[\sup _{f \in \mathcal{F}_{\text {target }}}\left|\mathbb{G}_n \xi f\right|\right].
\end{equation*}
The display above can be controlled easily for a sub-Gaussian process. Lemma~\ref{heavy tail multiplier process lemma} in Section~\ref{supp:empirical-process} provides a new result akin to Corollary 2.2.8 by \cite{van1996weak}, allowing for the dependence between $X_i$ and $\xi_i$. Applying Lemma~\ref{heavy tail multiplier process lemma} to analyze the local multiplier process above, we have that 
\begin{equation}
    \mathbb{E}\left[\sup_{f\in\mathcal{F}_{\textrm{target}}} \left|\mathbb{G}_n \xi f\right|\right] \lesssim \sigma J(1,\mathcal{F}_{\textrm{target}})\|F\|_{L_2(P)}.\nonumber
\end{equation}
where $F$ is an envelope function of the function class $\mathcal{F}_{target}$ and $J(1,\mathcal{F}_{\textrm{target}})$ is an entropy integral defined in Lemma~\ref{heavy tail multiplier process lemma}. In Lemma~\ref{local envelope lemma}, we will show that the envelope function $F(\cdot)$ of $\mathcal{F}_{\textrm{target}}$ satisfies 
\[\|F\|_{L_2(P)} \lesssim \sqrt{m}\tilde\delta_n \sqrt{\log(B_0/\tilde\delta_n)}\] 
and $\|F\|_{L_2(Q)} \geq \tilde\delta_n \geq n^{-1}$ for any probability measure $Q$. We use this characterization of the envelope function shortly. We also invoke the following bound on the metric entropy of the isotonic function space: 
\begin{lemma}[Lemma~8 of \cite{han2018robustness}]\label{lamma:entropy-integral}
We define $\mathcal{F}_0 \subset L_{\infty}(1)$ where $L_{\infty}(1)$ is a collection of functions uniformly bounded by $1$. Let $\mathcal{F}_0$ be a VC-major class defined on $\mathcal{X}$. Then, there exists some constant $C \equiv C_{\mathcal{F}_0}>0$ such that for any $\mathcal{F} \subset \mathcal{F}_0$, and any probability measure $Q$, the entropy estimate
\begin{equation}\nonumber
\log \mathcal{N}\left(\varepsilon\|F\|_{L_2(Q)}, \mathcal{F}, \|\cdot\|_{L_2(Q)}\right) \leq \frac{C}{\varepsilon} \log \left(\frac{C}{\varepsilon}\right) \log \left(\frac{1}{\varepsilon\|F\|_{L_2(Q)}}\right), \text { for all } \varepsilon \in(0,1)
\end{equation}
holds for any envelope $F$ of $\mathcal{F}$.
\end{lemma}

Applying Lemma~\ref{lamma:entropy-integral} to our class, we have
\begin{equation}\nonumber
    \log \mathcal{N}\left(\varepsilon\|F\|_{L_2(Q)}, \mathcal{F}_{\textrm{target}}, \|\cdot\|_{L_2(Q)}\right) 
    \leq \frac{C}{\varepsilon} \log \left(\frac{C}{\varepsilon}\right) \log \left(\frac{B_0}{\varepsilon\|F\|_{L_2(Q)}}\right).
\end{equation}
This implies
\begin{equation}\nonumber
\begin{aligned}
    J(1,\mathcal{F}_{\textrm{target}}) &= \int_0^1 \sqrt{ \log \mathcal{N}\left(\varepsilon\|F\|_{L_2(Q)}, \mathcal{F}_{target}, \|\cdot\|_{L_2(Q)}\right) } d\varepsilon \\
    & \lesssim \int_0^1 \sqrt{\frac{C}{\varepsilon} \log \left(\frac{C}{\varepsilon}\right) \log \left(\frac{B_0}{\varepsilon\|F\|_{L_2(Q)}}\right)} d\varepsilon\\
    & \lesssim \sqrt{\log(B_0/\tilde \delta_n)}.
\end{aligned}
\end{equation}
We used $\|F\|_{L_2(Q)} \geq \tilde{\delta}_n$ in the last inequality. Now we can apply Lemma~\ref{heavy tail multiplier process lemma} to obtain
\begin{equation}\nonumber
\mathbb{E}\left[ 
\sup _{f\in \mathcal{F}_{\textrm{target}}}
\left|\mathbb{G}_n \xi f\right|\right]
\lesssim \sqrt{m} \sigma \tilde{\delta}_n \log (B_0 / \tilde{\delta}_n).
\end{equation}
Plug this into \eqref{transform the supremum}, we obtain 
\begin{equation}\nonumber
    \begin{aligned}
    &\mathbb{E}\left[ \sup _{\substack{f \in \mathcal{F}(1, L):\left\|f-f^*\right\|_{L_2(P)} \leq \delta_n \\\left\|f-f^*\right\|_{\infty} \leq B}}\left|\mathbb{G}_n\xi\left(f-f^*\right)\right|\right] \nonumber\\
    &\qquad \lesssim \sqrt{m} \sigma \tilde{\delta}_n \log (B_0 / \tilde{\delta}_n) + \sigma \|f_m - f^*\|_{L_2(P)}\\
    &  \qquad\lesssim \sqrt{m}\sigma (\delta_n + \|f_m - f^*\|_{L_2(P)}) \log\{\delta_n^{-1}(B+2\|f^*\|_\infty)\},
    \end{aligned}
\end{equation}
recalling the definition $B_0 = B + 2\|f^*\|_{\infty}$ and $\tilde\delta_n = \delta_n + \|f_m - f^*\|_{L_2(P)}$ from earlier. We conclude the claim by providing a bound on $\|f^*\|_\infty$ in view of Lemma~\ref{lemma:boundedness_L2-projection}. 
\end{proof}

\begin{lemma}\label{lemma:boundedness_L2-projection}
Let $f_{L}^*$ be a fixed-$L$ oracle function given by \eqref{eq:fixed-L-oracle}. It then follows that $\|f^*_{L}\|_\infty \le \|f_0\|_\infty + 2|L|$.
\end{lemma}
\begin{proof}[Proof of Lemma~\ref{lemma:boundedness_L2-projection}]
	An $L_2(P)$-projection of any function onto $\mathcal{F}(1, L)$ can be decomposed as a sum of $g^*_{L}$ and $-Lx$ where $g^*_{L}$ is a non-increasing function from $[0,1]$ to $\mathbb{R}$. We then observe that  
	\begin{align}
		\|Y-f^*_{L}\|_{L_2(P)} = \|(Y+LX)-g^*_{L}\|_{L_2(P)}.\nonumber
	\end{align}
	Here, the monotonic component $g^*_{L}$ corresponds to the $L_2(P)$-projection of a conditional expectation $\eta_0(x) := \E[Z\mid X=x]$ where $Z:=Y+LX$ onto a space of monotone functions. The solution of the best projection onto the monotonic space is explicitly given by the max-min formula (see Equation 2.6 of \cite{mammen1991estimating} and Lemma 2 of \cite{anevski2011monotone}):
	\begin{align}
		g^*_{L}(x) = \sup_{x \le u}\, \inf_{x \ge l}\, \frac{1}{\Prob(X\in[l, u])}\int_{l}^u\eta_0(x)\, dP(x)\nonumber.
	\end{align}
	Furthermore, we have $\|g^*_{L}\|_\infty \le \|\eta_0\|_\infty \le \|f_0\|_\infty + |L|$.
	Hence, we conclude that 
	\begin{align}
		\|f^*_{L}\|_\infty \le \|g^*_{L}\|_\infty +|L| \le \|f_{0}\|_\infty +2|L|\nonumber.
	\end{align}
	\end{proof}

\subsection{Control the Supremum of Empirical Processes without Assuming Independence}
\label{supp:empirical-process}

The following lemmas are invoked during the proof of Lemma~\ref{multiplier process lemma} in order to relate the supremum of a multiplier process with the entropy integral of the studied class. Although this result is similar to Corollary~2.2.8 of \cite{van1996weak}, it is more general such that the error variable $\xi_i$ is not required to be independent of its corresponding covariate $X_i$.
\begin{lemma}
\label{heavy tail multiplier process lemma}
Let $(X_i,\xi_i)$ for $i=1, \dots ,n$ be IID random variable pairs such that $X_i\in \mathbb{R}^p\sim P_X$, $\E[\xi_i | X_i]=0$ and $\E[\xi_i^2 | X_i]\leq \sigma^2$. We assume that the index function space $\mathcal{F}$ satisfies the following metric entropy condition:
\begin{equation}
   J(1,\mathcal{F}) = \sup_{Q} \int_0^1 \sqrt{\log \mathcal{N}(\varepsilon\|F\|_{L_2(Q)},\mathcal{F}, \|\cdot\|_{L_2(Q)})} \, d\varepsilon< \infty,\nonumber
\end{equation}
where $F$ is the envelope function of $\mathcal{F}$ and the supremum is taken over all the probability measures whose support is identical to that of $X$. We further assume $0\in\mathcal{F}$. Then we have
\begin{equation}
    \mathbb{E}\left[\sup_{f\in\mathcal{F}} \left|\mathbb{G}_n \xi f\right|\right] \lesssim \sigma J(1,\mathcal{F})\|F\|_{L_2(P_X)}.\nonumber
\end{equation}
\end{lemma}

\begin{proof}[Proof of Lemma~\ref{heavy tail multiplier process lemma}]
We first relate the multiplier process of interest to a Rademacher process by the following symmetrization. Denoting by $\{(\tilde\xi_i,\tilde X_i)\}_{i=1}^n$ the jointly IID copies of $\{(\xi_i,X_i)\}_{i=1}^n$ and by $\{\epsilon_i\}_{i=1}^n$ the IID Rademacher variables, it follows that 
\begin{equation}\nonumber
    \begin{aligned}
    \E\left[\sup_{f\in\mathcal{F}}\, \left|\mathbb{G}_n \xi f\right|\right]
    & = \E\left[\sup_{f\in\mathcal{F}}\, \left|\frac{1}{\sqrt{n}}\sum_{i=1}^n \xi_i f(X_i) -\E \tilde\xi_i f(\tilde X_i)\right|\right]\\    
    &\leq n^{-1/2} \E\left[\sup_{f \in \mathcal{F}} \,\left|\sum_{i=1}^n \xi_i f(X_i) - \tilde \xi_i f(\tilde X_i)\right|\right]\\
    & = n^{-1/2} \E\left[\sup_{f \in \mathcal{F}}\, \left|\sum_{i=1}^n \epsilon_i\xi_i f(X_i) - \epsilon_i\tilde \xi_i f(\tilde X_i)\right|\right]\\    
    & \leq 2n^{-1/2}\E\left[\sup_{f\in \mathcal{F}} \,\left|\sum_{i=1}^n \epsilon_i \xi_i f(X_i)\right|\right]\\
    & = 2n^{-1/2}\E\left[\E\left[\sup_{f\in \mathcal{F}} \, \left|\sum_{i=1}^n \epsilon_i \xi_i f(X_i)\right| \mid \{(\xi_i,X_i)\}_{i=1}^n\right]\right].
    \end{aligned}
\end{equation}
We now define a stochastic process $W_f$ indexed by $f\in\mathcal{F}$:
\begin{equation}
\label{define wf}
    W_f := n^{-1/2}\sum_{i=1}^n \epsilon_i \xi_i f(X_i).
\end{equation}
We derive the upper bound of the supremum of $W_f$ conditioning on the observation $\{(\xi_i,X_i)\}_{i=1}^n$:
\begin{equation}\nonumber
    \E\left[\sup _{f \in \mathcal{F}}\,\left|W_f\right| \mid\left\{\left(\xi_i, X_i\right)\right\}_{i=1}^n\right].
\end{equation}
Lemma~\ref{wf is subgaussian lemma} below states that conditioning on $\{(\xi_i,X_i)\}_{i=1}^n$, the stochastic process $\{W_f : f \in \mathcal{F}\}$ is sub-Gaussian with respect to the following pseudo-metric $d$ on $\mathcal{F}$:
\begin{equation}\nonumber
    d(f,g) = \left[n^{-1}\sum_{i=1}^n \xi_i^2 \left\{f(X_i) - g(X_i)\right\}^2\right]^{1/2}.
\end{equation}

Then, we can bound the expected supremum of the $W_f$ by the following Dudley's integral (i.e., Corollary 2.2.8 of \cite{van1996weak}):
\begin{equation}
\label{apply dudley's integral}
\begin{aligned}
     \E\left[\sup_{f\in\mathcal{F}} |W_f|\mid\{(\xi_i,X_i)\}_{i=1}^n\right] 
     & \lesssim \int_0^{diam(\mathcal{F}, d)} \sqrt{\log \mathcal{N}(\delta, \mathcal{F}, d)} \, d\delta\\
     & \leq \int_0^{\|F\|_d} \sqrt{\log \mathcal{N}(\delta, \mathcal{F}, d)} \, d\delta. 
\end{aligned}
\end{equation}
The diameter $diam(\mathcal{F},d)$ is defined as $\sup_{f,g\in\mathcal{F}} d(f,g)$ and $\|f\|_d$ is the induced norm of $d$ where $\|f\|_d := d(f,0)$. 

We now express the upper bound in the above display using the metric entropy condition in the assumption. We define $Q_i := (n\sum_{j=1}^n \xi_j^2)^{-1} \xi_i^2 = n^{-1} \|\xi\|^{-2} \xi_i^2$ where $\|\cdot\|$ is an Euclidean norm. It then follows that for any $f,g\in\mathcal{F}$:
\begin{equation}\nonumber
\begin{aligned}
     d^2(f,g)& = n^{-1}\sum_{i=1}^n \xi_i^2 \{f(X_i) - g(X_i)\}^2 = \left(\sum_{j=1}^n \xi_j^2\right) \left(\sum_{i=1}^n \{f(X_i) - g(X_i)\}^2  Q_i \right).
\end{aligned}
\end{equation}

Conditioning on $\{(\xi_i,X_i)\}_{i=1}^n$, we can define a  discrete probability measure $Q(\cdot) = \sum_{i=1}^n Q_i\delta_{X_i}(\cdot)$. This measure further induces a metric: 
\begin{equation}\nonumber
   \|f - g\|^2_{L_2(Q)} = \int (f(x) - g(x))^2 dQ(x) = \sum_{i=1}^n \{f(X_i) - g(X_i)\}^2  Q_i.  
\end{equation}
It is also direct to verify that $ d(f,g) = \sqrt{\sum_{j=1}^n \xi_j^2}\|f-g\|_{L_2(Q)}$.
We can now continue on \eqref{apply dudley's integral} using this new measure $Q(\cdot)$ as follows:
\begin{equation}\nonumber
\begin{aligned}
     \E\left[\left.\sup_{f\in\mathcal{F}}\,  |W_f|\right|\{(\xi_i,X_i)\}_{i=1}^n\right] 
     & \lesssim \int_0^{\|\xi\|\|F\|_{L_2(Q)}} \sqrt{\log \mathcal{N}(\|\xi\|^{-1}\delta, \mathcal{F}, \|\cdot\|_{L_2(Q)})} \, d\delta\\
     & = \|\xi\|\|F\|_{L_2(Q)}\int_0^1 \sqrt{\log \mathcal{N}(\|F\|_{L_2(Q)}\tau, \mathcal{F}, \|\cdot\|_{L_2(Q)})} \, d\tau.
\end{aligned}
\end{equation}
We take expectation with respect to the joint distribution $\{(\xi_i,X_i)\}_{i=1}^n$ on both sides and obtain
\begin{equation}\nonumber
\begin{aligned}
 \E\left[\sup_{f\in\mathcal{F}} \, \left|W_f\right|\right] 
 & \lesssim J(1,\mathcal{F})\E[\|\xi\|\|F\|_{L_2(Q)} ]\\
 & = J(1,\mathcal{F})\E\left[\|\xi\|\left(\sum_{i=1}^n F^2(X_i) Q_i\right)^{1/2}\right]\\
 & =J(1,\mathcal{F})\E\left[\left(n^{-1}\sum_{i=1}^n F^2(X_i) \xi_i^2\right)^{1/2}\right]\\
 & \leq J(1,\mathcal{F})\left(\E\left[n^{-1}\sum_{i=1}^n F^2(X_i) \xi_i^2\right]\right)^{1/2} \\
 & \leq \sigma J(1,\mathcal{F}) \|F\|_{L_2(P_X)}.
\end{aligned}
\end{equation}
This concludes the claim.
\end{proof}

\begin{lemma}
\label{wf is subgaussian lemma}
Let $W_f :=  n^{-1/2}\sum_{i=1}^n \epsilon_i \xi_i f(X_i)$ and $\{W_f : f \in \mathcal{F}\}$ be the stochastic process defined in \eqref{define wf}. Conditioning on $\{\xi_i, X_i\}$, the above stochastic process is sub-Gaussian with respect to the following pseudo-metric:
\begin{equation}\nonumber
    d(f,g) = \left(n^{-1}\sum_{i=1}^n \xi_i^2 \left(f(X_i) - g(X_i)\right)^2\right)^{1/2}.
\end{equation}
\end{lemma}
\begin{proof}[Proof of Lemma~\ref{wf is subgaussian lemma}]
First, it is direct to see that $W_f$ has a mean zero for every $f$ as this is one of the requirements for a sub-Gaussian process. We recall that the 2-Orlicz-norm for a centered random variable $Z$ is defined as:
\begin{equation}
    \|Z\|_{\psi_2} = \inf \left\{ \lambda > 0:\,  \E\left[\exp\left(\frac{Z^2}{\lambda^2}\right) -1\right]\leq 1\right \}\nonumber.
\end{equation}
A stochastic process is called sub-Gaussian with respect to the metric $d$ if
\begin{align}
	\Prob(|W_f-W_g| > t) \le 2\exp(-t^2/2d(f,g))\nonumber
\end{align}
for every $f,g\in \mathcal{F}$ and $t > 0$. This is implied if we can provide the bound on 2-Orlicz norm such that 
\begin{align}
	\|W_f-W_g\|_{\psi_2} \le \sqrt{6}d(f,g)\nonumber
\end{align}
for any $f,g\in \mathcal{F}$. See Section 2.2.1 of \cite{van1996weak} for details. 
By definition, we have
\begin{equation}\nonumber
    \left(W_f-W_g\right)^2=n^{-1}\left(\sum_{i=1}^n \epsilon_i \xi_i\left\{f\left(X_i\right)-g\left(X_i\right)\right\}\right)^2.
\end{equation}
In the presentation below, we use $ \Prob_c(\cdot)$ to denote the conditional probability $\Prob(\cdot | \{(\xi_i,X_i)\}_{i=1}^n)$. 
Then we have
\begin{equation}\nonumber
    \begin{aligned}
    &\E[\exp((W_f - W_g)^2/\lambda^2)\mid \{(\xi_i,X_i)\}_{i=1}^n]\\
    &\qquad  \leq 1+\int_1^{\infty} \mathbb{P}_c\left(\lambda^{-2}\left(W_f-W_g\right)^2 \geq \log t\right) \, d t\\
    & \qquad= 1 + \int_1^\infty \mathbb{P}_c\left( \left|\sum_{i=1}^n \epsilon_i \xi_i \left(f(X_i) - g(X_i)\right)\right|\geq (n\lambda^2 \log t)^{1/2}\right) dt\\
    &\qquad\stackrel{(1)}{\leq} 1 + \int_1^\infty 2 \exp\left(-\left[2\sum_{i=1}^n \xi_i^2 \left\{f(X_i) - g(X_i)\right\}^2\right]^{-1} n\lambda^2 \log t\right) dt.
    \end{aligned}
\end{equation}
Step $(1)$ follows by the application of Hoeffding's inequality (i.e., Lemma 2.2.7 of \cite{van1996weak}). We denote by $\diamond := n^{-1}\sum_{i=1}^n \xi_i^2 \{f(X_i) - g(X_i)\}^2$, and take $\lambda^2 = 6\diamond$:
\begin{equation}\nonumber
    \begin{aligned}
    &\E[\exp((W_f - W_g)^2/\lambda^2)\mid \{(\xi_i,X_i)\}_{i=1}^n] -1\leq 2\int_1^\infty \exp(-3\log t) dt   = 1.
    \end{aligned}
\end{equation}
By the definition of $\|\cdot\|_{\psi_2}$, we deduce $\|W_f - W_g\|_{\psi_2} \leq \sqrt{6\diamond} = \sqrt{6}d(f,g)$. Therefore we conclude that $W_f$ is a sub-Gaussian process with respect to the norm $d$ on $\mathcal{F}$.
\end{proof}

\subsection{Finite Envelope Functions} \label{supp:envelope}
The aim of this subsection is to provide the norm of envelope functions associated with the piecewise monotonic functions. The corresponding result is used during the proof of \Cref{multiplier process lemma}.
\begin{lemma}
\label{local envelope lemma}
Let $g_m:[0,1]\rightarrow\mathbb{R}$ denote a fixed piecewise constant function of the form
\begin{equation}
    g_m(x) = \sum_{j=1}^m a_j 1(x\in I_j),\nonumber
\end{equation}
where $a_j\in \mathbb{R}$ and $\{I_j\}_{j=1}^m$ is a non-overlapping partition of $[0,1]$. We denote by $P_X$ a marginal probability distribution of $X$ over $[0,1]$. Consider the following function space:
\begin{equation}
    \mathcal{F} := \{q\in \mathcal{C} - g_m :\left\|q\right\|_{L_2(P_X)} \leq \delta, 
\|q\|_{\infty} \leq B\},\nonumber
\end{equation}
where $\mathcal{C}\subset L_2(P_X)$ is the space of monotone functions over $[0,1]$. It then follows that the envelope function $F$ of $\mathcal{F}$ satisfies:
\begin{equation}
    \|F\|_{L_2(P_X)} \lesssim \sqrt{m} \delta \sqrt{\log(B/\delta)}\nonumber
\end{equation}
\end{lemma}
\begin{proof}[Proof of Lemma~\ref{local envelope lemma}]

\begin{figure}
  \centering
  \includegraphics[width=\textwidth]{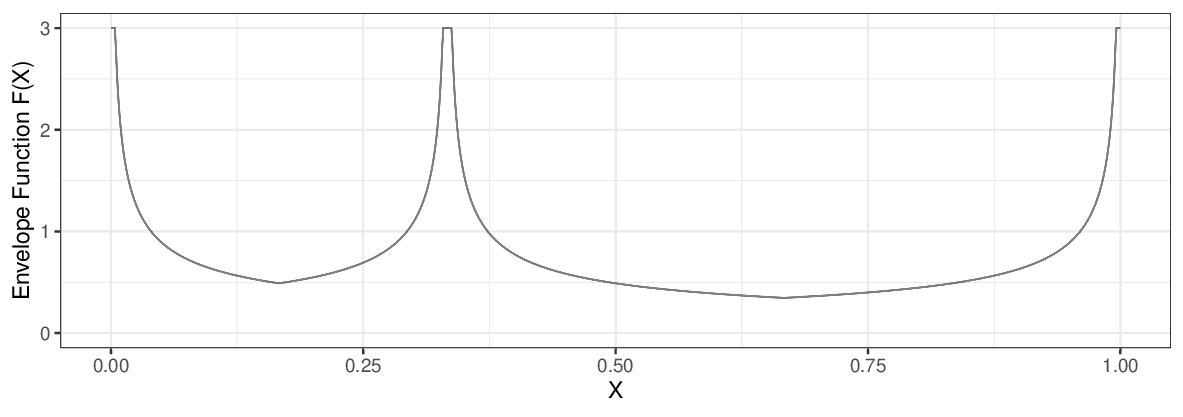}
  \caption{An illustration of the envelope function $F(x)$. We demonstrate the case when $B = 3$, $\delta = 0.2$, $I_1 = [0,1/3], I_2 = (1/3, 1]$ and $P_X$ is uniform over $[0,1]$.}
  \label{fig:env}
\end{figure}

We are going to construct an envelope function of $\mathcal{F}$ for each interval $I_j$. Combining all the envelopes together would give us an envelope function of $\mathcal{F}$ over $[0,1]$.

First, when the interval $I_j$ is measure zero under $P_X$, in other words, $P_X(I_j) = \int_{I_j} dP_X(x) =0$, we can only use the constant function at $B = \|q\|_\infty$ as the envelope for that interval $I_j$. Now we consider the non-trivial case for $I_j$ with $P_X(I_j)$ strictly larger than $0$. For a pre-specified $P_X$, we note that $P_j = P_X/P_X(I_j)$ forms a new probability measure for each interval $I_j$ under the assumption that $P_X(I_j) > 0$. We are going to use the following two facts:
\begin{enumerate}
    \item For any $q\in \mathcal{F}$, its restriction on $I_j$ ,$q|_{I_j}:I_j\rightarrow \mathbb{R}$, is an isotonic function.
    \item The function space 
    \begin{equation}
        \mathcal{H} := \{h \text{ is non-decreasing over }[0,1] \,| \, \|h\|_{L_2(\mu)} \leq \delta, \|h\|_{\infty} \leq B\}\nonumber
    \end{equation}
    has its corresponding envelope function
    \begin{equation}
H(x) := \min \left\{B, \delta \max \left\{\mu([0, x]), \mu([x, 1])\right\}^{-1 / 2}\right\}. \nonumber
\end{equation}
Here, the probability measure $\mu$ is defined over $[0,1]$ and $\mu([a, b])$ is $=\int_a^b \mu(x)$ for $0\leq a < b \leq 1$. Furthermore, we have $\left\|H\right\|_{L_2(\mu)} \leq C \delta \sqrt{\log (B / \delta)}$ (See Section~5.1 of \cite{kuchibhotla2022least}).
\end{enumerate}
The restriction of $\mathcal{F}$ on each interval $I_j$, denoted by $\mathcal{F}_{I_j}$, is a subset of the following function space:
\begin{align}
    &\mathcal{F}_{I_j} \subset 
        \mathcal{G}_j \nonumber\\
        &\qquad := \left\{g:[0,1]\mapsto \mathbb{R}\,\bigg|\, g \text{ is non-decreasing over }I_j, \|g\|_{L_2(P_j)} \leq \delta/P_X^{1/2}(I_j), \|g\|_{\infty} \leq  B\right\}.\nonumber
\end{align}

Combining the above two facts, we know the envelope function of $\mathcal{G}_j$ is
\begin{equation}
G_j(x)=\min \left\{B, P_X^{-1/2}(I_j)\delta \max \left(P_j[I^-_j, x], P_j[x, I^+_{j}]\right)^{-1 / 2}\right\},\text{ for } x\in I_j,\nonumber
\end{equation}
where $I^-_j$ is the left end and $I^+_j$ is the right end of interval $I_j$. We further have 
$$\|G_j\|_{L_2(P_j)} \lesssim P_X^{-1/2}(I_j)\delta\sqrt{\log(P_X^{1/2}(I_j)B/\delta)}.$$

Then we know the envelope function $F$ of $\mathcal{F}$ can be expressed as:
\begin{equation}
    F(x) = \sum_{j=1}^m 1(x\in I_j)\{G_j(x)1(P_X(I_j) > 0) + B1(P_X(I_j) = 0)\}.\nonumber
\end{equation}

We calculate the $\|\cdot\|_{L_2(P_X)}$-norm of $F$:
\begin{equation}\nonumber
\begin{aligned}
    \int_0^1 F^2(x) dP_X(x) 
    &\leq \sum_{j=1}^m \int_{I_j} G_j^2(x) dP_X(x) = \sum_{j=1}^m P_X(I_j)\int_{I_j} G_j^2(x) dP_j(x)\\
    & \lesssim \sum_{j=1}^m P_X(I_j) \left( P_X^{-1}(I_j) \delta^2 \log(P_X^{1/2}(I_j)B/\delta)\right)\\
    &\leq m\delta^2\log(B/\delta).
\end{aligned} 
\end{equation}
This concludes the claim. Figure~\ref{fig:env} below illustrates the derived envelope function. 
\end{proof}

\newpage
\section{Model Selection under Heteroscedastic and Heavy-tailed Errors}
This section derives general oracle inequalities for cross-validation under heteroscedastic and heavy-tailed errors. We first recall several notations. Let $(X_1, Y_1) \ldots (X_n, Y_n)$ be an IID observation from the joint distribution $P$ taking values in $\mathcal{X}\times\mathbb{R}$ where 
\begin{align}
	Y_i = f_0(X_i) + \xi_i \quad \textrm{for}\quad 1\le i \le n
\end{align} 
and $f_0(x) := \E[Y|X=x]$.

Let $\mathcal{F}$ be a subset of a metric space, corresponding to the set of candidate functions (i.e., estimators) of $f_0$. We assume that $\mathcal{F}$ does not depend on the observations. Typically, this is attained by first splitting observations into two sets, using the first half for constructing $\mathcal{F}$ and the other half for selecting the best candidate. We define a loss function $\ell(X,Y; f)$ as the squared error, corresponding to $\ell(X,Y; f):= (Y - f(X))^2$. We assume that $\E[\xi^2|X_i] \le \sigma^2$ for all $1\le i \le n$. 

\subsection{Oracle Inequality --- \texttt{MEAN}}
The first estimator $\widehat f^{\mathtt{MEAN}}$ is determined by the minimizer of the empirical mean of the loss function: 
\begin{align}\label{eq:LSE-estimator}
	\widehat f^{\mathtt{MEAN}} := \argmin_{f \in \mathcal{F}} \frac{1}{n}\sum_{i =1}^n \ell(X,Y; f).
\end{align}
We frequently use the notation $P g := \int g(X,Y) \, dP = \E[g(X,Y)]$. We denote by $\mathbb{P}_n$ the empirical distribution function of the observation, and by $\mathbb{G}_n$ the empirical process, specifically,
\begin{align}
	\mathbb{G}_n g := n^{1/2}\left(\mathbb{P}_n - P\right)g = n^{1/2}\left(\frac{1}{n}\sum_{i=1}^n g(X_i, Y_i) - \E[g(X, Y)]\right).
\end{align}
With the above definition in place, one can deduce the following algebraic inequality, which is a slight modification of \citep[Lemma~2.1]{vaart2006oracle}:

\begin{lemma}\label{lemma:almost-lse}
	Let $\widehat f := \argmin_{f \in \mathcal{F}} \mathbb{P}_n \ell(\cdot; f)$ be the minimizer of the empirical loss function and let $\widehat f^\dag \in \mathcal{F}_\varepsilon \subseteq \mathcal{F}$ be an approximate minimizer of the empirical loss, satisfying for any $\varepsilon, \delta > 0$, 
	\begin{align}\label{lemma:almost-lse-cond}
		 \mathbb{P}_n \ell(\cdot; \widehat f) \le \mathbb{P}_n \ell(\cdot; \widehat f^\dag) \le (1+\delta)\mathbb{P}_n \ell(\cdot; \widehat f) + r(\varepsilon, \delta)
	\end{align}
	where $r(\varepsilon, \delta)$ is a fixed function. Then for any $f \in \mathcal{F}$, it follows that 
		\begin{align*}
	\E[\ell((X, Y); \widehat f^\dag)] &\le (1+2\delta)(1+\delta)\E[\ell((X,Y); f)]\\
	&\qquad+(1+\delta)n^{-1/2}\left\{\sup_{\eta \in \mathcal{F}}\left((1+\delta)\mathbb{G}_n-\delta n^{1/2}P\right)\ell(\cdot; \eta)\right\}\\
    &\qquad + n^{-1/2}\left\{\sup_{\eta \in \mathcal{F}}\left((1+\delta)\mathbb{G}_n+\delta n^{1/2}P\right)-\ell(\cdot; \eta)\right\} \nonumber\\
    &\qquad+ (1+\delta)r(\varepsilon, \delta)
\end{align*}
where the expectations are over $X$ and $Y$. 
\end{lemma}
\begin{proof}[Proof of Lemma~\ref{lemma:almost-lse}]
By the basic inequality and the assumption that $\widehat f^\dag \in \mathcal{F}_\varepsilon \subseteq \mathcal{F}$ is an approximate minimizer of the empirical loss, we have  
	\begin{align}
		\eqref{lemma:almost-lse-cond} \implies (1+\delta)\mathbb{P}_n \ell(\cdot; \widehat f^\dag) &\le  (1+\delta)^2\mathbb{P}_n \ell(\cdot; \widehat f)+ (1+\delta)r(\varepsilon, \delta) \nonumber\\
		&\le (1+\delta)^2\mathbb{P}_n \ell(\cdot; f)  + (1+\delta)r(\varepsilon, \delta)  \quad\textrm{ for any }\quad  f \in \mathcal{F}\nonumber.
	\end{align}
	By adding and subtracting relevant terms, the above inequality implies that 
	\begin{align}
		&P \ell(\cdot; \widehat f^\dag) \le (1+\delta)^2\mathbb{P}_n \ell(\cdot; f) - (1+\delta)\mathbb{P}_n \ell(\cdot; \widehat f^\dag) +P \ell(\cdot; \widehat f^\dag)+ (1+\delta)r(\varepsilon, \delta) \nonumber\\
		&\qquad \Longrightarrow P \ell(\cdot; \widehat f^\dag) \le (1+\delta)^2(\mathbb{P}_n-P) \ell(\cdot; f) - (1+\delta)(\mathbb{P}_n-P) \ell(\cdot; \widehat f^\dag)\nonumber \\
		&\qquad\qquad\qquad+\delta P \ell(\cdot; \widehat f^\dag) + (1+\delta)^2P\ell(\cdot; f)+ (1+\delta)r(\varepsilon, \delta)\nonumber\\
		&\qquad\Longrightarrow P \ell(\cdot; \widehat f^\dag)  \le (1+\delta)^2P\ell(\cdot; f) + (1+\delta)\delta P \ell(\cdot; f) \nonumber \\
        &\qquad \qquad\qquad+ (1+\delta)n^{-1/2}\left\{(1+\delta)\mathbb{G}_n \ell(\cdot; f)-n^{1/2}\delta P \ell(\cdot; f)\right\} \nonumber\\
		&\qquad \qquad \qquad +n^{-1/2}\left\{(1+\delta)\mathbb{G}_n \{-\ell(\cdot; \widehat f^\dag)\}-n^{1/2}\delta P \{-\ell(\cdot; \widehat f^\dag)\}\right\}+ (1+\delta)r(\varepsilon, \delta)\nonumber\\
		&\qquad\Longrightarrow P \ell(\cdot; \widehat f^\dag)  \le (1+2\delta)(1+\delta)P\ell(\cdot; f) \nonumber\\
        &\qquad\qquad\qquad+ (1+\delta)\sup_{\eta\in\mathcal{F}}\, n^{-1/2}\left\{(1+\delta)\mathbb{G}_n \ell(\cdot; \eta)-n^{1/2}\delta P \ell(\cdot; \eta)\right\} \nonumber\\
		&\qquad \qquad \qquad +\sup_{\eta\in\mathcal{F}}\, n^{-1/2}\left\{(1+\delta)\mathbb{G}_n \{-\ell(\cdot;  \eta)\}-n^{1/2}\delta P \{-\ell(\cdot; \eta)\}\right\}+ (1+\delta)r(\varepsilon, \delta)\nonumber.
	\end{align}
        This concludes the claim.
\end{proof}

When $\widehat f^\dag$ is exactly the least-square estimator, we have $r(\varepsilon, \delta)=0$, which coincides with \citep[Lemma~2.1]{vaart2006oracle}. This result suggests that we need to analyze the two uncentered empirical process terms to derive oracle inequalities. \cite{vaart2006oracle} studies this problem under stronger assumptions for the distribution of $\xi$, which we aim to generalize here.

\subsubsection{Proof of \Cref{thm:oracle-cv}}
\begin{proof}[Proof of \Cref{thm:oracle-cv}]
For each $\eta > 0$, let $\mathcal{N}(\eta, \mathcal{F}, L_2(\mathbb{P}_n))$ be the $\eta$-covering number of $\mathcal{F}$ under $L_2(\mathbb{P}_n)$-norm. We denote by $f_j$ for $j = 1, \ldots, \mathcal{N}(\eta, \mathcal{F}, L_2(\mathbb{P}_n))$ the functions corresponding to the center of the $\eta$-covering. We construct a finite set of functions $\mathcal{F}_\eta$ as follows: for $j=1, \ldots, \mathcal{N}(\eta, \mathcal{F}, L_2(\mathbb{P}_n))$, if $f_j \in \mathcal{F}$, we include in $\mathcal{F}_\eta$. Otherwise, we include any function $f'_j \in \mathcal{F}$ such that $\|f_j'-f_j\|_{L_2(\mathbb{P}_n)} \le \eta$. We note that $f_j'$ always exists by the definition of $\eta$-covering. The resulting set $\mathcal{F}_\eta$ satisfies the following properties by construction: (i) $|\mathcal{F}_\eta|=\mathcal{N}(\eta, \mathcal{F}, L_2(\mathbb{P}_n))$, (ii) $\mathcal{F}_\eta \subset \mathcal{F}$ and (iii) for any function $f\in\mathcal{F}$, there exits $f' \in \mathcal{F}_\eta$ such that $\|f-f'\|_{L_2(\mathbb{P}_n)}\le 2\eta$ by the triangle inequality. 

We denote by $\widehat {f}$ the empirical risk minimizer over the original space $\mathcal{F}$.
	By the property (iii) of $\mathcal{F}_\eta$, there exists $\widehat f^\diamond \in \mathcal{F}_\eta$ such that 
	\begin{align}
		\| \widehat f^\diamond - \widehat f \|_{L_2(\mathbb{P}_n)} = \left\{\frac{1}{n}\sum_{i=1}(\widehat f^\diamond(X_i) - \widehat f(X_i))^2\right\}^{1/2} \le 2\eta.\nonumber
	\end{align}   
In other words, the function $\widehat f^\diamond$ is the element of $\mathcal{F}_\eta$ that is closest to $\widehat f$ in $L_2(\mathbb{P}_n)$ metric. This implies the following basic inequality:

\begin{equation*}
\begin{aligned}
n^{-1} \sum_{i=1}^n \ell(\cdot ; \widehat{f}^{\diamond}) & =n^{-1} \sum_{i=1}^n \ell(\cdot ; \widehat{f})+n^{-1} \sum_{i=1}^n \ell(\cdot ; \widehat{f}^{\diamond})-n^{-1} \sum_{i=1}^n \ell(\cdot ; \widehat{f}) \\
& =n^{-1} \sum_{i=1}^n \ell(\cdot ; \widehat{f})+n^{-1} \sum_{i=1}^n(Y_i-\widehat{f}^{\diamond})^2-(Y_i-\widehat{f})^2 \\
& =n^{-1} \sum_{i=1}^n \ell(\cdot ; \widehat{f})+n^{-1} \sum_{i=1}^n-2 Y_i(\widehat{f}^{\diamond}-\widehat{f})+(\widehat{f}^{\diamond}-\widehat{f})(\widehat{f}^{\diamond}+\widehat{f}) \\
& =n^{-1} \sum_{i=1}^n \ell(\cdot ; \widehat{f})+n^{-1} \sum_{i=1}^n(\widehat{f}^{\diamond}-\widehat{f}+2 \widehat{f}-2 Y_i)(\widehat{f}^{\diamond}-\widehat{f}) \\
& =n^{-1} \sum_{i=1}^n \ell(\cdot ; \widehat{f})+n^{-1} \sum_{i=1}^n 2(\widehat{f}-Y_i)(\widehat{f}^{\diamond}-\widehat{f})+n^{-1} \sum_{i=1}^n(\widehat{f}^{\diamond}-\widehat{f})^2 \\
& \stackrel{(I)}{\leq} n^{-1} \sum_{i=1}^n \ell(\cdot ; \widehat{f})+2\left(n^{-1} \sum_{i=1}^n \ell(\cdot ; \widehat{f})\right)^{1 / 2} (2\eta)+4 \eta^2 \\
& \stackrel{(II)}{\leq}(1+\delta) n^{-1} \sum_{i=1}^n \ell(\cdot ; \widehat{f})+(1+1 / \delta) 4 \eta^2,
\end{aligned}
\end{equation*}
where in step $(I)$ uses the Cauchy–Schwarz inequality. Step $(II)$ is Young's inequality 
\begin{equation*}
    2AB = 2 (\sqrt{\delta} A) \cdot (\sqrt{1/\delta} B) \leq \delta A^2 + B^2/\delta
\end{equation*}
where $\delta>0$ can be any fixed number. We now define $\widehat f^\dag$ as the empirical risk minimizer over the finite covering $\mathcal{F}_{\eta}$. Since $\widehat f$ is the empirical risk minimizer over $\mathcal{F}$ and $\mathcal{F}_{\eta}\subset \mathcal{F}$, we obtain the following chain of inequalities:
\begin{equation*}
\begin{aligned}
n^{-1} \sum_{i=1}^n \ell(\cdot ; \widehat{f}^{\diamond}) & \leq(1+\delta) n^{-1} \sum_{i=1}^n \ell(\cdot ; \widehat{f})+(1+1 / \delta) 4 \eta^2 \\
& \leq(1+\delta) n^{-1} \sum_{i=1}^n \ell(\cdot ; \widehat{f}^{\dagger})+(1+1 / \delta) 4 \eta^2.
\end{aligned}
\end{equation*}
We thus have shown that $\widehat f^\diamond$ approximately minimizes the empirical risk over $\mathcal{F}_\eta$. In view of Lemma~\ref{lemma:almost-lse}, for any $\delta >0$ and any $f^\diamond\in\mathcal{F}_\eta$, 
\begin{align*}
	\E[\ell((X, Y); \widehat f^\diamond)] &\le (1+2\delta)(1+\delta)\E[\ell((X,Y); f^\diamond)]\\
	&\qquad+(1+\delta)^2n^{-1/2}\left\{\max_{f \in \mathcal{F}_\eta}\left(\mathbb{G}_n-\frac{\delta}{(1+\delta)} n^{1/2}P\right)\ell(\cdot; f)\right\}\\
    &\qquad + (1+\delta)n^{-1/2} \left\{\max_{f \in \mathcal{F}_\eta}\left(\mathbb{G}_n+\frac{\delta}{(1+\delta)} n^{1/2}P\right)-\ell(\cdot; f)\right\} \nonumber\\
    &\qquad + (1+\delta)(1+1/\delta)4\eta^2.\nonumber
\end{align*}
Let $\ell_0$ be the excess squared loss $\ell_0(f) = \ell_0((X,Y); f) := (Y-f(X))^2 - (Y-f_0(X))^2$, where $(X,Y)\sim P$ is independent from the potentially random function $f$. It is direct to show the marginal expectation $\mathbb{E}[\ell_0(f)] = \mathbb{E}[(f - f_0)^2]$. 

The result thus far implies for $\delta \in (0,1)$,
    \begin{equation}\label{eq: after bound the non-centered}
\begin{aligned}
\mathbb{E}[\ell_0(\widehat{f}^{\diamond})] \leq & (1+2 \delta)(1+\delta) \mathbb{E}[\ell_0(f^{\diamond})] +8(1+1 / \delta)  \eta^2 +R_{n, \delta, \eta}
\end{aligned}
\end{equation}
where 
\begin{equation}\label{eq:R-definition}
\begin{aligned}
R_{n, \delta, \eta} &= (1+\delta)^2n^{-1/2}\left\{\max_{f \in \mathcal{F}_\eta}\left(\mathbb{G}_n-\frac{\delta}{(1+\delta)} n^{1/2}P\right)\ell_0(\cdot; f)\right\}\\
    &\qquad + (1+\delta)n^{-1/2} \left\{\max_{f \in \mathcal{F}_\eta}\left(\mathbb{G}_n+\frac{\delta}{(1+\delta)} n^{1/2}P\right)-\ell_0(\cdot; f)\right\}.
\end{aligned}
\end{equation}

\Cref{eq: after bound the non-centered} resembles the object of interest, except we hope to replace $\widehat{f}^{\diamond}$ by $\widehat f$ on the LHS and replace $f^{\diamond}$ by general $f\in\mathcal{F}$ on the RHS. We can see
\begin{equation*}
\begin{aligned}
\left|\mathbb{E}[\ell_0(\widehat{f}^{\diamond})]-\mathbb{E}[\ell_0(\widehat{f})]\right| & =\left|\mathbb{E}[(\widehat{f}^{\diamond}-f_0)^2]-\mathbb{E}[(\widehat{f}-f_0)^2]\right| \\
& =\left|\mathbb{E}[(\widehat{f}^{\diamond}+\widehat{f}-2 f_0)(\widehat{f}^{\diamond}-\widehat{f})]\right| \\
& =\left|\mathbb{E}[(\widehat{f}^{\diamond}-\widehat{f}+2 \widehat{f}-2 f_0)(\widehat{f}^{\diamond}-\widehat{f})]\right| \\
& \leq \mathbb{E}[(\widehat{f}^{\diamond}-\widehat{f})^2]+2\left|\mathbb{E}[(\widehat{f}-f_0)(\widehat{f}^{\diamond}-\widehat{f})]\right| .
\end{aligned}
\end{equation*}
Furthermore, we have that 
\begin{align*}
    \E[(\widehat f^\diamond -\widehat f)^2] = \E\|\widehat f^\diamond-\widehat f\|^2_{L_2(\mathbb{P}_n)} \le 4\eta^2
\end{align*}
and 
\begin{align*}
    |\E[(\widehat f - f_0)(\widehat f^\diamond -\widehat f)]| \le \sqrt{\E[(\widehat f - f_0)^2]}\sqrt{\E[(\widehat f^\diamond -\widehat f)^2]} \le 2\eta \sqrt{\E[(\widehat f - f_0)^2]}.
\end{align*}
Putting together, we obtain 
\begin{align*}
    \E[\ell_0(\widehat f)]\leq \E[\ell_0(\widehat f^\diamond)] + 4\eta^2 + 2\eta \sqrt{\E[(\widehat f - f_0)^2]}.
\end{align*}
Now the inequality \eqref{eq: after bound the non-centered} can be written as:
\begin{equation}\label{eq: LHS is clean}
    \begin{aligned}
      \mathbb{E}[\ell_0(\widehat{f})] & \leq (1+2 \delta)(1+\delta) \mathbb{E}[\ell_0(f^{\diamond})] + 4 \eta^2+2 \eta \sqrt{\mathbb{E}[\ell_0(\widehat f)]} + 8(1+1 / \delta)  \eta^2 + R_{n, \delta, \eta}\\
    & \leq (1+2 \delta)(1+\delta) \mathbb{E}[\ell_0(f^{\diamond})]+4(3+2\delta^{-1}) \eta^2+ \delta\mathbb{E}[\ell_0(\widehat{f})]+R_{n, \delta, \eta}  \\
    \Rightarrow (1-\delta) \mathbb{E}[\ell_0(\widehat{f})]& \leq (1+2 \delta)(1+\delta) \mathbb{E}[\ell_0(f^{\diamond})]+4(3+2\delta^{-1}) \eta^2 + R_{n, \delta, \eta}.
    \end{aligned}
\end{equation}

Similarly, by the property (iii) of $\mathcal{F}_\eta$, for any $f\in\mathcal{F}$, there exists $ f^\diamond \in \mathcal{F}_\eta$ such that 
\begin{align}
    \mathbb{E}[(f^{\diamond}-f)^2] = \E\| f - f^\diamond \|^2_{L_2(\mathbb{P}_n)} \le 4\eta^2\nonumber
\end{align}  
and 
we have
\begin{align*}
\mathbb{E}[\ell_0(f^\diamond)] - \mathbb{E}[\ell_0(f)]& = \E[( f^\diamond - f_0)^2]-\E[( f - f_0)^2]\\
&= \left|\E[( f^\diamond - f)^2]+ 2\E[( f - f_0)( f^\diamond - f)]\right|\\
&\le 4\eta^2 + 4\eta\left(\E( f - f_0)^2\right)^{1/2}.
\end{align*}
We continue \eqref{eq: LHS is clean} as:
\begin{equation*}
    \begin{aligned}
        (1-\delta) \mathbb{E}[\ell_0(\widehat{f})] & \leq (1+2 \delta)(1+\delta)(\mathbb{E}[\ell_0(f)] + 4 \eta^2+4 \eta\sqrt{\mathbb{E}[\ell_0(f)]})+(12+8\delta^{-1}) \eta^2+R_{n, \delta, \eta}\\
        & \leq (1+\delta)^2(1+2\delta)\mathbb{E}\left[\ell_0(f)\right] + (4(1+2\delta)(1+\delta)(1+\delta^{-1}) + 12 + 8\delta^{-1})\eta^2 + R_{n, \delta, \eta}.
    \end{aligned}
\end{equation*}
When $\delta \leq 1/2$, we bound $(1-\delta)^{-1}$ by $(1+2\delta)$:
\begin{equation*}
\begin{aligned}
\mathbb{E}[\ell_0(\widehat{f})] & \leq (1+2\delta)^4 \mathbb{E}[\ell_0(f)] + C\delta^{-1}\eta^2 + R_{n, \delta, \eta}
\end{aligned}
\end{equation*}
where $C$ is a universal constant. Since this inequity holds for any $f \in \mathcal{F}$, we obtain 
\begin{equation*}
\begin{aligned}
\mathbb{E}[\ell_0(\widehat{f})] & \leq (1+2\delta)^4 \inf_{f \in \mathcal{F}}\mathbb{E}[\ell_0(f)] + C\delta^{-1}\eta^2 + R_{n, \delta, \eta}.
\end{aligned}
\end{equation*}
Finally, recalling the definition of $R_{n, \delta, \eta}$ in \eqref{eq:R-definition} and \Cref{lemma:bound-on-uncentered-empirical-process} under \ref{as:q-moment}, we have 
\begin{align*}
    \mathbb{P}\left(|R_{n, \delta, \eta}| \ge C_\epsilon \log (|\mathcal{F_\eta}|)\left(n^{-1/2+1/q}F + n^{-1/2}F^2/\delta\right) \right) \le \epsilon
\end{align*}
by Markov's inequality. Let $\eta_*$ be the element that attains the infimum defined as 
\begin{align*}
    \eta_* = \argmin_{\eta > 0 }\left\{\log (|\mathcal{F_\eta}|)\left(n^{-1/2+1/q}F + n^{-1/2}F^2/\delta\right) + \eta^2\right\}.
\end{align*}
We thus conclude with probability at least $1-\epsilon$,
\begin{align*}
    \mathbb{E}[\ell_0(\widehat{f})] & \leq (1+2\delta)^4 \inf_{f \in \mathcal{F}}\mathbb{E}[\ell_0(f)] + C\delta^{-1}\eta_*^2 + C_\epsilon \log (|\mathcal{F_{\eta_*}}|)\left(n^{-1/2+1/q}F + n^{-1/2}F^2/\delta\right).
\end{align*}
The result under \ref{as:beta-weibull} can be derived analogously by instead applying \Cref{lemma:bound-on-uncentered-empirical-process} under \ref{as:beta-weibull}. 
\end{proof}
\begin{remark}
    The leading constant $(1+2\delta)^4$ can be improved to $1$, but requires a separate result tailored for known $L_0$. This is possible when the class f functions $\mathcal{F}$ is convex, corresponding to the result in the literature known as a sharp oracle inequality \cite{bellec2018sharp}. The current result includes $(1+2\delta)^4$ as we perform model selection over a possibly non-convex class of estimators. See also the difference between Theorems 5.1 and 5.2 of \cite{koltchinskii2011oracle}.
\end{remark}
We now appeal to the following bound on the supremum of the expectation of the uncentered empirical process over a finite set of functions. 
\begin{lemma}\label{lemma:bound-on-uncentered-empirical-process}
 Suppose that the observations $\{(X_i, Y_i)\}_{i=1}^n$ are generated from the regression model $Y=f_0(X) + \xi$ where $\E[\xi_i|X_i]=0$ and $\E[\xi_i^2 |X_i]\le \sigma^2$ for all $1\le i \le n$ and additionally either \ref{as:q-moment} or \ref{as:beta-weibull} holds. Let $\ell_0((x,y); f)$ be the excess squared loss such that $\ell_0((x,y); f):=(y-f(x))^2 - (y-f_0(x))^2$. Then, for a finite class of functions $\mathcal{F}$, uniformly bounded by $F$, and for any $\delta \in (0,1)$, there exists a constant $C$ only depending on $\|f_0\|_\infty$, $q$, $C_q$ and $\sigma^2$ such that
 \begin{align}
     \E\left[\max_{f \in \mathcal{F}}\left(\mathbb{G}_n -\delta n^{1/2}P\right)\ell_0(\cdot; f)\right]&\le C\log (|\mathcal{F}|)\left(n^{-1/2+1/q}F + n^{-1/2}F^2/\delta\right) 
 \end{align}
 under \ref{as:q-moment}. Similarly, for any $\delta \in (0,1)$, there exists a constant $C$ only depending on $\|f_0\|_\infty$, $\beta$, $C_\beta$ and $\sigma^2$ such that
 \begin{align}
     \E\left[\max_{f \in \mathcal{F}}\left(\mathbb{G}_n -\delta n^{1/2}P\right)\ell_0(\cdot; f)\right]&\le Cn^{-1/2}\log (|\mathcal{F}|) \left(F(\log n)^{1/\beta} + F^2/\delta\right) 
 \end{align}
  under \ref{as:beta-weibull}. The same upper bounds hold for $\E\left[\max_{f \in \mathcal{F}}\left(\mathbb{G}_n +\delta n^{1/2}P\right)\{-\ell_0(\cdot; f)\}\right]$.
\end{lemma}
\begin{proof}[Proof of Lemma~\ref{lemma:bound-on-uncentered-empirical-process}]
First, we observe that 
	\begin{align}
		\ell_0((X_i,Y_i); f) &= (Y_i-f(X_i))^2 - (Y_i-f_0(X_i))^2\nonumber\\
		&= (Y_i-f_0(X_i)+f_0(X_i)-f(X_i))^2 - (Y_i-f_0(X_i))^2\nonumber\\
		&=2\xi_i(f_0(X_i)-f(X_i))+(f_0(X_i)-f(X_i))^2. \nonumber
	\end{align}

When $\mathcal{F}=\{f\}$, or a singleton set, it immediately follows that 
\begin{align*}
    \E\left[\max_{f \in \mathcal{F}}\left(\mathbb{G}_n -\delta n^{1/2}P\right)\ell_0(\cdot; f)\right] = \E\left[\left(\mathbb{G}_n -\delta n^{1/2}P\right)\ell_0(\cdot; f)\right] \le 0.
\end{align*}
Thus Lemma~\ref{lemma:bound-on-uncentered-empirical-process} trivially holds since $\log(|\mathcal{F}|)=\log(1)=0$. 

We thus focus on the cases with $|\mathcal{F}|\ge 2$. It follows that 
	\begin{align}
		&\E\left[\max_{f \in \mathcal{F}}\,\left(\mathbb{G}_n -\delta n^{1/2} P\right)\ell_0(\cdot; f)\right]\nonumber\\
		&\qquad = \E\left[\max_{f \in \mathcal{F}}\left(\mathbb{G}_n -\delta n^{1/2} P\right)\{2\xi(f_0(X)-f(X))+(f_0(X)-f(X))^2\}\right]\nonumber\\
		&\qquad = \E\left[\max_{f \in \mathcal{F}}\, n^{-1/2}\sum_{i=1}^n 2\xi_i(f_0(X_i)-f(X_i))+\left(\mathbb{G}_n -\delta n^{1/2} P\right)(f_0(X)-f(X))^2\right]\nonumber
	\end{align}
 where the last step follows from the assumption $\E[\xi| X]=0$ and the random variable $\xi_i(f_0(X_i)-f(X_i))$ is already centered for all $f \in \mathcal{F}$. We now use the following conditional symmetrization argument. Let $\xi' := (\xi'_1, \dots, \xi'_n)$ be a conditionally independent copy of $\xi :=(\xi_1, \dots, \xi_n)$ given $(X_1, \dots, X_n)$, meaning that $\xi'$ and $\xi$ follow the same conditional distribution. We also denote by $\varepsilon := (\varepsilon_1,\ldots, \varepsilon_n)$ an IID Rademacher random variable. We then have
\begin{align}
    &\E\left[\max_{f\in\mathcal{F}}\,n^{-1/2}\sum_{i=1}^n 2\xi_i (f_0(X_i)- f(X_i))+\left(\mathbb{G}_n -\delta n^{1/2} P\right)(f_0(X)-f(X))^2\right] \nonumber\\
    &\qquad = \mathbb{E}_{X_1 \dots X_n}\left[\mathbb{E}_{\xi \mid X_1 \dots X_n}\left[\max_{f\in\mathcal{F}}\,n^{-1/2}\sum_{i=1}^n 2\xi_i (f_0(X_i)-  f(X_i))\right.\right.\\
    &\qquad\qquad\left.\left.-2\mathbb{E}_{\xi' \mid X_1 \dots X_n}[\xi'_i (f_0(X_i)- f(X_i))]\right]\right.\nonumber\\
    &\qquad\qquad\left. + \left(\mathbb{G}_n -\delta n^{1/2} P\right)(f_0(X)-f(X))^2\right]\nonumber\\
     &\qquad = \mathbb{E}_{X_1 \dots X_n}\left[\mathbb{E}_{\xi', \xi \mid X_1 \dots X_n}\left[\max_{f\in\mathcal{F}}\,n^{-1/2}\sum_{i=1}^n 2(\xi_i-\xi'_i) (f_0(X_i)-  f(X_i)) \right]\right.\nonumber\\
    &\qquad\qquad\left. + \left(\mathbb{G}_n -\delta n^{1/2} P\right)(f_0(X)-f(X))^2\right]\nonumber\\
    &\qquad = \mathbb{E}_{X_1 \dots X_n}\left[\mathbb{E}_{\varepsilon, \xi', \xi \mid X_1 \dots X_n}\left[\max_{f\in\mathcal{F}}\,n^{-1/2}\sum_{i=1}^n 2\varepsilon_i|\xi_i-\xi'_i| (f_0(X_i)-  f(X_i))\right]\right.\nonumber\\
    &\qquad\qquad\left. + \left(\mathbb{G}_n -\delta n^{1/2} P\right)(f_0(X)-f(X))^2\right]\nonumber\nonumber\\
    &\qquad \le  \E\left[\max_{f\in\mathcal{F}}\,n^{-1/2}\sum_{i=1}^n 4\varepsilon_i|\xi_i| (f_0(X_i)-  f(X_i))+\left(\mathbb{G}_n -\delta n^{1/2} P\right)(f_0(X)-f(X))^2\right].\nonumber
\end{align}
Using this result, we can split the analysis into two parts after truncating the processes by $B$:
\begin{align}
	&\E\left[\max_{f \in \mathcal{F}}\left(\mathbb{G}_n -\delta n^{1/2} P\right)\ell(\cdot; f)\right]\nonumber\\
&\qquad \le \E\left[\max_{f\in\mathcal{F}}\,n^{-1/2}\sum_{i=1}^n 4\varepsilon_i|\xi_i|1\{|\xi_i| \le B\} (f_0(X_i)-  f(X_i))\right.\nonumber\\
&\left.\qquad\qquad +\left(\mathbb{G}_n -\delta n^{1/2} P\right)(f_0(X)-f(X))^2\right]\nonumber\\
&\qquad \qquad + \E\left[\max_{f\in\mathcal{F}}\,n^{-1/2}\sum_{i=1}^n 4\varepsilon_i|\xi_i|1\{|\xi_i| > B\} (f_0(X_i)-  f(X_i))\right]\label{term2}
\end{align}
where $B$ is a positive real number that may change with $n$. We can use the standard technique to provide the bounds on the first two terms as they are (uncentered) empirical processes for bounded random variables. Toward this task, we first introduce the following Bernstein numbers. 
\begin{definition}[Bernstein numbers]
	Given a measurable function $f: \mathcal{X} \rightarrow \mathbb{R}$, call $(M(f), v(f))$ a pair of Bernstein numbers of $f$ if
	\begin{align}\nonumber
		M(f)^2 P\left(e^{|f| / M(f)}-1-\frac{|f|}{M(f)}\right) \leq \frac{1}{2} v(f).
	\end{align}
\end{definition}
\cite[Section 8.1]{vaart2006oracle} provides the following useful properties of Bernstein numbers, which we present in the following corollary:

\begin{corollary}\label{cor:bernstein-properties} The following statements regarding Bernstein numbers hold:
\begin{enumerate}
\item[(i)]If $f$ is uniformly bounded, then $\left(\|f\|_{\infty}, 1.5 P f^2\right)$ is a pair of Bernstein numbers.
\item[(ii)] If $|f| \leq g$, then a Bernstein pair for $g$ is also a Bernstein pair for $f$.
\item[(iii)]If $(M(f), v(f))$ and $(M(g), v(g))$ are Bernstein pairs for $f$ and $g$, then $2(M(f) \vee$ $M(g), v(f)+v(g))$ is a Bernstein pair for $f+g$.
\item[(iv)]If $(M(f), v(f))$ is a Bernstein pair for $f$ and $c>0$, then $\left(c M(f), c^2 v(f)\right)$ is a Bernstein pair for $c f$. \end{enumerate}
\end{corollary} 
We then use the following bounds on the uncentered empirical process in terms of Bernstein numbers:
\begin{lemma}[Lemma 2.2 of \cite{vaart2006oracle}]\label{lemma:sup-Gn-bernstein}
	Let $\mathbb{G}_n$ be the empirical process of an IID sample of size $n$ from the distribution $P$ and assume that $P f \geq 0$ for every $f \in \mathcal{F}$. Then, for any Bernstein pairs $(M(f), v(f))$ and for any $\delta>0$ and $1 \leq p \leq 2$,
	\begin{align}
		&\E\left[\max_{f \in \mathcal{F}}\, (\mathbb{G}_n-\delta n^{1/2} P)f\right] \le \frac{8}{n^{1 / p-1 / 2}} \log (1+|\mathcal{F}|)\max_{f \in \mathcal{F}}\left[\frac{M(f)}{n^{1-1 / p}}+\left(\frac{v(f)}{(\delta P f)^{2-p}}\right)^{1 / p}\right]\nonumber
	\end{align}
	where $|\mathcal{F}|$ denotes the cardinality of the set $\mathcal{F}$. The same upper bound holds for \[\E\left[\max_{f \in \mathcal{F}}(\mathbb{G}_n+\delta n^{1/2} P)(-f)\right].\]
\end{lemma}

We now apply Lemma~\ref{lemma:sup-Gn-bernstein} to the class of functions 
\[\mathcal{H} := \{(x, \xi, \varepsilon) \mapsto 4\varepsilon|\xi|I\{|\xi|\le B\}(f_0(x)-f(x))+ (f_0(x)-f(x))^2 : f \in \mathcal{F}\}.\nonumber\]
First, we observe that for each $\eta \in \mathcal{H}$, we have $P \eta = \|f_0-f\|_{L_2(P)}^2 \ge 0$ from earlier symmetrization. We now derive a Bernstein pair. For each $\eta \in \mathcal{H}$, we split functions into two parts:
\begin{align}
	\underbrace{4\varepsilon|\xi|I\{|\xi|\le B\}(f_0(x)-f(x))}_{:=\eta_1}+ \underbrace{(f_0(x)-f(x))^2}_{:=\eta_2}\nonumber
\end{align}
and use (iii) of Corollary~\ref{cor:bernstein-properties}. Since $\eta_1$ is uniformly bounded, $M(\eta_1) = 4B\|f_0-f\|_\infty$ and the variance is given by 
\begin{align}
	&\E\left[\left\{4\varepsilon|\xi|I\{|\xi|\le B\}(f_0(x)-f(x))\right\}^2\right]\nonumber \\
 &\qquad = \E\left[\E\left[\left\{4\varepsilon|\xi|I\{|\xi|\le B\}\right\}^2\mid X\right](f_0(x)-f(x))^2\right]\nonumber\\
	&\qquad \le 16 \sigma^2\|f_0-f\|^2_{L_2(P)}.\nonumber
\end{align}
Similarly for $\eta_2$, we have 
\begin{align}
	M(\eta_2) = \|(f_0-f)^2\|_\infty \le (\|f_0\|_\infty + F)\|f_0 - f\|_\infty\nonumber
\end{align}
and 
\begin{align}
	\E\left[(f_0(x)-f(x))^4\right] \le (\|f_0\|_\infty + F)^2\|f_0-f\|^2_{L_2(P)}.\nonumber
	\end{align}
By (iii) of Corollary~\ref{cor:bernstein-properties}, we have
	\begin{align}
		&\left\{M(\eta), v(\eta)\right\}\nonumber\\
  &\qquad = \left\{(8B \vee 2\|f_0\|_\infty + 2F)\|f_0-f\|_\infty, (16\sigma^2+2\|f_0\|_\infty^2 + 2F^2)\|f_0-f\|^2_{L_2(P)}\right\}.\nonumber
	\end{align} 
By Lemma~\ref{lemma:sup-Gn-bernstein} with $p=1$, for $\delta > 0$, 
\begin{align}
		&\E\left[\max_{f \in \mathcal{F}}\, (\mathbb{G}_n-\delta n^{1/2} P)(f_0(X)-f(X))^2\right] \nonumber\\
		&\qquad \le 8n^{-1/2}\log (1+|\mathcal{F}|)\left[(8B \vee 2\|f_0\|_\infty + 2F)\|f_0-f\|_\infty+\frac{(16\sigma^2+2\|f_0\|_\infty^2 + 2F^2)}{\delta}\right].\nonumber
	\end{align}
	
Next, we turn to \eqref{term2}. Under the finite $q$th moment of $\xi$, we have 
\begin{align*}
    &\E\left[\max_{f \in \mathcal{F}}\, n^{-1/2}\sum_{i=1}^n 4\varepsilon_i|\xi_i| I(|\xi_i|> B)(f_0(X_i) - f(X_i))\right]\\
    &\qquad \le \E\left[\max_{f \in \mathcal{F}}\,n^{-1/2}\sum_{i=1}^n |\xi_i| |\xi_i|^{q-1}/B^{q-1}\|f_0-f\|_\infty\right]\\
    & \qquad \le n^{-1/2}\max_{f \in \mathcal{F}}\,\frac{\|f_0-f\|_\infty}{B^{q-1}}\E\left[\sum_{i=1}^n |\xi_i|^{q}\right].
\end{align*}
Putting together, we have
\begin{align}
	&\E\left[\max_{f \in \mathcal{F}}\left(\mathbb{G}_n -\delta n^{1/2}P\right)\ell(\cdot; f)\right]\nonumber\\
&\qquad \le 8n^{-1/2}\log (1+|\mathcal{F}|)\bigg[\{8B \vee 2(\|f_0\|_\infty + F)\}(\|f_0\|_\infty+F) +\frac{(16\sigma^2+2\|f_0\|_\infty^2 + 2F^2)}{\delta}\bigg]\nonumber\\
&\qquad \qquad + n^{-1/2}\frac{(\|f_0\|_\infty+F)}{B^{q-1}}\E\left[\sum_{i=1}^n |\xi_i|^{q}\right].\nonumber
\end{align}
As the choice of $B$ is arbitrary, we optimize and obtain that 
\begin{align}
	B = \left(\frac{(q-1)\E\left[\sum_{i=1}^n |\xi_i|^{q}\right]}{64\log(1+|\mathcal{F}|)}\right)^{1/q}\nonumber.
\end{align}
Plugging this in, we have 
\begin{align}
	&\E\left[\max_{f \in \mathcal{F}}\left(\mathbb{G}_n -\delta n^{1/2}P\right)\ell_0(\cdot; f)\right]\nonumber\\
  &\qquad\le 2n^{-1/2}(\|f_0\|_\infty + F)\left((q-1)\E\left[\sum_{i=1}^n |\xi_i|^{q}\right]\right)^{1/q}\left(64\log (1+|\mathcal{F}|)\right)^{1-1/q}\nonumber \\
&\qquad \qquad+ 8n^{-1/2}\log (1+|\mathcal{F}|)\left[2(\|f_0\|_\infty + F)^2+\frac{(16\sigma^2+2\|f_0\|_\infty^2 + 2F^2)}{\delta}\right].\nonumber
\end{align}
Finally by \ref{as:q-moment}, we have for any $f\in\mathcal{F}$ and $\delta \in (0,1)$,
\begin{align}
	&\E\left[\max_{f \in \mathcal{F}}\left(\mathbb{G}_n -\delta n^{1/2}P\right)\ell_0(\cdot; f)\right]\nonumber\\
  &\qquad\le 2C_qn^{-1/2}(\|f_0\|_\infty + F)\left((q-1)n\right)^{1/q}\left(64\log (1+|\mathcal{F}|)\right)^{1-1/q}\nonumber \\
&\qquad \qquad+ 8n^{-1/2}\log (1+|\mathcal{F}|)\left[2(\|f_0\|_\infty + F)^2+\frac{(16\sigma^2+2\|f_0\|_\infty^2 + 2F^2)}{\delta}\right]\nonumber\\
  &\qquad\le 128e C_q  n^{-1/2 + 1/q}(\|f_0\|_\infty + F)\log (1+|\mathcal{F}|)\nonumber \\
&\qquad \qquad+ 8n^{-1/2}\log (1+|\mathcal{F}|)\left(\frac{16\sigma^2+6\|f_0\|_\infty^2 + 6F^2}{\delta}\right).\nonumber
\end{align}
since $(64 \log (1+|\mathcal{F}|))^{1/q} \ge 1$ for $|\mathcal{F}| \ge 2$ and $q \ge 2$. Thus we conclude that there exists a constant $C$ depending only on $\|f_0\|_\infty$, $q$, $C_q$ and $\sigma^2$ such that 
\begin{align}
	\E\left[\max_{f \in \mathcal{F}}\left(\mathbb{G}_n -\delta n^{1/2}P\right)\ell_0(\cdot; f)\right]&\le C\log (1+|\mathcal{F}|)\left(n^{-1/2+1/q}F + n^{-1/2}F^2/\delta\right). \nonumber \end{align}
When $|\mathcal{F}| \ge 2$, it follows that $\log (1+|\mathcal{F}|) \le 2\log (|\mathcal{F}|)$, which concludes the claim.
 
Next, under \ref{as:beta-weibull}, we have 
\begin{align*}
	&\E\left[\max_{f \in \mathcal{F}}\left(\mathbb{G}_n -\delta n^{1/2}P\right)\ell_0(\cdot; f)\right] \nonumber\\
 &\qquad \le 2C_\beta n^{-1/2}(\|f_0\|_\infty + F)\left((q-1)nq^{q/\beta}\right)^{1/q}\left(64\log (1+|\mathcal{F}|)\right)^{1-1/q}\nonumber \\
    &\qquad \qquad + 8n^{-1/2}\log (1+|\mathcal{F}|)\left[2(\|f_0\|_\infty + F)^2+\frac{(16\sigma^2+2\|f_0\|_\infty^2 + 2F^2)}{\delta}\right]\\
    &\qquad \le 128 eC_\beta n^{-1/2}(\|f_0\|_\infty + F)n^{1/q}q^{1/\beta}\log (1+|\mathcal{F}|)\nonumber \\
    &\qquad \qquad + 8n^{-1/2}\log (1+|\mathcal{F}|)\left(\frac{16\sigma^2+6\|f_0\|_\infty^2 + 6F^2}{\delta}\right)\\
    &\qquad \le Cn^{-1/2}\log (1+|\mathcal{F}|) \left(Fn^{1/q}q^{1/\beta} + F^2/\delta\right)\nonumber 
\end{align*}
for some constant $C$ depending only on $\|f_0\|_\infty$, $C_\beta$ and $\sigma^2$. The only term involving $q$ is $n^{1/q}q^{1/\beta}$. By choosing the optimal choice of $q_*=\beta \log n$, we get
\begin{align}
	n^{1/q_*}q_*^{1/\beta} = n^{1/(\beta \log n)}(\beta \log n)^{1/\beta} = e^{1/\beta} \beta^{1/\beta} (\log n)^{1/\beta} \nonumber
\end{align} 
We thus conclude 
\begin{align}
	\E\left[\max_{f \in \mathcal{F}}\left(\mathbb{G}_n -\delta n^{1/2}P\right)\ell_0(\cdot; f)\right]&\le Cn^{-1/2}\log (1+|\mathcal{F}|) \left(F(\log n)^{1/\beta} + F^2/\delta\right). \nonumber \end{align}
As in the case with \ref{as:q-moment}, we conclude by observing that $\log (1+|\mathcal{F}|) \le 2\log (|\mathcal{F}|)$ for $|\mathcal{F}| \ge 2$. In view of Lemma~\ref{lemma:sup-Gn-bernstein}, the bound for $\E\left[\max_{f \in \mathcal{F}}\left(\mathbb{G}_n +\delta n^{1/2}P\right)\{-\ell_0(\cdot; f)\}\right]$ is identical.
\end{proof}
\subsection{Oracle Inequality --- \texttt{ROBUST}}\label{supp:oracle-robust}
\begin{definition}[Median of Means]\label{def:mom}
    Let $W_1, \ldots, W_n$ denote $n$ random variables. We split the observations into $B$ disjoint batches of the same size $(n/B)$ (assuming $n/B$ is an integer). Specifically, we select the partition uniformly at random over all the equipartitions. 
    
    Let $\overline{\mu}_1, \ldots, \overline{\mu}_B$ be the sample means computed on each of them. The median of means operator is defined as 
    \begin{equation*}
        \mathtt{MOM}_{B}(\{W_i\}_{i=1}^n) := \mathrm{median}(\overline{\mu}_1, \ldots, \overline{\mu}_B).
    \end{equation*}
\end{definition}
Below, we adopt the notation $\|\cdot\|= \|\cdot\|_{L_2(P_X)}$.
\subsubsection{Proof of \Cref{thm:oracle-cv-mom}}
As the method requires a discretized function space $\mathcal{F}$, we index each element as $f_1, f_2, \ldots, f_J$ for $J \ge 2$. Let $\mathcal{J} = \{1,2,\ldots, J\}$. For any $i, j \in \mathcal{J}$, we define 
\begin{align*}
    \widehat \theta_{i,j} = \mathtt{MOM}_{B}(\ell(\cdot; f_j) -  \ell(\cdot; f_i)) \text{ and } \theta_{i,j} = P \ell(\cdot; f_j) - P\ell(\cdot; f_i).
\end{align*}
With this notation, we have 
\begin{align}
    \widehat f^{\mathtt{MOM}}_n = f_{\widehat J} \text{ where } \widehat J = \argmin_{j \in \mathcal{J}}\, \max_{i \in \mathcal{J}} \, \widehat \theta_{i,j}.
\end{align}
We observe that 
\begin{align*}
&(Y_k - f_j(X_k))^2 - (Y_k - f_i(X_k))^2\\
&\qquad = (Y_k - f_0(X_k))^2 + 2(Y_k - f_0(X_k))(f_0(X_k) - f_j(X_k)) + (f_0(X_k) - f_j(X_k))^2\\
&\qquad\qquad - (Y_k - f_0(X_k))^2 - 2(Y_k - f_0(X_k))(f_0(X_k) - f_i(X_k)) - (f_0(X_k) - f_i(X_k))^2\\
&\qquad= 2(Y_k - f_0(X_k))(f_i(X_k) - f_j(X_k)) + (f_0(X_k) - f_j(X_k))^2 - (f_0(X_k) - f_i(X_k))^2.
\end{align*}
Since $f_0(x) = \E[Y|X=x]$, we have 
\begin{align*}
    \theta_{i,j} = \E[(f_0(X) - f_j(X))^2 - (f_0(X) - f_i(X))^2] = \|f_0-f_j\|^2 - \|f_0 - f_i\|^2.
\end{align*}
We also define the variance of difference of losses as
\begin{align*}
    \mathcal{S}^2_{i,j} = \mathrm{Var}[\ell(\cdot; f_j) - \ell(\cdot; f_i)].
\end{align*}
Using the EV-formula, that is $\mbox{Var}(V) = \mbox{Var}(\mathbb{E}[V|U]) + \mathbb{E}[\mbox{Var}(V|U)]$, we obtain
\begin{align*}
\mbox{Var}(\ell(\cdot; f_j) - \ell(\cdot; f_i)) &= \mbox{Var}((f_j(X) - f_i(X))(2Y_k - f_i(X) - f_j(X)))\\
&= \mbox{Var}((f_j(X_k) - f_i(X_k))(2f_0(X_k) - f_i(X_k) - f_j(X_k))) \\
&\qquad + 4\mathbb{E}[(f_j(X_k) - f_i(X_k))^2\mbox{Var}[\xi|X_k]]\\
&\le \|f_j - f_i\|^2\|2f_0 - f_i - f_j\|_{\infty}^2 + \|f_j - f_i\|^2\sigma^2\\
&= \mathfrak{C}^2\|f_j - f_i\|^2,
\end{align*}
where
\[
\mathfrak{C}^2 := \sigma^2 + \|2f_0 - f_i - f_j\|_{\infty}^2 \le \sigma^2  + 2(\|f_0\|_\infty + F)^2.
\]
For any $k \in \mathcal{J}$, it follows from the basic inequality that
\begin{align*}
    \max_{i \in \mathcal{J}} \,\widehat \theta_{i,\widehat J} \le \max_{i \in \mathcal{J}} \,\widehat \theta_{i, k}.
\end{align*}
This implies that 
\begin{equation}\label{eq:algebraic-manipulation}
    \begin{aligned}
        \max_{i \in \mathcal{J}} \,\left(\widehat \theta_{i,\widehat J} + \theta_{i,\widehat J}- \theta_{i,\widehat J}\right) &\le \max_{i \in \mathcal{J}} \, \left(\widehat \theta_{i, k} + \theta_{i,k}- \theta_{i,k} \right)\\
    \Rightarrow \max_{i \in \mathcal{J}} \,\theta_{i,\widehat J}-\mathcal{S}_{i,\widehat J}\left|\frac{\widehat \theta_{i,\widehat J}  - \theta_{i,\widehat J}}{\mathcal{S}_{i,\widehat J}}\right| &\le \max_{i \in \mathcal{J}} \,  \theta_{i,k}+\mathcal{S}_{i,k}\left|\frac{\widehat \theta_{i, k}- \theta_{i,k}}{\mathcal{S}_{i,k}}\right|\\
    \Rightarrow \max_{i \in \mathcal{J}} \,\theta_{i,\widehat J}-\mathfrak{C}\|f_i-f_{\widehat J}\|R_{n}& \le \max_{i \in \mathcal{J}} \,  \theta_{i,k}+\mathfrak{C}\|f_i - f_k\|R_{n}
    \end{aligned}
\end{equation}
where 
\begin{align*}
    R_{n} := \max_{1\le i, j\le J}\, \left|\frac{\widehat{\theta}_{ij} - \theta_{ij}}{\mathcal{S}_{ij}}\right|.
\end{align*}
Observe now that for any $1 \le i, j \le J$,
\begin{align*}
    \theta_{i j} - \mathfrak{C}\|f_i - f_j\|R_n~&=~ \| f_j - f_0 \|^2 - \| f_i - f_0 \|^2 - R_n \mathfrak{C} \| f_i - f_j \| \\
    ~&\geq~ \| f_j - f_0 \|^2 - R_n \mathfrak{C} \| f_j - f_0 \| - \| f_i - f_0 \|^2 - R_n \mathfrak{C} \| f_i - f_0 \| \\
    ~&=~ \left( \| f_j - f_0 \| - \frac{R_n \mathfrak{C}}{2} \right)^2 - \left( \| f_i - f_0 \| + \frac{R_n \mathfrak{C}}{2} \right)^2, \text{ and}\\
    \theta_{i j} + \mathfrak{C}\|f_i - f_j\| R_n ~&=~ \| f_j - f_0 \|^2 - \| f_i - f_0 \|^2 + R_n \mathfrak{C} \| f_i - f_j \| \\
    ~&\leq~ \left( \| f_j - f_0 \|^2 + R_n \mathfrak{C} \| f_j - f_0 \| \right) - \left( \| f_i - f_0 \|^2 - R_n \mathfrak{C} \| f_i - f_0 \| \right) \\
    ~&=~ \left( \| f_j - f_0 \| + \frac{R_n \mathfrak{C}}{2} \right)^2 - \left( \| f_i - f_0 \| - \frac{R_n \mathfrak{C}}{2} \right)^2.
\end{align*}
Hence, the last inequality of~\eqref{eq:algebraic-manipulation} yields
\begin{align*}
    &\left( \| f_{\widehat{J}} - f_0 \| - \frac{R_n \mathfrak{C}}{2} \right)^2 - \min_{i\in \mathcal{J}}\left( \| f_i - f_0 \| + \frac{R_n \mathfrak{C}}{2} \right)^2 \\
    &\qquad \le \left( \| f_{k} - f_0 \| + \frac{R_n \mathfrak{C}}{2} \right)^2 - \min_{i\in \mathcal{J}}\left( \| f_i - f_0 \| - \frac{R_n \mathfrak{C}}{2} \right)^2,
\end{align*}
which in turn yields
\begin{align*}
    &\left( \| f_{\widehat{J}} - f_0 \| - \frac{R_n \mathfrak{C}}{2} \right)^2 \\
    &\qquad \le \left( \| f_{k} - f_0 \| + \frac{R_n \mathfrak{C}}{2} \right)^2 + \min_{i\in \mathcal{J}}\left( \| f_i - f_0 \| + \frac{R_n \mathfrak{C}}{2} \right)^2 -\min_{i\in \mathcal{J}}\left( \| f_i - f_0 \| - \frac{R_n \mathfrak{C}}{2} \right)^2 \\
    \Rightarrow &\,\left( \| f_{\widehat{J}} - f_0 \| - \frac{R_n \mathfrak{C}}{2} \right)^2 \le 2\left( \| f_{k} - f_0 \| + \frac{R_n \mathfrak{C}}{2} \right)^2  \\
    \Rightarrow &\, \|f_{\widehat{J}} - f_0\| \le \frac{R_n\mathfrak{C}}{2} + \sqrt{2}\left(\|f_{k} - f_0\| + \frac{R_n\mathfrak{C}}{2}\right).
\end{align*}
Therefore, 
\[
\|f_{\widehat{J}} - f_0\|^2 \le 2(1 + \delta)\|f_{k} - f_0\|^2 + (1+\delta^{-1})(1/2 + 1/\sqrt{2})^2\mathfrak{C}^2R_n^2
\]
by Young's inequality for any $\delta > 0$. Since the inequality hold for any $k \in \mathcal{J}$, 
\[
\|f_{\widehat{J}} - f_0\|^2 \le 2(1 + \delta)\inf_{k\in\mathcal{J}}\|f_{k} - f_0\|^2 + (1+\delta^{-1})(1/2 + 1/\sqrt{2})^2\mathfrak{C}^2R_n^2.
\]

The claim is concluded by establishing the upper bound to $\mathbb{E}[R_n^2]$ and by Markov's inequality. \Cref{lemma:Rn-bound} below states that with $d = J^2$ and $n \ge B = 4 \lceil \ln J \rceil$,
\begin{align*}
    \mathbb{E}[R_n^2] \le 4(3.5)^2 e \frac{\lceil \ln J \rceil}{n}.
\end{align*}
This concludes the result.
\begin{lemma}\label{lemma:Rn-bound}
    Suppose $Z_1, \ldots, Z_n$ are IID random vectors in $\mathbb{R}^d$ with mean $\mu\in\mathbb{R}^d$. Define $\widehat{\mu}_B\in\mathbb{R}^d$ as 
    \[
    e_j^{\top}\widehat{\mu}^{\mathtt{MOM}}_B := \mathrm{Median}(e_j^{\top}\overline{\mu}_1, \ldots, e_j^{\top}\overline{\mu}_B),
    \]
    where $\overline{\mu}_1, \ldots, \overline{\mu}_B$ are the sample means computed based on $B$ disjoint batches of the same size $n/B$ from $Z_1, \ldots, Z_n$.
    If $\sigma_j^2 = \mathrm{Var}(e_j^{\top}Z_1)$, then for any $B \ge 4$,
    \[
    \mathbb{E}\left[\max_{1\le j\le d}\,\left|\frac{e_j^{\top}(\widehat{\mu}^{\mathtt{MOM}}_B - \mu)}{\sigma_j}\right|^2\right] \le (3.5)^2 d^{2/(B-2)}\frac{B}{n}.
    \]
    In particular, if $n \ge 2\lceil\ln(d)\rceil$, then  
    \[
    \mathbb{E}\left[\max_{1\le j\le d}\,\left|\frac{e_j^{\top}(\widehat{\mu}^{\mathtt{MOM}}_{2\lceil\ln(d)\rceil} - \mu)}{\sigma_j}\right|^2\right] \le 2(3.5)^2 e\frac{\lceil\ln(d)\rceil}{n}.
    \]
\end{lemma}
\begin{proof}[Proof of \Cref{lemma:Rn-bound}]
    By Jensen's inequality, with assumption that $B\ge 4$, 
    \begin{align*}
     \mathbb{E}\left[\max_{1\le j\le d}\,\left|\frac{e_j^{\top}(\widehat{\mu}^{\mathtt{MOM}}_B - \mu)}{\sigma_j}\right|^2\right] &\le \left(\mathbb{E}\left[\max_{1\le j\le d}\,\left|\frac{e_j^{\top}(\widehat{\mu}^{\mathtt{MOM}}_B - \mu)}{\sigma_j}\right|^{B-2}\right] \right)^{2/(B-2)} \\
     &\le \left(\mathbb{E}\left[\sum_{j=1}^d\,\left|\frac{e_j^{\top}(\widehat{\mu}^{\mathtt{MOM}}_B - \mu)}{\sigma_j}\right|^{B-2}\right] \right)^{2/(B-2)}\\
     &\le d^{2/(B-2)}\max_{1\le j \le d}\left(\mathbb{E}\left[\left|\frac{e_j^{\top}(\widehat{\mu}^{\mathtt{MOM}}_B - \mu)}{\sigma_j}\right|^{B-2}\right] \right)^{2/(B-2)} \\
     &\le (3.5)^2 d^{2/(B-2)}\frac{B}{n}
    \end{align*}
    where the last inequality follows from \Cref{thm:median-of-means-moment-version}. In particular, taking $B = 2\lceil\ln(d)\rceil$, we get 
    $$
    d^{1/(B - 2)} \le \exp(\ln(d)/(2(\lceil\ln(d)\rceil - 1))) \le e,\quad\mbox{for}\quad d\ge3,
    $$
    and consequently, 
    \[
    \mathbb{E}\left[\max_{1\le j\le d}\,\left|\frac{e_j^{\top}(\widehat{\mu}^{\mathtt{MOM}}_{2\lceil\ln(d)\rceil} - \mu)}{\sigma_j}\right|^2\right] \le (3.5)^2e\frac{\lceil\ln(d)\rceil}{n}.
    \]
\end{proof}
\begin{lemma}\label{thm:median-of-means-moment-version}
Suppose $W_1, \ldots, W_n$ are IID random variables with mean $\mu$. Consider the estimator
\[
\widehat{\mu}_B^{\mathtt{MOM}} = \mathrm{Median}(\overline{\mu}_1, \ldots, \overline{\mu}_B),
\]
where $\overline{\mu}_1, \ldots, \overline{\mu}_B$ are the sample means computed based on $B$ disjoint batches of the same size $(n/B)$ from $W_1, \ldots, W_n$. 
Then, for any number $B \ge 4$,
\[
\left(\mathbb{E}|\widehat{\mu}_B^{\mathtt{MOM}} - \mu|^{B - 2}\right)^{1/(B-2)} ~\le~ 3.5(\mathbb{E}[|\overline{\mu}_1 - \mu|^2])^{1/2} ~\le~ 3.5\sqrt{\frac{\mathrm{Var}(W_1)B}{n}}. 
\]
\end{lemma}
\begin{proof}[Proof of \Cref{thm:median-of-means-moment-version}]
Let $\mathrm{Be}(\cdot)$ be the Beta function.  Then Eq (2) of~\cite{gribkova1995bounds} (with $i = B/2, \rho = B/2 - 1,$ $\delta = 2$, and $k = \rho\delta$) yields
    \begin{equation}\label{eq:moment-order-statistic}
    \begin{split}
    \mathbb{E}[|\widehat{\mu}_B^{\mathtt{MOM}} - \mu|^k] &\le (\mathbb{E}[|\overline{\mu}_1 - \mu|^2])^{\rho}\left(\frac{\text{Be}(1, B/2 + 1) + \text{Be}(B/2, 2)}{\text{Be}(B/2, B/2 + 1)}\right)\\
    &\le (\mathbb{E}[|\overline{\mu}_1 - \mu|^2])^{\rho}\left(\frac{B!}{(B/2 + 1)!(B/2 - 1)!} + \frac{B!}{(B/2)!(B/2 + 1)!}\right)\\
    &\le 2(\mathbb{E}[|\overline{\mu}_1 - \mu|^2])^{\rho}\binom{B}{B/2 + 1}.
    \end{split}
    \end{equation}
    From the main Theorem of~\cite{agievich2022upper}, we get
    \[
    \binom{B}{B/2 + 1} \le \frac{2^B}{\sqrt{\pi B/2}}\exp\left(-\frac{2}{B} + \frac{23}{18B}\right) = \frac{2^B}{\sqrt{\pi B/2}}\exp\left(-\frac{13}{18B}\right).
    \]
    Putting together, we obtain 
    \begin{align*}
        \mathbb{E}[|\widehat{\mu}_B^{\mathtt{MOM}} - \mu|^k]  \le (\mathbb{E}[|\overline{\mu}_1 - \mu|^2])^{\rho}\frac{2^{B+1}}{\sqrt{\pi B/2}}\exp\left(-\frac{13}{18B}\right).
    \end{align*}
    Equivalently, for all $B \ge 4$,
    \begin{align*}
         \left(\mathbb{E}|\widehat{\mu}_B^{\mathtt{MOM}} - \mu|^{B - 2}\right)^{1/(B - 2)} &\le (\mathbb{E}[|\overline{\mu}_1 - \mu|^2])^{1/2}\left(\frac{2^{B+1}}{\sqrt{\pi B/2}}\right)^{1/(B - 2)} \left\{\exp\left(-\frac{13}{18B}\right)\right\}^{1/(B-2)}\\
         &\le 3.5(\mathbb{E}[|\overline{\mu}_1 - \mu|^2])^{1/2} = 3.5\sqrt{\frac{\mathrm{Var}(W_1)B}{n}}.
    \end{align*}
    This concludes the result.
\end{proof}

\newpage
\section{Decomposition and approximation results}\label{appsec:auxiliary-results}

\begin{proposition}\label{lemma: F_nested}For a given $k\in\mathbb{N}$ and $L'\geq L \geq 0$, we have 
\begin{equation*}
    \mathcal{F}(k,L) \subset \mathcal{F}(k,L'). 
\end{equation*}
\end{proposition}
\begin{proof}[Proof of \Cref{lemma: F_nested}]
    Let $f \in \mathcal{F}(1, L)$. Then, for any $L' \ge L$,
    \begin{align*}
        f = g(x) - Lx =  g(x) + (L' - L) x-L'x 
    \end{align*}
    where $g(x) + (L' - L) x$ is also non-decreasing since $L' \ge L$. Therefore, $f \in \mathcal{F}(1, L)$ implies $f \in \mathcal{F}(1, L')$ for $L' \ge L$. For general $k$, the argument is analogous where we repeat the construction over the $(k-1)$th weak derivative. 
\end{proof}

\begin{proof}[Proof of \Cref{prop:k-monotone-decomposition}]
Consider a function $f\in \mathrm{BL}_1(k,L;\Omega)$. By definition, it is $(k-1)$-times differentiable. It follows that $g_L(x):=f(x)+(L / k!) x^k$ is also is also $(k-1)$ times differentiable with the $(k-1)$th derivative $f^{(k-1)}(x) + L x$. 
    
It thus remains to show $g_{L}^{(k-1)}(x)= f^{(k-1)}(x)+ L x$ is non-decreasing, which implies that $g_{L}$ is $k$-monotone. For any $y \ge x$,
    \begin{align*}
        g^{(k-1)}_{L}(y) - g^{(k-1)}_{L}(x) &= \left(f^{(k-1)}(y) + L y\right) - \left(f^{(k-1)}(x) + L x\right) \\
        &\ge -L |y-x| +L(y-x) \ge 0,
    \end{align*}
    which follows by the Lipschitz continuity of $f^{(k-1)}$. So $f\in \mathcal{F}^\dagger(k,L;\Omega)$.
    
    To show the inclusion is strict, we note that it is possible to find a function $f\in \mathcal{F}^\dagger(k, L)$ whose $f^{(k-1)}$ is piecewise constant, non-decreasing, and discontinuous. Such a $f^{(k-1)}$ is not Lipschitz.
\end{proof}

\begin{figure}[!t]
  \centering
  \includegraphics[width=3.2in]{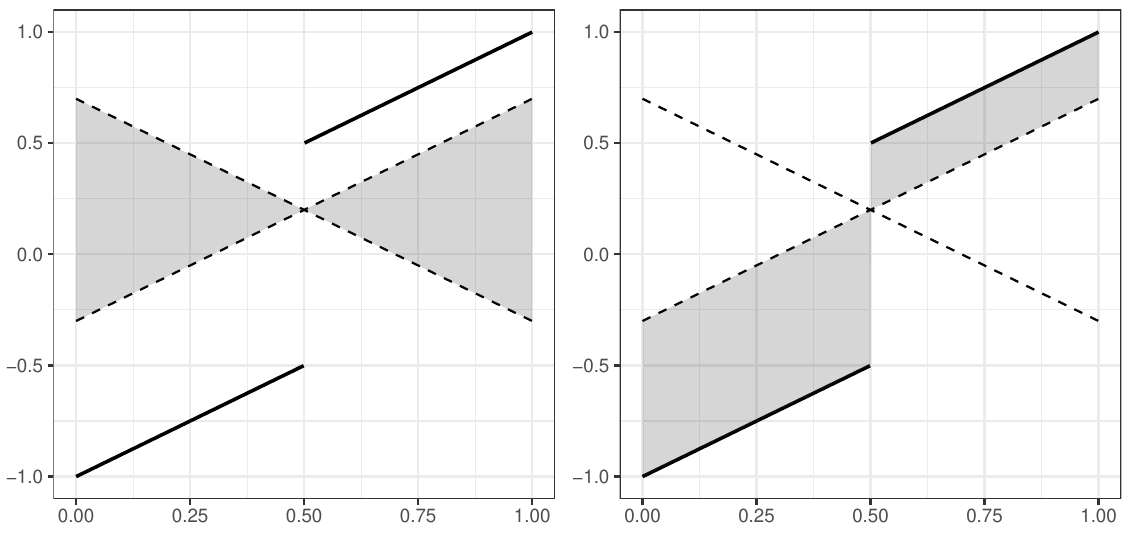}
  \caption{$\mathrm{BL}_1(1, L)$ is not dense in $\mathcal{F}^{\dagger}(1, L)$.}
  \label{fig:Lipschitz-dense}
\end{figure}

\begin{proposition}\label{prop:dense-subset}
$\mathrm{BL}_1(1,L)$ is not dense in  $\mathcal{F}^\dagger(1,L)$ under $L_2$-norm. 
\end{proposition}
\begin{proof}[\Cref{prop:dense-subset}]
We take $L=1$ without loss of generality. We will show that there exist functions in $\mathcal{F}^\dagger(1,1)$ that cannot be arbitrarily approximated by functions in $\mathrm{BL}_1(1, 1)$. Specifically, consider
\begin{align*}
    f^\diamond(x) =  1( x > 1/2) -1 + x.
\end{align*}
Figure~\ref{fig:Lipschitz-dense} shows this function in the solid black line. We have $f^\diamond \in \mathcal{F}^\dagger(1, 1)$ by construction. Let $g$ be any fixed function in $\mathrm{BL}_1(1, 1)$, we denote $c = g(1/2)$. 

By the $1$-Lipschitz continuity of $g$, we know
\begin{equation*}
    g(x) \in [c - |x - 1/2|, c+|x-1/2|].
\end{equation*}
This is demonstrated in the left panel of Figure~\ref{fig:Lipschitz-dense} when $c=0.2$. We claim that we only need to consider $c\in [-1/2, 1/2]$: the reason is given at the end of this proof.

In this case:
\begin{equation*}
\begin{aligned}
   \left\|f^{\diamond}-g\right\|_{L_2([0,1])}^2 & = \int_0^{1/2} (g(x) - f^\diamond(x))^2 dx + \int_{1/2}^1 (f^\diamond(x) - g(x))^2 dx \\
   & \geq \int_0^{1/2} (c + 1/2)^2 dx + \int_{1/2}^1 (1/2 - c)^2 dx \geq 1/4.
\end{aligned}
\end{equation*}

This means we can find a $f^{\diamond} \in \mathcal{F}^\dagger(1,1)$ and a constant $\epsilon > 0$ such that for any function $g\in \mathrm{BL}_1(1,1)$, $\left\|f^{\diamond}-g\right\|_{L_2([0,1])} > \epsilon$. For general $L >0$, consider $f^{\diamond}(x)=L(1(x>1 / 2)-1+x)$ instead.

For any function $g \in\mathrm{BL}_1(1, L)$ such that $g(1/2) > 1/2$, we have for $x\in[0,1/2]$
\begin{equation*}
    g(x) \geq c -1/2 + x > x > f^\diamond(x),
\end{equation*}
so
\begin{equation*}
\left\|f^{\diamond}-g\right\|_{L_2([0,1])}^2 \geq \int_0^{1 / 2}\left(g(x)-f^{\diamond}(x)\right)^2 d x \geq \int_1^{1/2}(x - x +1)^2 = 1/2.
\end{equation*}
The argument for $g(1/2) < -1/2$ is similar.
\end{proof}

As each function in an integer-order $\mathrm{BL}$ space admits an exact decomposition $f(x)=g_L(x)-(L / k!) x^k$ for a $g_{L}\in \mathcal{C}(k)$, we can derive some ``approximate-decompositions'' for functions in general $\mathrm{BL}$ spaces.

\begin{proposition}\label{prop:k-monotone-approximation}
    There exists a universal constant $\mathfrak{C} > 1$ such that for any $\varepsilon,  s, L > 0$, we have
    \[
    \sup_{f\in\mathrm{BL}_1( s, L)}\,\inf_{g\in\mathcal{F}(\lceil s\rceil, \overline{L}_{\varepsilon})}\,\|f - g\|_{L_2([0, 1])} ~\le~ \sup_{f\in\mathrm{BL}_1( s, L)}\, \inf_{g\in\mathrm{BL}_1(\lceil s\rceil, \overline{L}_{\varepsilon})}\,\|f - g\|_{L_2([0, 1])}~\le~ \varepsilon,
    \]
    where $\overline{L}_{\varepsilon} = \mathfrak{C}L^{\lceil s\rceil/ s}\varepsilon^{1 - \lceil s\rceil/ s}$. Here, the domain of the functions in $\mathrm{BL}_1( s, L)$ and $\mathcal{F}(\lceil s\rceil, \overline{L}_{\varepsilon})$ are assumed to be $[0,1]$.
\end{proposition}

\begin{proof}[Proof of \Cref{prop:k-monotone-approximation}]
The result is trivial for $ s\in\mathbb{N}$, because $\overline{L}_{\varepsilon} = \mathfrak{C}L > L$ and $\mathrm{BL}_1( s, L) \subset \mathrm{BL}_1(\lceil s\rceil, \overline{L}_{\varepsilon})$. Now, we consider non-integer $ s > 0$. The first inequality is obvious because $\mathrm{BL}_1(\lceil s\rceil, L) \subsetneq \mathcal{F}(\lceil s\rceil, L)$ by \Cref{prop:k-monotone-decomposition}. The second inequality follows from Lemma~\ref{prop:Holder-Zygmund} presented below.
\end{proof}

For $ s > 0$, define the H{\"o}lder-Zygmund space of smoothness order $ s$ on $[0, 1]$ as
\begin{align*}
\mathcal{Z}_{1}( s, L) &= \left\{f\in L_2([0, 1])\,\bigg| \,\|f\|_{\infty} + \sup_{t > 0}\,t^{- s}\omega_{\lfloor s\rfloor+1}(f, t)_{\infty} \le L\right\},\\
&= \left\{f\in L_2([0, 1])\,\bigg| \,\|f\|_{\infty} + \sup_{h\in[0, 1/2]}\, \sup_{x\in[0, 1-2h]}\,|\Delta_h^{\lfloor s\rfloor + 1}(f, x)| \le L\right\},
\end{align*}
where for any $t > 0$ and integer $r \ge 1$,
\[
\omega_r(f, t)_{\infty} := \sup_{h\in[0, t]}\sup_{x\in[0, 1-rh]}\left|\Delta_h^r(f, x)\right|,\quad\Delta_h^r(f, x) := \sum_{k=0}^r \binom{r}{k}(-1)^{r-k}f(x + kh).
\]
See equation (B.18) and Appendix B.11 of~\cite{johnstone2017gaussian}. Also, see Chapter 2, Section 9 of~\cite{devore1993constructive}. The H{\"o}lder-Zygmund space is a generalization of H{\"o}lder spaces and such a generalization is useful in characterizing these spaces in terms of Besov spaces and wavelet coefficients. Moreover, if $ s$ is a non-integer, then $\mathcal{Z}_1( s, L) = \Sigma_1( s, L)$~\citep[Proposition 4.3.23]{gine2021mathematical}, but for $ s$ integer, $\Sigma_1( s, L)\subsetneq \mathcal{Z}_1( s, L)$ (and this is a strict inclusion); see, for example,~\cite[page 92]{devore1993constructive}. The difference can be understood by examining the definition of $\mathcal{Z}_1(1, L)$.
\begin{align*}
    &\mathcal{Z}_1(1, L) \\
    &\qquad := \left\{f\in L_2([0, 1])\big|\|f\|_{\infty} + \sup_{h\ge 0, x\in[0, 1-2h]}\,\frac{|f(x) - 2f(x + h) + f(x + 2h)|}{h} \le L\right\}.
\end{align*}
It is clear that H{\"o}lder-Zygmund class of order 1 is defined in terms of second-order divided difference rather than the first-order divided difference as in the classical H{\"o}lder spaces. 
From the discussion around equation (4.113) of~\cite{gine2021mathematical}, we get that $\mathcal{Z}_1( s, L) = \mathcal{B}_{\infty, \infty}^{ s}(L; [0, 1])$ 
(i.e., $\mathcal{Z}_1( s, L)$ is a Besov space) and following equation (9.45) of~\cite{johnstone2017gaussian}, we can represent functions in $\mathcal{Z}_1( s, L)$ in terms of the Wavelet bases functions with certain constraints on the coefficients. With this background, we now state an approximation result for general H{\"o}lder-Zygmund spaces.
\begin{lemma}\label{prop:Holder-Zygmund}
For any $\varepsilon, L > 0$ and $ s' \ge  s > 0$, we have
\[
\sup_{f\in\mathcal{Z}_1( s, L)}\,\inf_{g\in\mathcal{Z}_1( s', 2L^{ s'/ s}\varepsilon^{1- s'/ s})}\, \|f - g\|_{L_2([0, 1])} \le \varepsilon.
\]
Moreover, there exists a universal constant $\mathfrak{C} \ge 1$ such that, for any non-integer $ s > 0$,
\[
\sup_{f\in\Sigma_{1}( s, L)}\inf_{g\in\Sigma_1(\lceil s\rceil, \overline{L}_{\varepsilon})}\,\|f - g\|_{L_2([0, 1])} \le \varepsilon,
\]
where $\overline{L} = \mathfrak{C}L^{\lceil s\rceil/ s}\varepsilon^{1-\lceil s\rceil/ s}$.
\end{lemma}
\begin{proof}[Proof of Lemma~\ref{prop:Holder-Zygmund}]
From equation (9.45) of~\cite{johnstone2017gaussian}, we can write any function $f\in\mathcal{Z}_1( s, L)$ as
\[
f(x) = \sum_{k=0}^{2^{K_0} - 1} \gamma_k \varphi_{K_0, k}(x) + \sum_{j\ge K_0}\sum_{k=0}^{2^j - 1} \theta_{j,k} \psi_{j,k}(x),
\]
with coefficients satisfying $\max_{j\ge 1}2^{( s+1/2)j}\max_{k\ge1}|\theta_{j,k}| \le L$. The Zygmund space is characterized by the Wavelet representation along with this constraint on the coefficients as discussed on page 270 of~\cite{johnstone2017gaussian}. For any $\varepsilon > 0$, set $N_{\varepsilon} = \lceil\log_2(L/\varepsilon)/ s\rceil$ and define
\[
f_{\varepsilon}(x) = \sum_{k=0}^{2^{K_0} - 1} \gamma_k\varphi_{K_0, k}(x) + \sum_{j=K_0}^{N_{\varepsilon}}\sum_{k=0}^{2^{j} - 1} \theta_{j,k}\psi_{j,k}.
\]
By the orthogonality of the wavelet basis, we get
\begin{align*}
    &\|f - f_{\varepsilon}\|_{L_2([0, 1])}^2 \\
    &\qquad = \sum_{j \ge N_{\varepsilon} + 1}\sum_{k=0}^{2^{j} - 1} \theta_{j,k}^2  \sum_{j\ge N_{\varepsilon} + 1} \sum_{k=0}^{2^j - 1} L^22^{-(2 s + 1)j} = \sum_{j\ge N_{\varepsilon} + 1} L^22^{-2 s j} \le L^22^{-2 s N_{\varepsilon}} \le \varepsilon^2.
\end{align*}
Moreover, for any $ s' \ge  s$, we have
\[
\max_{1\le j\le N_{\varepsilon}}2^{( s' + 1/2)j}\max_{k\ge1} |\theta_{j,k}| \le L\times\max_{1\le j\le N_{\varepsilon}} 2^{( s' -  s)j} \le 2L(L/\varepsilon)^{ s'/ s - 1} = 2L^{ s'/ s}\varepsilon^{1 -  s'/ s}.
\]
Therefore, $f_{\varepsilon} \in \mathcal{Z}_1( s', 2L^{ s'/ s}\varepsilon^{1- s'/ s})$ and $\|f - f_{\varepsilon}\|_{L_2([0, 1])} \le \varepsilon.$

To prove the second part, recall that for any non-integer $ s > 0$, there exists universal constants $\underline{C}, \overline{C} > 0$ such that  $\Sigma_1( s, \underline{C}L) \subseteq \mathcal{Z}_1( s, L) \subseteq \Sigma_1( s, \overline{C}L)$. Hence, applying the first part with $ s$ and $ s' \equiv  s_m = \lceil s\rceil + 1/\log(m)$, we get
\[
\sup_{f\in\Sigma_1( s, L)}\inf_{g\in\Sigma_1( s_m, L_{m,\varepsilon})}\,\|f - g\|_{L_2([0, 1])} \le \varepsilon,
\]
where $L_{m,\varepsilon} = 2\mathfrak{C}L(\varepsilon/L)^{1 -  s_m/ s}$. Note that this holds true for any $m\ge1$ such that $1/\log(m)$ is not an integer. 
Note that any $g\in\Sigma_1( s_m, L_{m,\varepsilon})$ satisfies
\[
\sup_{x\neq y}\frac{|g^{(\lceil s\rceil)}(x) - g^{\lceil s\rceil}(y)|}{|x - y|^{1/\log(m)}} \le L_{m,\varepsilon}\quad\Longrightarrow\quad \sup_{x\neq y}|g^{\lceil s\rceil}(x) - g^{\lceil s\rceil}(y)| \le L_{m,\varepsilon},
\]
because $x, y\in[0, 1]$. Therefore, $\Sigma_1( s_m, L_{m,\varepsilon}) \subseteq \Sigma_1(\lceil s\rceil, L_{m,\varepsilon}).$ Rewrite $L_{m,\varepsilon}$ as 
\[
L_{m,\varepsilon} = 2\mathfrak{C}L(\varepsilon/L)^{1-\lceil s\rceil/ s}\{(\varepsilon/L)^{-1/ s}\}^{1/\log(m)}.
\]
For $m = (L/\varepsilon)^{1/ s}$, we get $2\mathfrak{C}L(\varepsilon/L)^{1-\lceil s\rceil/ s} \le L_{m,\varepsilon} \le 2e\mathfrak{C}L(\varepsilon/L)^{1-\lceil s\rceil/ s}$. So, choose an $m$ such that $1/\log(m)$ is not an integer and $L_{m,\varepsilon} \le 6\mathfrak{C}L^{\lceil s\rceil/ s}\varepsilon^{1 - \lceil s\rceil/ s}$. Hence, we get
\[
\sup_{f\in\Sigma_1( s, L)}\inf_{g\in\Sigma_1(\lceil s\rceil, \overline{L}_{\varepsilon})}\,\|f - g\|_{L_2([0, 1])} \le \varepsilon.
\]
Renaming $6\mathfrak{C}$ to $\mathfrak{C}$, we get the result.
\end{proof}

\subsection{Results Related to Bounded Variation Spaces}\label{app: bv proof}

We now prove the inclusion of $\mathcal{F}^\dagger(k, L)$ in the $k$th bounded variation class. Toward this aim, we recall several properties of the $k$th convex functions. 
\begin{lemma}[Theorems 83A and 83B of \cite{roberts1993convex}; Theorem A of \cite{kopotun1998approximation}]\label{lemma:k-convex-derive}
    Assume $g: [0,1]\mapsto \mathbb{R}$ is uniformly bounded and $k$-monotone for some $k\ge2$, i.e., $\Delta_h^k(g, x) \ge 0$ for all $x \in [0,1]$. Then $g^{(k-2)}(x)$ exists and is convex.
\end{lemma}
\begin{lemma}[Theorems 11B and 11C of \cite{roberts1993convex}]\label{lemma:derivative-of-convex}
    Assume $g: [0,1]\mapsto \mathbb{R}$ is convex. Then the left and right derivatives, denoted by $g^{(1)}_-$ and $g^{(1)}_+$ respectively, exist and non-decreasing on $[0,1]$\footnote{The original statement is provided for any open interval $I \subset [0,1]$. This result was extended to the entire (closed) interval in Problem C (page 7) of \cite{roberts1993convex}.}.
\end{lemma}

\begin{proof}[Proof of \Cref{prop:k-bv}]
\begin{figure}
  \centering
  \includegraphics[width=3.2in]{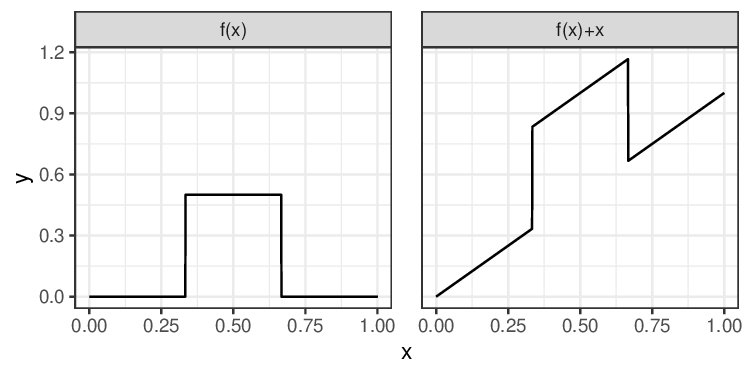}
  \caption{A counterexample of a function $f$ whose total variation is $5L$ but which is not in $\mathcal{F}(1, L)$. The figure is shown for $L = 1/5$.}
  \label{fig:counter-example}
\end{figure}

    First, we provide the corresponding result for $k=1$. For any partition $\{z_i\}_{i=1}^p$ of $[a,b]$ and $f\in\mathcal{F}^\dagger(1,L;[a,b])$, it follows that 
    \begin{align*}
        &\sum_{i=1}^p |f(z_{i+1})-f(z_i)| \\
        &\qquad\le \sum_{i=1}^p |g(z_{i+1})-g(z_i)| + L(z_{i+1} - z_{i}) = \sum_{i=1}^p g(z_{i+1})-g(z_i) + L(z_{i+1} - z_{i})
    \end{align*}
    where the absolute value around $g$ is removed since $g(x+h)-g(x) \ge 0$ for $h \ge 0$. By telescoping, we obtain $\operatorname{TV}(f)  \le g(1)-g(0)+(b-a)L$. Since $\|f\|_\infty \leq L$ for $f \in \mathcal{F}^\dagger(1,L)$, we deduce that 
    \begin{align*}
        \|g\|_{\infty} \leq \|f\|_{\infty} + \|Lx\|_{\infty} \leq (1+|a|+ |b|)L. 
     \end{align*}
     Hence, we have 
     \begin{align*}
\operatorname{TV}(f) \le 2\|g\|_\infty + (b-a)L \le (2 +b-a+
2|a| + 2|b|)L.
     \end{align*}
     
    This concludes that $f\in \mathcal{F}^\dagger(1,L)$ has a bounded variation. 
    
    For $k\ge 2$, the argument is similar. Lemma~\ref{lemma:k-convex-derive} implies that any function $g$ such that $g \in \mathcal{C}(k)$ is $(k-2)$-times differentiable. Importantly, the $(k-2)$th derivative is convex. By the monotonicity result of Lemma~\ref{lemma:derivative-of-convex},
    we have $\operatorname{TV}(f^{(k-1)}) \le g_-^{(k-1)}(1)-g_+^{(k-1)}(0)+(b-a)L$. Since we take the intersection between $\mathcal{F}(k,L)$ and $\{f: \|f^{(k-1)}\|_{\infty} \le L\}$, which is stronger than assuming bounded essential supremum, we have 
    \[\operatorname{TV}(f^{(k-1)}) \le g_-^{(k-1)}(1)-g_+^{(k-1)}(0)+ (b-a)L \le (2+b-a+2|a| + 2|b|) L.\]
    This concludes the inclusion. 
    
    For the strict inclusion, we provide a counterexample. When $k=1$, $[a,b] = [0,1]$, consider the following function:
    \begin{align*}
        f(x) = 5L/2\times 1(x\in [1/3, 2/3]).
    \end{align*}
    This function has a total variation of $5L$ with ``jumps" at $x=1/3$ and $x=2/3$ (see the left panel of Figure~\ref{fig:counter-example} for $L=1$). However, $g := x\mapsto f(x)+5Lx/2$ is not monotone in the neighborhood of $x=2/3$ (see the right panel of Figure~\ref{fig:counter-example} for $L=1/5$). More specifically, we let $x_- \in (5/12, 2/3)$ and $x_+ \in (2/3, 11/12)$, and then we have 
    \[\Delta_{x_+-x_-}^1(g, x_-) = g(x_+)-g(x_-) < 0.\]
    Hence, $g$ is not included in $\mathcal{C}(1)$ and thus $f \notin \mathcal{F}(1,L)$. For $k\ge 2$, we consider the case where the $(k-1)$th weak derivative is given by $f$. By a similar argument, this contradicts the $k$-monotonicity of $g$ in view of Lemma~\ref{lemma:derivative-of-convex}. This concludes the proposition.
    
    Another intuition is that any function with bounded total variation can be expressed as a difference between two arbitrary non-decreasing functions. On the other hand, our class can only be expressed as a difference of non-decreasing and linear functions. 
\end{proof}

\begin{proof}[Proof of \Cref{prop: BV approximation}]
    We first apply \cite[Section 6.6.2, Theorem 2]{evans2018measure} to claim that there exists a $L$-Lipschitz function $f_L$ such that $f_L = f_0$ except for a set $A\subset [a,b]$ whose Lebesgue measure is no larger than $\epsilon(L) = C\operatorname{TV}(f)L^{-1}$. Here, $C>0$ is an absolute constant. This implies
    \begin{equation}\label{eq: bound L2 BV}
\int_a^b\left(f_L(x)-f_0(x)\right)^2 d x=\int_A\left(f_L(x)-f_0(x)\right)^2 d x \leq C\left(\left\|f_L\right\|_{\infty}+\left\|f_0\right\|_{\infty}\right)^2 \mathrm{TV}(f_0) L^{-1}
\end{equation}
Now we show that $f_L$ is bounded. 

Since the Lebesgue measure of $A$ is bounded by $\epsilon(L)$, we claim that for any $x\in A$, there exists a $y \notin A$ such that $|x-y|\leq \epsilon(L)/2$ (otherwise one can construct a large interval centered at $x$ not containing any $y\notin A$). 

We analyze the McShane extension \citep[Section 3.1, Theorem 1]{evans2018measure} of $f_L$ to $A$. It states that given a function $f_L$ that is $L$-Lipschitz on $[a,b]\backslash A$, we can construct a $L$-Lipschitz function over the whole $[a,b]$ by defining
$$
f_L(x) = \inf _{y \notin A}\{f_L(y)+L|x-y|\}.
$$

This McShane extension $f_L$ is also $L$-Lipschitz and satisfies
\begin{equation*}
\begin{aligned}
\left\|f_L\right\|_{\infty} &\leq \max _{y \notin A}\left|f_L(y)\right|+L \cdot \epsilon(L) / 2 \\
& \leq M+C \operatorname{TV}(f_0) \leq C_1 M.
\end{aligned}
\end{equation*}
Continue \eqref{eq: bound L2 BV}, we conclude 
$$
\int_a^b\left(f_L(x)-f_0(x)\right)^2 d x \leq C_2 M^2 \operatorname{TV}(f_0) L^{-1}.
$$
\end{proof}

\begin{corollary}[BV truth for \texttt{MEAN}]\label{cor: BV case} 
Assume the same setting as in \Cref{cor:low-complexity}, except that $f_0 \in \operatorname{BV}(1, M; [0,1])$. Additionally, suppose that $P_X$ admits a bounded density with respect to the Lebesgue measure on $[0,1]$. Let $\mathcal{L} = [0,n^{2/9} \log^{-4/9} n]$.

Then, for any $\epsilon\in(0,1)$ and large enough $n$, we have
\begin{equation*}
\left\|\widehat{f}_n-f_0\right\|^2 \leq C_\epsilon\left(\frac{\log ^{4 / 9} n}{n^{2 / 9}}+\log ^{5 / 9} n \cdot n^{2 / 9} \mathrm{R}_n\right)
\end{equation*}
with probability greater than $1-\epsilon$. Here, $C_\epsilon \in (0, \infty)$ is a constant depending on $\epsilon, \|f_0\|_\infty, \sigma^2$ as well as constants in \ref{as:q-moment} or \ref{as:beta-weibull}, and $\mathrm{R}_n$ is defined in \Cref{eq:rate_under_LSE}.
\end{corollary}
\begin{proof}
The proof of \Cref{cor: BV case} is similar to that of \Cref{cor:worst-case}, both need \Cref{th: general_isotonic}:
\begin{equation*}
\left\|\widehat{f}_n-f_0\right\|^2 \leq \inf _{L \in \mathcal{L}} \inf _{1 \leq m \leq n}\left\{\inf _{f \in \mathcal{F}_m(1, L)}\left\|f-f_0\right\|^2+\mathrm{R}_{L,m}^{\mathrm{Est}}\right\}+\mathrm{R}_{\mathrm{iso}}^{\mathrm{CV}}.
\end{equation*}
Consider an $l_n$-Lipschitz function $f_{l_n}$, we bound $\left\|f-f_0\right\|^2$ by
\begin{equation*}
\left\|f-f_0\right\|^2 \lesssim\left\|f-f_{l_n}\right\|^2+\left\|f_{l_n}-f_0\right\|^2.
\end{equation*}
Then we have 
\begin{equation}\label{eq: bv before plug in}
\left\|\widehat{f}_n-f_0\right\|^2 \lesssim \inf _{L \in \mathcal{L}} \inf _{1 \leq m \leq n}\left\{\inf _{f \in \mathcal{F}_m(1, L)}\left\|f-f_{l_n}\right\|^2+\mathrm{R}_{L,m}^{\mathrm{Est}}\right\}+\mathrm{R}_{\mathrm{iso}}^{\mathrm{CV}}+\left\|f_{l_n}-f_0\right\|^2.
\end{equation}
In the proof of \Cref{cor:worst-case} we have established that 
\begin{equation*}
\left\|f-f_{l_n}\right\|^2 \leq\left(\left\|f_{l_n}\right\|_{\infty}^2+l_n\right)^2 m^{-2}.
\end{equation*}
Applying \Cref{prop: BV approximation}, we also know that 
\begin{equation*}
\left\|f_{l_n}-f_0\right\|^2 \leq C\left(f_0, P_X\right) l_n^{-1}
\end{equation*}
for $P_X$ with a bounded density function.

Plug these quantities into \Cref{eq: bv before plug in} (omitting $\mathrm{R}_{\text {iso }}^{\mathrm{CV}}$ for now), we need to find $m,l_n$ that minimizes
\begin{equation*}
\left(\left\|f_{l_n}\right\|_{\infty}^2+l_n\right)^2 m^{-2}+C_\epsilon m \sigma^2 B_n^2 \log ^2(n) n^{-1}+C\left(f_0, P_X\right) l_n^{-1} .
\end{equation*}

One should use $m = n^{1/3}$, $L_+ = l_n = n^{2/9}\log^{-4/9} n$ to balance the three terms above. These choices give
\begin{equation*}
\begin{aligned}
\left\|\widehat{f}_n-f_0\right\|^2 & \lesssim C_\epsilon \log ^{4 / 9} n \cdot n^{-2 / 9}+\mathrm{R}_{\mathrm{iso}}^{\mathrm{CV}} \\
& \lesssim C_\epsilon\left(\log ^{4 / 9} n \cdot n^{-2 / 9}+\log ^{5 / 9} n \cdot n^{2 / 9} \mathrm{R}_n\right).
\end{aligned}
\end{equation*}
\end{proof}

\newpage
\section{Additional proofs for multivariate decomposition}\label{app: proof of coordinate-deriv}
\begin{proof}[Proof of \Cref{prop:coordinate-deriv}]
    For each $j \in [d]$, it follows that 
    \begin{align*}
        \frac{\partial^{k_j}}{\partial t^{k_j}} g_L (x+te_j) &=\frac{\partial^{k_j}}{\partial t^{k_j}} \left(f(x+te_j) +  \sum_{i=1}^d \frac{L_i}{(k+1)!}(x+te_j)^{k+1}\right)\\
        &=\frac{\partial^{k_j}}{\partial t^{k_j}} \left(f(x+te_j) +  \sum_{i\neq j} \frac{L_i}{(k_i+1)!}x_{[i]}^{k_i+1}+ \frac{L_j}{(k_j+1)!}(x_{[j]}+t)^{k_j+1}\right)\\
        &= f_{j,x}(t) + L_j (x_{[j]}+t).
    \end{align*}
    It remains to show that the univariate function $t \mapsto \frac{\partial^{k_j}}{\partial t^{k_j}} g_L (x+te_j)$ is non-decreasing in $t \in\mathbb{R}$. From the derivation above, $\frac{\partial^{k_j}}{\partial t^{k_j}} g_L (x+te_j)$ equals $f_{j,x}(0) + L_j x_{[j]}$ when $t=0$. Thus for $t > 0$, we have  
    \begin{align*}
         f_{j,x}(t) + L_j (x_{[j]} + t) - (f_{j,x}(0) +  L_j x_{[j]})  \ge -L_j |t| + L_j t \ge 0.
    \end{align*}
    Hence, this concludes that $g_L$ is coordinate-wise $(k+1)$-monotone. 
\end{proof}

\begin{proof}[Proof of \Cref{prop:multivariate-convex}]
A multivariate function $g : \Omega\mapsto \mathbb{R}$ is convex if and only if 
\[(y-x)^{\top}\left(D^1 g(y) - D^1 g(x)\right) \ge 0\]
for any $x, y \in \Omega$. By definition of $g_L$ and $f$, it follows that 
    \begin{align*}
        (y-x)^{\top}\left(D^1 g_L(y) - D^1 g_L(x)\right) &=(y-x)^{\top}\ \left(D^1 f(y) - D^1 f(x)\right) + L \|x-y\|_2^2 \\
        &\ge - \|y-x\|_2 \left\|D^1 f(y) - D^1 f(x)\right\|_2 + L \|x-y\|_2^2 \\
        &\ge - L \|y-x\|_2^2+ L \|x-y\|_2^2  \ge 0
    \end{align*}
which follows from H\"{o}lder's inequality and the assumed Lipschitz continuity. Thus, the function $g_L$ is convex.
\end{proof}

\newpage
\section{Additional details for additive models}\label{supp:additive}
This section presents more details for the additive model discussed in \Cref{example:additive-index}. 
\subsection{A Backfitting Algorithm}\label{supp: additive algorithm}
In Algorithm~\ref{algo:additive}, we present an estimation procedure of $f_0$, assuming it is additive and each component is Lipschitz. Its overall structure remains identical to the one in \Cref{section:general-estimation}. We implement backfitting---an iteration over $d$---in step 2 until the empirical risk converges. 

\begin{algorithm}
\caption{Additive Lipschitz Function Estimation using Monotone Estimators}\label{algo:additive}
\begin{algorithmic}
\Require Observation sequence $\{(X_i, Y_i)\}_{i=1}^n \subset \mathbb{R}^d \times \mathbb{R}$. A candidate vector set $\mathcal{L} \subset \mathbb{R}^d$.
\Ensure An estimator of $f_0(x) = \E[Y|X=x]$. 
\begin{enumerate}
    \item Initialize estimators $\widehat g_{i}(x)$ for $i = 1, 2, \ldots, d$. These can, for instance, be constant functions. Randomly split $\{1,2,\ldots, n\}$ into two disjoint subsets: $\mathcal{I}_1$ and $\mathcal{I}_2$.
    \item \label{step:coordinate-decent}For each $\ell \in \{1, 2, \ldots, d\}$ and $L \in \mathcal{L}$, compute an isotonic regression estimator based on $(X_{i,[\ell]}, \, Z_{i, L} -\sum_{j\neq \ell}\hat g_{j}(X_{i,[j]}))_{i=1}^n$ where $Z_{i, L} := Y_i +L^{\top} X_i$.

    Specifically, we minimize the LSE for each coordinate
    \[\widehat g_{\ell, L}:= \argmin_{g\in\mathcal{C}(1)}\sum_{i\in \mathcal{I}_1}\left\{\left(Z_{i, L}-\sum_{j\neq \ell}\widehat g_{j}(X_{i,[j]})\right)-g(X_{i,[\ell]})\right\}^2\]
iteratively while keeping the estimators of the remaining coordinates fixed.
    \item Repeat step 2 until the empirical risk associated with $\mathcal{I}_1$ given by 
    \[\sum_{i\in \mathcal{I}_1}\left(Z_{i, L}-\sum_{j=1}^d \widehat g_{j,L}(X_{i,[j]})\right)^2\]
    converges. Define the resulting estimator for each $L$ as $\widehat f_{L}(x) = \sum_{j=1}^d \widehat g_{j,L}(x) - L^{\top}x$.
     \item Select a vector $\widehat L$ minimize LSE associated with $\mathcal{I}_2$:
    \[\widehat L := \argmin_{L \in \mathcal{L}} \sum_{i\in\mathcal{I}_2} (Y_i-\widehat f_{L}(X_i))^2.\]
    \item Return the final estimator $\widehat f_{\widehat L}(x) :=  \sum_{j=1}^d \widehat g_{j, \widehat L}(x) - \widehat L^{\top}x$.
\end{enumerate}
\end{algorithmic}
\end{algorithm}

\subsection{Oracle property of the additive estimator}\label{suppsec:oracle-additive}
Below, we claim that, under the additive model, the convergence rate of each additive component informs the convergence rate of overall procedure. This follows from the so-called oracle property of additive estimator, which reduces its analysis to each univariate component \citep{mammen2007additive, guntuboyina2018nonparametric}. In \Cref{example:additive-index}, we discussed conditional mean functions with an additive structure that can be decomposed as: 
\begin{align*}
    f_0(x) = \sum_{j=1}^d f_j(x_{[j]}) &= \sum_{j=1}^d g_{j, \alpha_j}(x_{[j]}) -  \{\alpha_j/(k_j+1)!\} x_{[j]}^{k_j+1}\\
    &= \mu^* + \sum_{j=1}^d  g^*_{j, \alpha_j}(x_{[j]}) - \{\alpha_j/(k_j+1)!\} x_{[j]}^{k_j+1},
\end{align*}
where each shape-restricted $g_{j,\alpha_j}^*$ is a function in $\mathcal{C}(k_j)$ for some $k_j \geq 1$. The constant offset $\mu^*$ is introduced to ensure that $\int  g^*_{j, \alpha_j}(x)\, dP_X(x)=0$ for all $j = 1, \ldots, d$. This is a common requirement for the identification \citep{mammen2007additive}. The additive components of the estimator based on Algorithm~\ref{algo:additive} can be shown to converge (see Theorem 2 of \cite{mammen2007additive}) to the following empirical risk minimizer:
\begin{align*}
    (\widehat \mu,  \widehat g_1, \ldots \widehat g_d) := \argmin\sum_{i=1}^n\left\{Y_i-\mu+\sum_{j =1}^d\{\alpha_j/k_j!\} x_{[j]}^{k_j}-\sum_{j =1}^d g_{j}(X_{i,[j]})\right\}^2
\end{align*}
where $\argmin$ is taken over $\mu \in \mathbb{R}$, $g_{j}\in \mathcal{C}(k_j)$ under the constraint $\sum_{i=1}^n g_j(X_{i,[j]}) = 0$ for $j = 1, \ldots, d$. Analyzing the performance of each $\widehat g_\ell$ may seem challenging since all components are dependent, sharing the same observed data. However, its behavior can be analyzed as if it was constructed with the knowledge of $\mu^*$ and $g_{j,\alpha_j}^*$ for $j \neq \ell$. 

We define the oracle estimator of the $\ell$-th component as:
\begin{align*}
    \widetilde g_{\ell} := \argmin_{g\in\mathcal{C}(k_\ell)}\,\sum_{i=1}^n\left\{Y_i-\mu^*+\sum_{j =1}^d\{\alpha_j/k_j!\} x_{[j]}^{k_j}-\sum_{j \neq \ell} g^*_{j, \alpha_j}(X_{i,[j]})- g(X_{i,[\ell]})\right\}^2.
\end{align*}
This estimator is not practically feasible as it assumes the knowledge of $\mu^*$ and $g^*_{j, \alpha_j}$ for $j \neq \ell$.  Under the assumption that the covariate space $\Omega \subseteq \mathbb{R}^d$ forms a Cartesian product set $\mathcal{X}_1 \times \ldots \times \mathcal{X}_d$ and $\mathcal{X}_i \subseteq \mathbb{R}$ for all $i=1,\ldots,d$, Lemma 3.1 of \cite{guntuboyina2018nonparametric} states that $\widehat g_\ell = \widetilde g_\ell\, \text{ for each } \,\ell=1,\ldots, d.$ This is known as the oracle property of backfitting. With this general result, we can extend the identical proof of \Cref{th: general_isotonic} to additive model for $k_j=1$: our theoretical analysis can focus on $\widetilde g_{j}$ instead of $\widehat g_j$.
We then apply the proof of \Cref{th: general_isotonic} to each univariate component. Finally, the overall estimation error can be bounded by the summation of each component. When the covariate dimension $d$ does not vary with sample size $n$, the overall convergence rate is identical to the univariate results.
Similarly, for $k = 2$ and beyond, we expect the additive estimator's convergence rate to follow once the univariate result akin to \Cref{th: general_isotonic} for general $k$ is obtained.

\newpage
\section{Additional numerical studies}\label{supp:simulation}
\subsection{Details on numerical studies from the main text}\label{supp: numerical study detail}
The following estimators are considered in the numerical study from the main text.
\begin{enumerate}
    \item Kernel ridge regression (\texttt{KRR}): The kernel function is given by $K(x,z):= 1+\min(x,z)$, which corresponds to the first-order Sobolev space (see, for instance, Example 12.16 of \cite{wainwright2019high}). We select the penalty parameter of ridge regression among $10$ candidates. Due to the computationally intensive nature of KRR, we do not explore finer choices of the tuning parameters. 
    \item Gradient boosting machines (\texttt{GBM}): The shrinkage parameter is set to $0.01$, and we choose the maximum depth of each tree from the set $\{2, 5\}$. The total number of trees is selected from $\{100, 1000, 2000, 4000, 8000\}$. The remaining parameters are set to their default values according to the \texttt{GBM} library in \texttt{R}.
    \item Random forest regression (\texttt{RF}): The number of trees is selected from $\{50, 100, 500, 1000,$ $5000\}$. We use the estimator implemented by R package \texttt{randomForest} \citep{randomforrestPackage}.
    \item Penalized sieve estimator with cosine basis (\texttt{Sieve}): We employ 50 basis functions and select the penalization tuning parameter from (approximately) 100 default grids. The estimators are realized using R package \texttt{Sieve} \citep{zhang2022regression}.
\end{enumerate}

\subsection{Details on regression functions}\label{supp: regression description}
\subsubsection{Univariate cases}
We provide the details on the specification of the regression functions in the numerical study from the main text.
\subsection*{Scenario 1} 
We examine a Lipschitz function defined as:
\begin{align}
    f_1(x) &:= (1-3x)\times 1(x\in [0,1/3]) \label{eq:f1} \\
    &\qquad + (-1+3x)\times 1(x\in [1/3,2/3]) + (3-3x)\times  1(x\in [2/3,1]).\nonumber
\end{align}
The proposed estimator, along with other nonparametric regression estimators, is expected to converge at a rate of $n^{-2/3}$ in terms of MSE.
\subsection*{Scenario 2}
We consider a low complexity case where the proposed estimators are expected to be adaptive, converging at a parametric rate of $n^{-1}$ (up to a logarithmic term). Define an $m$-piecewise constant function $M_m(x) := \sum_{i=1}^m i \times  1(x\in [(i-1)/m,i/m])$. The proposed estimator is anticipated to achieve an adaptive rate when the true function is
\begin{equation}
    f_2(x; m, \beta) := M_m(x) + \beta x.\label{eq:f2}
\end{equation}
We choose $m=3$ and $\beta=1$.

\subsection*{Scenario 3}
The next two scenarios focus on the application of the convex regression. We implement the algorithm in \Cref{section:general-estimation} to estimate these functions with $k=2$ and $d=1$. The estimator for the $2$-monotone function is simply a convex regression estimator, which we use the implementation based on the \texttt{cobs} library in \texttt{R}. The following example is a smooth function, defined as:
\[f_3(x; \gamma) := \sin(\gamma(2x-1)).\]
We select $\gamma = 4$. The proposed estimator is anticipated to converge at a rate of $n^{-4/5}$ in view of Theorem 3.1 of \cite{kuchibhotla2022least} and Theorem 3 of \cite{han2018robustness}. 
\subsection*{Scenario 4}
The final example illustrates another scenario where the proposed estimator is expected to be adaptive. See \cite{guntuboyina2015global, han2018robustness} on the low complexity adaption of the convex LSE. Specifically, the estimator is designed to adapt to any function that can be decomposed as a sum of a convex $m$-piecewise linear function and a quadratic function. We define $C_m(x)$ as a convex $m$-piecewise linear function with $1/m$ equally sized segments over $X$, and the slopes are $(-1, 0, 1, \dots, m-2)$. Additionally, we enforce the condition $C_m(0)=0$, thereby defining a unique convex $m$-piecewise linear function. We generate observations from one such regression defined as:
\[f_4(x; m, \beta) :=C_m(x) + \beta x^2.\]
We consider the case $m=3$ and $\beta=1$. 

\subsubsection{Additive cases}
We now provide the regression functions from the additive case.
\begin{itemize}
    \item Scenario 1 (2d): Each additive component is identical to Scenario 1 for the univariate case, which corresponds to the ``worst case'' for $\mathcal{F}(1,L)$ class.
    \[f(x) := f_1(x_{[1]}) - f_1(x_{[2]}),\]
    \item Scenario 2 (2d): 
    \[f(x) := f_2(x_{[1]}; 3, 1) - f_2(x_{[2]}; 3, 1),\]
\end{itemize}
where the component functions $f_1,f_2$ are defined in \eqref{eq:f1} and \eqref{eq:f2}. Similar to the univariate cases, we anticipate that the proposed method will converge essentially at a rate of $n^{-2/3}$ for Scenario 1 and $n^{-1}$ for Scenario 2. We discuss the justification behind this argument in \Cref{suppsec:oracle-additive}.

We also consider two 5-dimensional examples:

\begin{itemize}
    \item Scenario 3 (5d): 
    \[f(x) := f_1(x_{[1]}) - f_1(x_{[2]}) + x_3- x_4 + 1,\]
    \item Scenario 4 (5d): 
    \[f(x) := f_2(x_{[1]}; 1, 0) + f_2(1-x_{[2]}; 3, 3) +f_2(x_{[3]}; 3, 3) + f_2(1-x_{[4]}; 1, 3) + f_2(x_{[5]}; 1, 3).\]
\end{itemize}

\subsection{Is the estimator tuning-parameter free in practice?}\label{sec:tuning-parameter-free}
\begin{figure}[!htbp]
    \centering
    \includegraphics[width=5.5in]{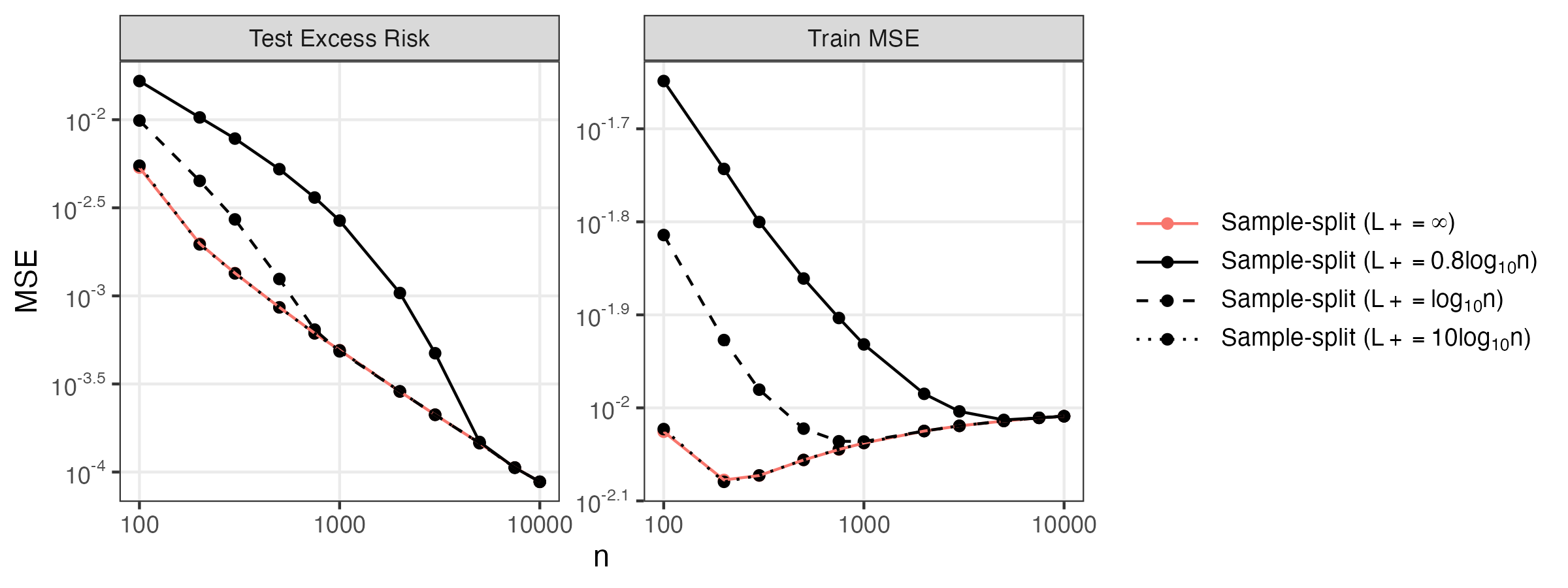}
    \caption{The performance of the proposed estimator with sample-splitting, evaluated as the upper bound $L_+$ varies. The X-axis represents the sample size $n$, and the Y-axis shows the MSEs computed on both training and test sets; both axes are on a logarithmic scale. Data are generated under Scenario 1. The right panel demonstrates that setting $L_+ = \infty$ does not result in overfitting relative to $L_+ = 10\log_{10} n$. The left panel suggests that larger values of $L_+$ may reduce out-of-sample error. While our theoretical guarantees assume $L_+ = O(\log n)$, we recommend using $L_+ = \infty$ in practice.}\label{fig:L-infinite}
\end{figure}

For technical reasons, we assume that $\mathcal{L}$ is a bounded set. This introduces an additional parameter $L_+ := \sup_{L \in \mathcal{L}}|L|$, which may appear as a tuning parameter. In practice, we find that the choice of $L_+$ has little impact on performance, and the set $\mathcal{L}$ can effectively be unbounded. To demonstrate this, we generate $n$ observations under Scenario 1 and display both in-sample and out-of-sample MSEs of the proposed method across varying values of $L_+$. Figure~\ref{fig:L-infinite} shows the average MSEs over $500$ replications. Here test excess risk is the mean-squared distance between the estimator and the truth. Notably, the estimator with $L_+ = \infty$ performs almost identically to $L_+ = 10$. This suggests that once $L_+$ becomes sufficiently large, further increases do not lead to overfitting. This robustness relies critically on selecting the value of $\widehat{L}$ through a sample-splitting procedure, and this conclusion no longer holds when $\widehat{L}$ is chosen without sample-splitting.

\subsection{On the effect of sample-splitting in the choice of $\widehat L$}\label{sec:est-wo-splitting}
As discussed in the main text, sample-splitting is a crucial aspect of the proposed estimation procedure. Without sample-splitting, the estimator eventually interpolates all observations and hence achieve zero training error. One might still wonder what happens if we purposely select a small value of $L_+ \ge 0$ where $\mathcal{L} = [0, L_+]$, and remove sample-splitting entirely from the procedure. We demonstrate that this leads to degenerate behavior where the estimator always picks the largest value of $\mathcal{L}$, effectively reducing to the case where the value of Lipschitz constant is known. To demonstrate this, we generate $n$ observations under Scenario 1 and plot the distribution of selected $\widehat L$ without performing sample-splitting. The different choices of $L_+$ are shown  as solid, dotted, and dashed black lines. For comparison, the behavior of the estimator \emph{with} sample-splitting is shown in red, with the red dashed line indicating the ``true'' Lipschitz constant of 3. Results are based on 500 replications. We observe that without sample-splitting, the estimator always selects the maximum $L_+$ across all 500 runs (hence explains the appearance of points rather than boxplots). Interestingly, with sample-splitting, the selected $\widehat{L}$ tends to concentrate around the true Lipschitz constant. We do not have a theoretical explanation for this favorable property.
\begin{figure}[!htbp]
    \centering
    \includegraphics[width=3.5in]{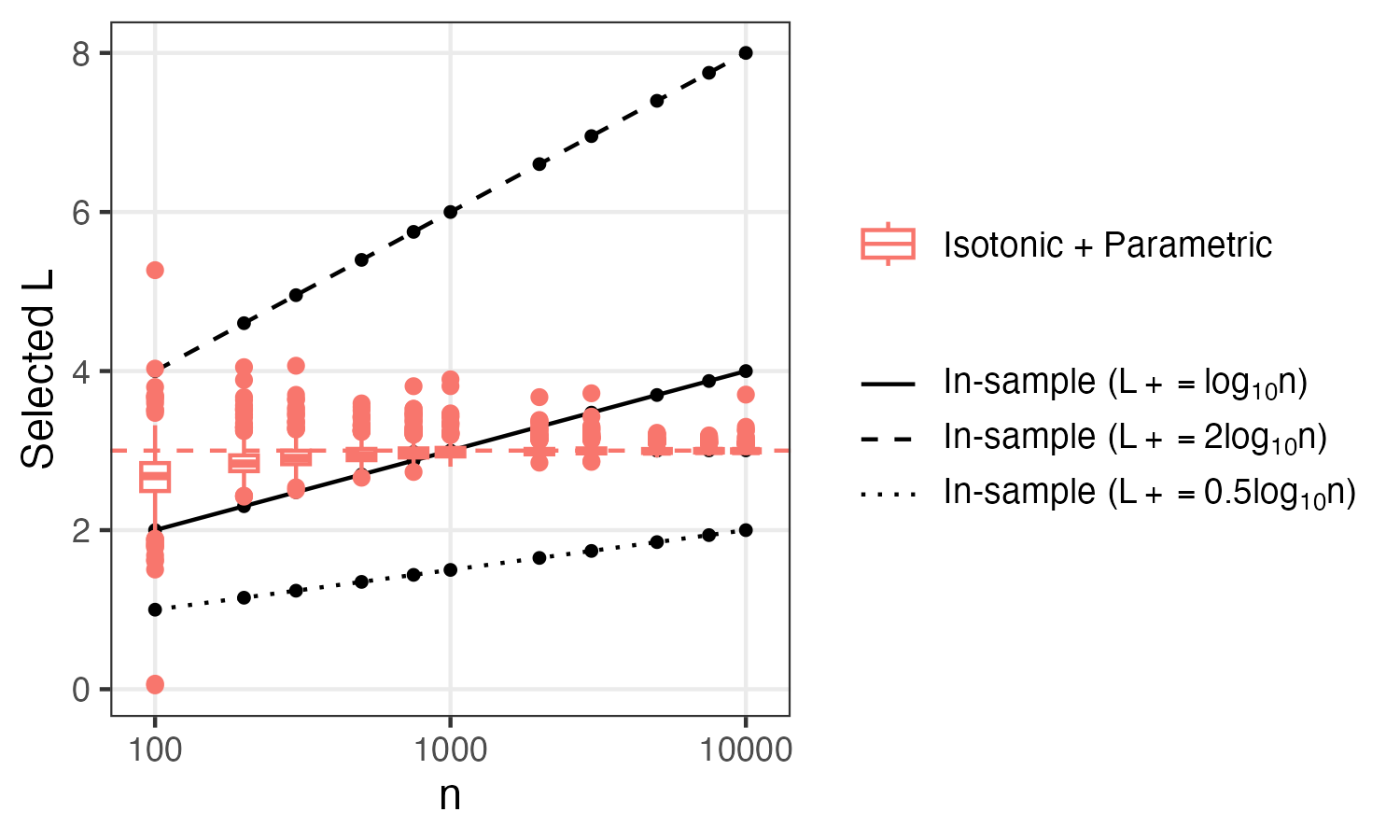}
    \caption{Box plots of the selected values of $\widehat{L} \in [0, L_+]$ without the sample-splitting procedure. The X-axis shows the sample size $n$ (on a logarithmic scale), and the Y-axis shows the selected values of $\widehat{L}$. Observations are generated under Scenario 1, where the true Lipschitz constant is 3. The red box plot corresponds to the estimator with sample-splitting. The black solid, dashed, and dotted lines represent the estimators without sample-splitting, each with a different value of $L_+$. Results are based on 500 replications. For the estimators without sample-splitting, we observe black dots rather than full box plots because $\widehat{L}$ is always equal to the largest allowable value, $L_+$.}\label{fig:L-in-sample-box}
\end{figure}

\begin{figure}[!htbp]
    \centering
    \includegraphics[width=5.5in]{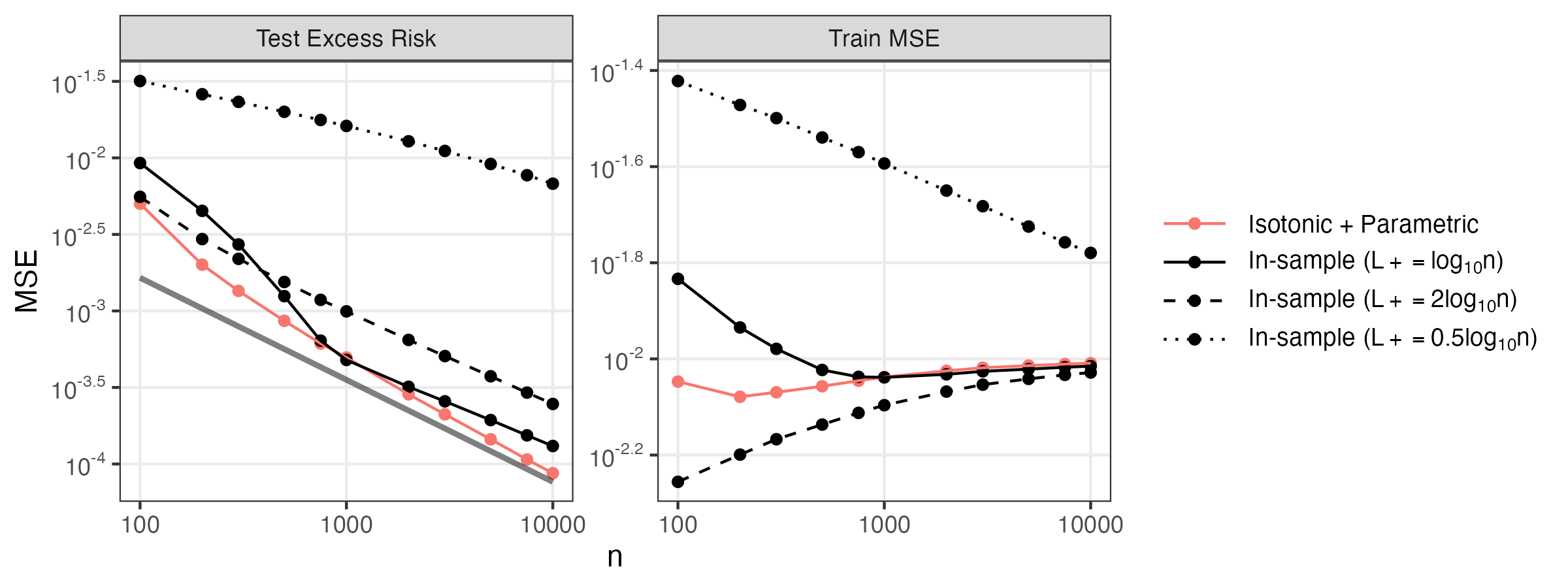}
    \caption{The performance of the estimator without sample-splitting as the values of $L_+$ change. The X-axis shows the sample size $n$ and the Y-axis shows the mean squared errors based on both training and test data. Both X- and Y-axes are on the logarithmic scale. The observation is generated under Scenario 1. The estimator without sample-splitting is highly sensitive to the choice of $L_+$, performing worse than the method with sample-splitting unless the true Lipschitz constant is known.}\label{fig:L-in-sample-rate}
\end{figure}

Next, we demonstrate that without sample-splitting, the procedure becomes highly sensitive to the choice of $L_+$. We generate $n$ observations under Scenario 1 and display both in-sample and out-of-sample MSEs of the estimator without sample-splitting, as $L_+$ varies among $0.5 \log_{10} n, \log_{10} n, 2 \log_{10} n$. Since the true Lipschitz constant is 3, the choice $0.5 \log_{10} n$ is underspecified for $n \le 10^6$, while $2 \log_{10} n$ is overspecified for $n \ge 10^{3/2}$. The intermediate choice $\log_{10} n$ exactly identifies the true constant when $n=1000$, and is otherwise either under- or over-specified. \Cref{fig:L-in-sample-rate} shows that the estimator without sample-splitting consistently underperforms the proposed method with sample-splitting, except in the case where $L_+ = \log_{10} n$ and $n = 1000$. This is expected, as it corresponds to the idealized setting where the true Lipschitz constant is known, whereas the sample-splitting approach estimates this constant from the data. Base on this observation, we recommend the use of sample-splitting in practice.

\subsection{On the effect of cross-fitting}

\begin{figure}[!htbp]
    \centering
    \includegraphics[width=3.5in]{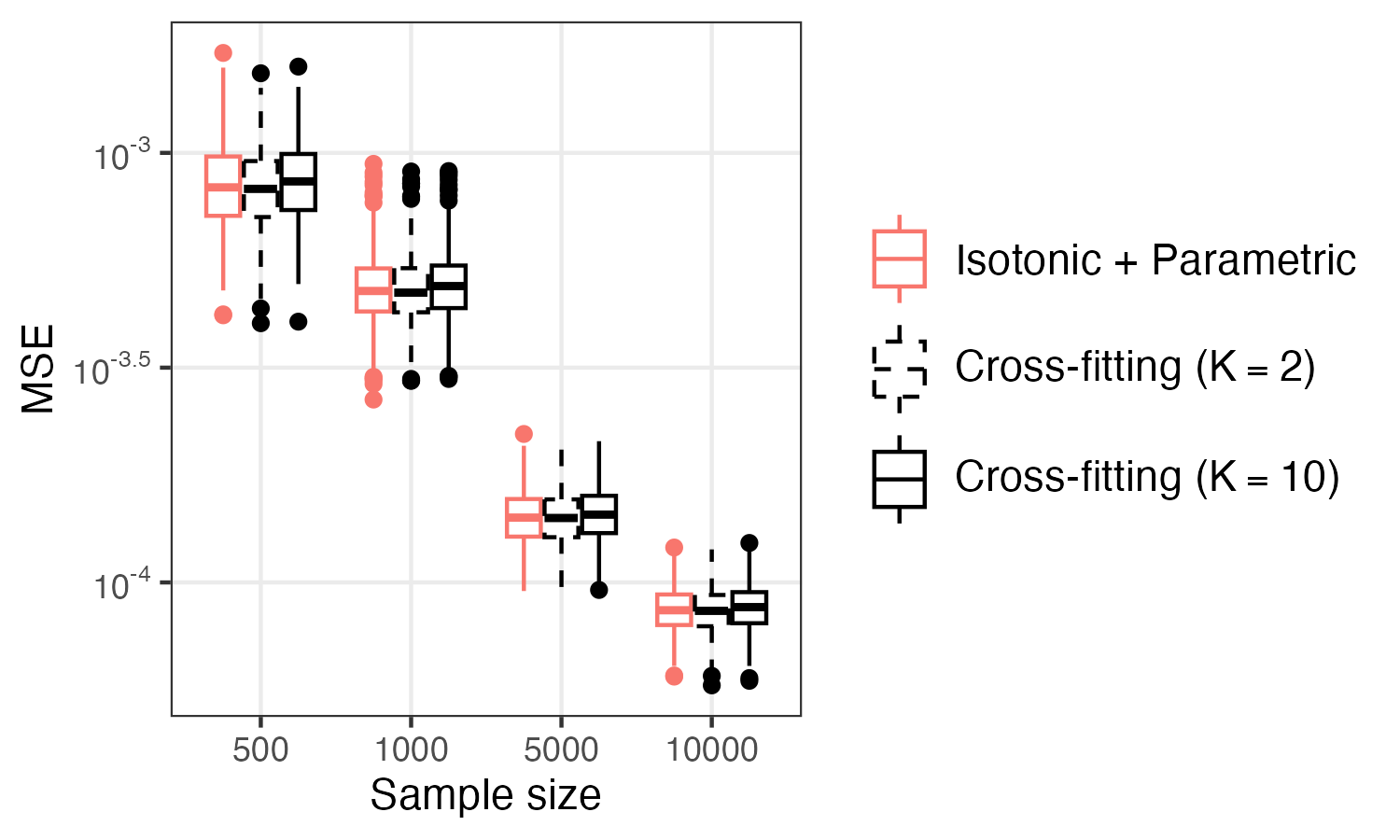}
    \caption{The excess risks of the proposed estimator based on cross-fitting as the number of folds varies. The $X$-axis shows the sample size $n$, and the $Y$-axis shows the excess MSEs evaluated on test data. The data-generating distribution corresponds to Scenario 1. The results are over $500$ replications. While cross-fitting provides a slight reduction in variance, especially for smaller sample sizes, the improvement becomes negligible as $n$ increases. }\label{fig:cross-fitting}
\end{figure}

Despite the aforementioned benefits of sample-splitting, one may still be concerned about a potential loss in efficiency due to reduced effective sample size. To address this question, we implement a cross-fitting version of the proposed method \citep{Bickel1982adaptive}. The procedure is defined as follows:
\begin{enumerate}
    \item Given an integer $K \ge 2$, randomly split $\{1,2,\ldots, n\}$ into $K$ disjoint subsets: $\mathcal{J}_1, \mathcal{J}_2, \ldots, \mathcal{J}_K$.
    \item For each $k \in \{1,2,\ldots, K\}$, obtain the proposed estimator by setting $\mathcal{I}_1 = \cup_{j \neq k} \mathcal{J}_j$ and $\mathcal{I}_2 = \mathcal{J}_k$. Denote the resulting estimator by $\widehat f_{n, k}$.
    \item Return the estimator as $\widehat f_n(x) = K^{-1} \sum_{i=1}^K \widehat f_{n, k}(x)$.
    \end{enumerate}
    We generate $n$ observations under Scenario 1 and evaluate the out-of-sample MSEs as $K$ varies between $2$ and $10$. \Cref{fig:cross-fitting} displays the results based on $500$ replications. For comparison, we include the estimator from the main text (without cross-fitting). The empirical evidence suggests that cross-fitting offers a modest reduction in variance, although the gain is marginal particularly when the sample size exceeds $1000$. In this setting, the improvement from variance reduction is relatively minor and may not justify the additional computational cost in practice. 

\subsection{Additional robustness of the proposed procedures}

As discussed in \Cref{sec:estimator}, the proposed estimator possesses more desirable properties than the standard cross-validation procedures. We provide additional numerical results that demonstrate its robustness. We examine the robustness of the proposed method regarding two aspects: (1) the specific values of the parameter $L$ and (2) the randomness from cross-validation (i.e., data-splitting). 

\begin{figure}[!tbp]
    \centering
    \includegraphics[width=3.5in]{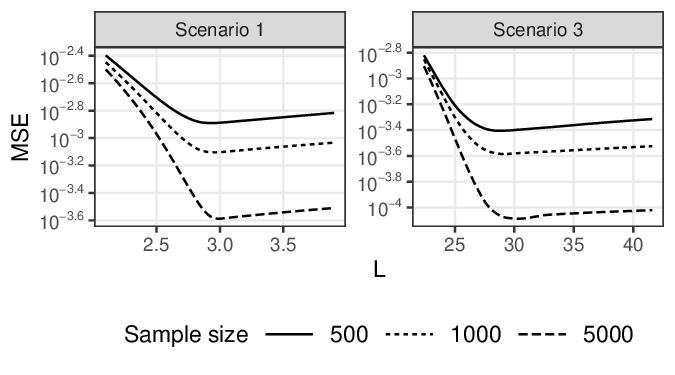}
    \caption{The MSEs of the proposed estimator as the values of $L$ change. The X-axis shows different values of $L$ and the $Y$-axis shows the mean squared errors based on the new data. Only the $Y$-axis is on the logarithmic scale. Different line types represent results for different sample sizes. Two scenarios correspond to the univariate settings defined in \Cref{sec:univariate-setting}. In Scenario 1, the Lipschitz constant is 3, while in Scenario 3, the Lipschitz constant is 32. }\label{fig:L-robust}
\end{figure}

First, we investigate the MSEs of the proposed methods when the parameter $L$ is pre-specified. This means the resulting estimator $\widehat g_L - L x^k/k!$ depends on the data only through the estimated $k$-monotone function $\widehat g_L$. We consider the case with univariate covariates and Scenarios 1 and 3 defined in \Cref{sec:univariate-setting}. Figure~\ref{fig:L-robust} displays the MSEs (in logarithmic scale), based on the new data, as the value of $L$ changes. The results are presented for sample sizes of $n=500, 1000$, and $5000$. To recall, \Cref{prop:k-monotone-decomposition} holds for any $L \ge L_0$ where $L_0$ is the true Lipschitz constant, implying that we expect the MSEs to be robust once the value of $L$ surpasses a certain threshold. In other words, we expect the performance of the estimator to be less sensitive to an over-specified $L$ and thus the overall procedure has a certain degree of robustness. Indeed, Figure~\ref{fig:L-robust} displays that the MSEs decrease as $L$ increases until a certain point, then they begin to increase again. The MSEs for larger values of $L$, however, are not dramatically worse than the underspecified $L$. 

\begin{figure}[!htbp]
    \centering
    \includegraphics[width=5.5in]{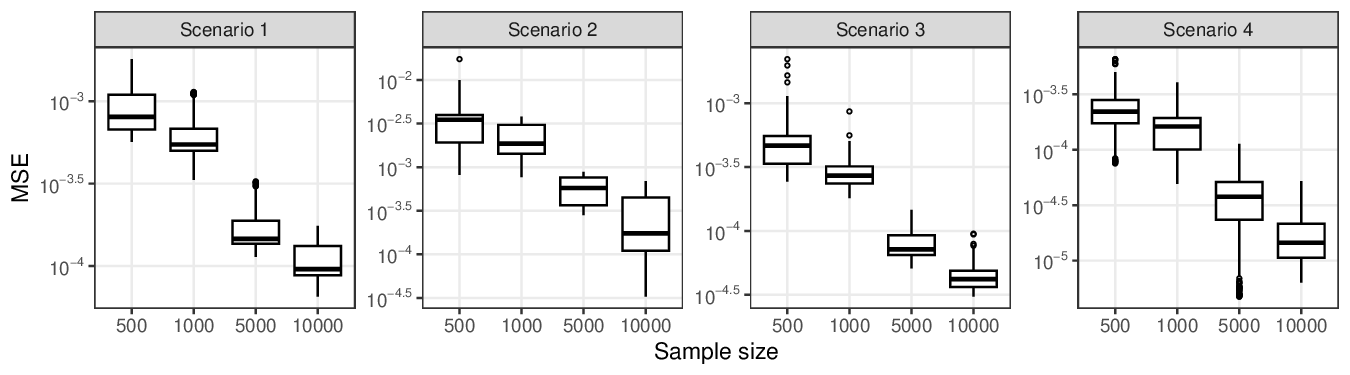}
    \includegraphics[width=5.5in]{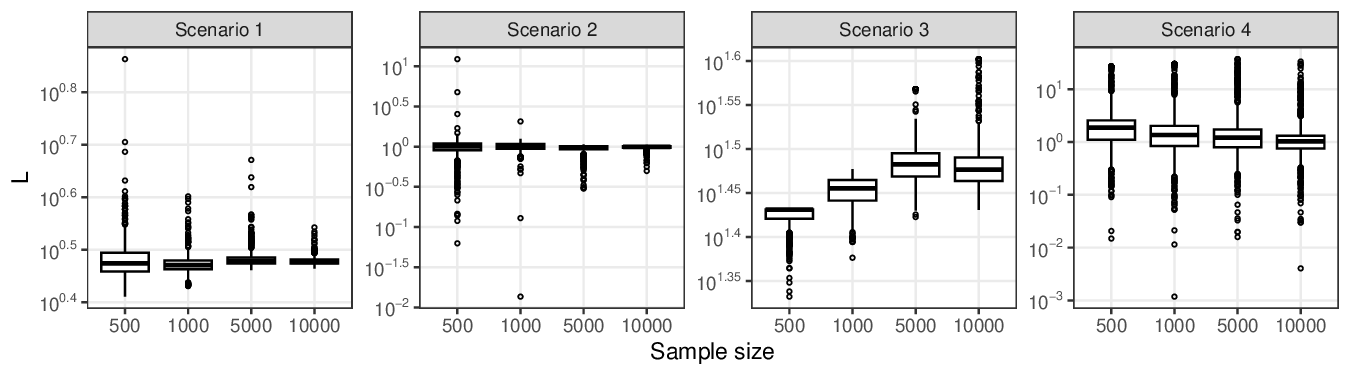}
    \caption{The box plots of the MSEs and the estimated $\alpha$ parameter across different random splits during the cross-validation step. Four scenarios correspond to the univariate settings defined in \Cref{sec:univariate-setting} and the $Y$-axes are on the logarithmic scales. The top row displays that the MSEs of the proposed estimator are fairly concentrated over the variability from the random splits. The bottom row shows the distribution of the $\alpha$ parameter selected by the cross-validation. 
    }\label{fig:CV-robust}
\end{figure}

We also investigate the impact of random split on the performance of the proposed methods. To study this, we generate $n$ observations for $n=500,1000,5000$ and $10000$ from univariate Scenarios 1 to 4. For each given dataset, we compute the proposed estimator 300 times for different cross-validation splits $\mathcal{I}_1$ and $\mathcal{I}_2$ where $\mathcal{I}_2$ contains $\lfloor n/2 \rfloor$ observations. To be precise, between each $300$ replications, the estimator uses the identical observations $\{(X_i, Y_i)\}_{i=1}^n$ and only the random splits $\mathcal{I}_1$ and $\mathcal{I}_2$ differ. The top plot of Figure~\ref{fig:CV-robust} displays the distribution of the MSEs on new data while the bottom plot shows that of $L$ obtained from the cross-validation splits. The $Y$-axes for both plots are on the logarithmic scale. We observe that the MSEs across different splits are concentrated even for small sample sizes. For example, in Scenario 1 with $n=500$, the majority of MSEs fall within the range of $10^{-3.25}$ to $10^{-3}$, indicating very small variability between cross-validation splits. Similar conclusions can be drawn for the other scenarios as well. The bottom plot of Figure~\ref{fig:CV-robust} shows that the interquartile range of the distribution of selected $\alpha$ is very small. Occasionally, we would observe outliers but they do not significantly affect the MSE of the estimators as shown in the top plot of Figure~\ref{fig:CV-robust}.

\subsection{Additional results on adaptive rates}
\begin{figure}[!htbp]
    \centering
    \includegraphics[width=3.5in]{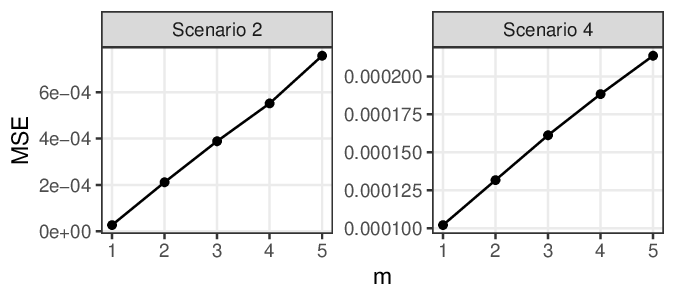}
    \caption{The average MSEs of the proposed estimator for Scenarios 2 and 4 from \Cref{sec:univariate-setting} with $n=5000$ as the number of $m$ pieces increases. The $Y$-axis shows the sample mean of the empirical MSEs over  $300$ repetitions and it is \textit{not} on the logarithmic scale. For scenario 2, the convergence rate based on \Cref{cor:low-complexity} is expected to exhibit a linear relationship with respect to the number of segments $m$. We expect a similar behavior for the convex case in view of the existing results in the literature. See, for instance, \cite{guntuboyina2015global, han2018robustness} among others.}\label{fig:m-pieacewise}
\end{figure}

Finally, we investigate the behavior of the proposed estimator in scenarios where the method is expected to be adaptive to a $m$-piecewise polynomial truth. In particular, Scenario 2 represents non-decreasing $m$-piecewise constant functions with linear functions, and Scenario 3 represents $m$-piecewise convex affines a quadratic function. \Cref{tab:convergence-rate} shows that the rate of convergence of the proposed estimator in these scenarios is expected to behave like $O(m/n)$ (without a logarithmic term), displaying a linear relationship as the number of $m$ increases. Similar behavior is expected for the convex case in view of the result in the literature. See, for instance, \cite{guntuboyina2015global, han2018robustness} among others. To verify this property, we generated $5000$ observations from Scenarios 2 and 4 over the values of $m$ in $\{1,2,3,4,5\}$. The results are presented in Figure~\ref{fig:m-pieacewise}. As expected, the average MSEs over 300 repetitions increase linearly as the value of $m$ increases, which further confirms our theoretical understanding of the estimator.

\newpage
\section{Extension to density estimation}\label{supp:density}

Below, we describe how to integrate the proposed framework into nonparametric density estimation. Let $X_1, \ldots, X_n \in \mathcal{X}\subseteq\mathbb{R}$ be IID observations from an unknown density function $f_0$. We begin by defining $\mathcal{I}_{1}$ and $\mathcal{I}_{2}$ as disjoint index sets such that $\mathcal{I}_{1} \cup \mathcal{I}_{2} = \{1, 2, \ldots, n\}$ with cardinalities $n_1$ and $n_2$, respectively. Two datasets are obtained as follows: $D_1 := \{X_i : i \in \mathcal{I}_{1}\}$ and $D_2 := \{X_i : i \in \mathcal{I}_{2}\}$.
 
We denote by $X_{(1)} < \ldots < X_{(n_1)} < X_{(n_1+1)}=\infty$ the order statistics of the observations in $D_1$. The proposed estimator takes the following form: 
\begin{align}
	\widehat f_L(x) = \exp\left( g(x)- Lx \right) \quad \text{where } g \textrm{ is non-decreasing and } \int \widehat f_L(x) \, dx = 1.\nonumber
\end{align}
We propose an esimation procedure for $L > 0$. We consider the nonparametric maximum likelihood estimator over $\widehat f_L$, corresponding to 
\begin{align}
    g_{\textrm{MLE}} := \argmax_{g \in \mathcal{C}}\, \log \widehat f_L(X_i)\label{eq:original-mle-objective} 
\end{align}
where $\mathcal{C}$ is the set of all non-decreasing functions. 

We claim that it suffices to consider the space of non-decreasing piecewise constant functions that take at most $n_1$ distinct values at each observation $X_1, \ldots,  X_{n_1}$. For any $g \in \mathcal{C}$, we define the following function:
\begin{align*}
    g^*(x) := \begin{cases}
        -\infty & \textrm{when $x < X_{(1)}$} \\
        g(X_{(i)}) & \textrm{when $ X_{(i)} \le x < X_{(i+1)}$ for any $i \in \{1,2, \ldots, n\}$}\\
        g(X_{(n_1)}) & \textrm{when $x \ge X_{(n_1)}$}.
    \end{cases}
\end{align*}
The resulting function $g^*$ is a non-decreasing piecewise constant.  We then define the corresponding density estimator as 
\begin{align}
	\widehat f^*_L(x) := C\exp\left( g^*(x)- Lx \right) \quad \text{where} \quad C \textrm{ is a  normalizing constant.}\nonumber 
\end{align}
The constant $C$ is introduced to ensure that $\widehat f^*_L$ is a density function. Now, we observe that for any $X_i \in D_1$, it follows
\begin{align}
    \log \widehat f^*_L(X_i) = \log C + g^*(X_i)- LX_i  = \log C + g(X_i)- LX_i =  \log C + \log \widehat f_L(X_i).
\end{align}
Therefore, the solution of the original maximum likelihood problem, given by \eqref{eq:original-mle-objective}, is not unique but attained by a non-decreasing piecewise constant function with an appropriate normalizing constant $C \ge 1$.

The original optimization problem is thus equivalent to determining the sequence of numbers $w = (w_1, w_2,  \ldots,  w_{n_1})$ such that:
\begin{align}
	&\argmax_{w}\, \sum_{i \in \mathcal{I}_1} \log \widehat f_L(X_i; w) = \sum_{i \in \mathcal{I}_1} w_i- LX_i \nonumber \quad  \text{subject to} \\
 &\qquad (1) \quad w_1 \le w_2  \le  \ldots  \le  w_{n_1} \quad \text{and} \nonumber\\
 &\qquad (2) \quad \int \widehat{f}_L(x ; w) d x=1 \Longleftrightarrow \sum_{i=1}^{n_1} \exp \left(w_i\right)\left(\frac{\exp \left(-L X_{(i)}\right)-\exp \left(-L X_{(i+1)}\right)}{L}\right)=1.\nonumber
\end{align}
We note that $\exp(-LX_{(n_1+1)})=0$ for any $L > 0$. 

We denote by $w_{\textrm{MLE}}$ the solution of the above optimization, and the estimator for each fixed $L$ is given by $\widehat f_L(x; w_{\textrm{MLE}})$. From now on, we suppress $w_{\textrm{MLE}}$. 

We perform a data-adaptive selection procedure for $L \in \mathcal{L}$, where  $\mathcal{L}$ is a prespecified set that contains positive real. Specifically, let $d F_{n_2}$ be the empirical probability measure for $D_2$ defined as $dF_{n_2} := n_2^{-1}\sum_{i\in \mathcal{I}_2}\delta_{x_i}$, where $\delta_x$ denotes the Dirac measure, which places unit mass at a point $x$. The data-adaptive choice of $L$ is defined as the minimizer of the Kullback-Leibler (KL) divergence between $dF_{n_2}$ and $\widehat f_{L}$. Formally, we define
\begin{align}
	\widehat L := \argmin_{L \in \mathcal{L}}\, \mathrm{KL}(dF_{n_2}, \widehat f_{L}) &= \argmin_{L \in \mathcal{L}}\, \int  \log \left(\frac{dF_{n_2}(x)}{\widehat f_{L}(x)}\right)dF_{n_2}(x)\, dx\nonumber\\
 &=\argmax_{L\in\mathcal{L}}\, \sum_{i \in \mathcal{I}_2} \log \widehat f_L(X_i). \nonumber
	\end{align} 
The final estimator is then given by $\widehat f_{\widehat L}(x)$.
	
Based on the theoretical results presented in this manuscript, we anticipate the following properties of the corresponding estimator. First, this procedure yields a consistent nonparametric density estimator for log-Lipschitz $f_0$, that is, the logarithm of $f_0$ is a Lipschitz function. Second, the estimator is expected to exhibit adaptive convergence rates when the true density is of lower complexity, such that 
\begin{align}
	f_0(x) \propto \exp\left(\sum_{i=1}^m w_i\times 1\{x \in I_i\} - Lx \right)\nonumber,
\end{align}
for $L > 0$, unknown $m \in \mathbb{N}$ and $\{I_i\}_{i=1}^m$ is a non-overlapping partition of the domain of $X$. This includes cases where $f_0$ is an exponential density ($m=0$), and any piecewise exponential densities. 

The examples in the main text are generated from the following true density functions:
\subsection*{Scenario 1}
The data is generated from a Laplace distribution where the true density is defined as $f_0(x) := \frac{1}{2}\exp(-|x|)$. The estimator is expected to converge at the minimax rate. 

\subsection*{Scenario 2}
The data is generated from an exponential distribution where the true density is defined as $f_0(x) := \exp(-x)$. The estimator is expected to converge at the adaptive rate. 
\end{appendices}
\end{document}